\documentclass[draft]{agujournal2019}
\usepackage{url} 
\usepackage{lineno}
\usepackage[inline]{trackchanges} 
\usepackage{soul}

\usepackage{amssymb}
\usepackage{color}
\usepackage{lscape}
\usepackage{comment}

\draftfalse

\journalname{Journal of Geophysical Research: Planets}

\begin{document}

\title{A probabilistic approach to determination of Ceres' average surface composition from Dawn VIR and GRaND data}

\authors{H. Kurokawa\affil{1}, B. L. Ehlmann\affil{2,3}, M. C. De Sanctis\affil{4}, M. G. A. Lap\^{o}tre\affil{5}, T. Usui\affil{6}, N. T. Stein\affil{2}, T. H. Prettyman\affil{7}, A. Raponi\affil{8}, M. Ciarniello\affil{8}}

\affiliation{1}{Earth-Life Science Institute, Tokyo Institute of Technology, Ookayama, Meguro-ku, Tokyo, Japan}
\affiliation{2}{Division of Geological and Planetary Sciences, California Institute of Technology, Pasadena, California, USA}
\affiliation{3}{Jet Propulsion Laboratory, California Institute of Technology, Pasadena, California, USA}
\affiliation{4}{INAF-IAPS, Via del Fosso del Cavaliere, Rome, Italy}
\affiliation{5}{Department of Geological Sciences, Stanford University, Stanford, California, USA}
\affiliation{6}{Institute of Space and Astronautical Science, Japan Aerospace Exploration Agency,Yoshinodai, Chuo-ku, Sagamihara, Kanagawa, Japan}
\affiliation{7}{Planetary Science Institute, Tucson, Arizona, USA}
\affiliation{8}{Istituto di Astrofisica e Planetologia Spaziali-Istituto Nazionale di Astrofisica, Roma, Italy}

\correspondingauthor{H. Kurokawa}{hiro.kurokawa@elsi.jp}

\begin{keypoints}
\item We derive Ceres' composition using a Bayesian Hapke radiative-transfer model for infrared spectra, coupled with elemental constraints
\item We reconcile compositional estimates of Ceres' mineralogy from reflectance spectra and elemental abundances from gamma ray and neutron data
\item Ceres is modeled by a $\sim$50\% carbonaceous chondrite-like composition with 10\% excess carbon and additional carbonates and phyllosilicates
\end{keypoints}

\begin{abstract}
The Visible-Infrared Mapping Spectrometer (VIR) on board the Dawn spacecraft revealed that aqueous secondary minerals -- Mg-phyllosilicates, NH$_4$-bearing phases, and Mg/Ca carbonates -- are ubiquitous on Ceres. Ceres' low reflectance requires dark phases, which were assumed to be amorphous carbon and/or magnetite ($\sim$80 wt.\%). In contrast, the Gamma Ray and Neutron Detector (GRaND) constrained the abundances of C ($8$--$14$ wt.\%) and Fe ($15$--$17$ wt.\%). Here, we reconcile the VIR-derived mineral composition with the GRaND-derived elemental composition. First, we model mineral abundances from VIR data, including either meteorite-derived insoluble organic matter, amorphous carbon, magnetite, or combination as the darkening agent and provide statistically rigorous error bars from a Bayesian algorithm combined with a radiative-transfer model. Elemental abundances of C and Fe are much higher than is suggested by the GRaND observations for all models satisfying VIR data. We then show that radiative transfer modeling predicts higher reflectance from a carbonaceous chondrite of known composition than its measured reflectance. Consequently, our second models use multiple carbonaceous chondrite endmembers, allowing for the possibility that their specific textures or minerals other than carbon or magnetite act as darkening agents, including sulfides and tochilinite. Unmixing models with carbonaceous chondrites eliminate the discrepancy in elemental abundances of C and Fe. Ceres' average reflectance spectrum and elemental abundances are best reproduced by carbonaceous-chondrite-like materials ($40$--$70$ wt.\%), IOM or amorphous carbon ($10$ wt.\%), magnetite ($3$--$8$ wt.\%), serpentine ($10$--$25$ wt.\%), carbonates ($4$--$12$ wt.\%), and NH$_4$-bearing phyllosilicates ($1$--$11$ wt.\%).

\vspace{0.2in}
\end{abstract}

\section*{Plain Language Summary}

Ceres is a dwarf planet and the largest object in the main asteroid belt, consisting of ice and assemblages of hydrous minerals that incorporate water, ammonium, and carbon. An open question about Ceres is its surface composition and the nature of the materials that make its surface very dark. Whereas iron oxides and amorphous carbon have been suggested from the analyses of spectra at infrared wavelengths of light and mixture modeling using pure mineral or organic endmembers, elemental analysis independently found that C and Fe are not as abundant as posited by these analysis. Thus, another phase or set of phases must be responsible. The dark nature of Ceres is similar to the dark nature of carbonaceous chondrite meteorites, measured in laboratory. We find that the tiny nanometer- and micrometer-scale of darkening agents in these meteorites is not well-accounted for by existing physics-based mixture models of mineral and organic endmembers. Consequently, we use a spectral unmixing model that involves minerals, organics and carbonaceous chondrite meteorites to show that Ceres' surface contains multiple darkening agents of the style found in carbonaceous-chondrite meteorites and additional carbon, hydrous minerals, carbonates, and NH$_4$-bearing minerals.

\section{Introduction}
\label{sec:intro}

Ceres, a dwarf planet and the largest body in the asteroid belt, is thought to have a unique mineral composition. Ceres itself has not been directly sampled but parent bodies of carbonaceous chondrite meteorites (CCs), especially CM and CI groups have been considered the closest analogs for Ceres in existing meteorite collections \cite<e.g.,>{Feierberg+1985,Rivkin+2006,Mcsween+2017,Schafer+2018}. That analogy is largely based on qualitative similarities in their spectral properties including an overall low albedo and $\sim$3 $\mu$m absorption. Nevertheless, telescopic measurements prior to the arrival of the Dawn spacecraft indicated that Ceres somewhat differed from CM and CI meteorites, notably through an unusual spectral feature at 3.06 $\mu$m. Such a feature is not found in most other dark asteroids, and was attributed to multiple possible causes \cite{Rivkin+2006,Takir+Emery2012} with the most likely candidate considered to be ammoniated Mg-smectite clay \cite{King+1992}.

The suite of instruments onboard the Dawn spacecraft enables us to determine Ceres' mineral and elemental composition and compare it with those of other bodies, such as CCs. Specifically, reflectance spectra of Ceres' surface, unobscured by Earth's atmosphere, were obtained by the Visible-Infrared Mapping Spectrometer (VIR) \cite{DeSanctis+2010}. Analyses of VIR-derived spectra from Ceres' surface have identified several features that distinctly differentiate the dwarf planet from CCs, such as the ubiquity of NH$_4$-bearing clay mineral-absorptions  \cite<3.06 $\mu$m,>{DeSanctis+2015,Ammannito+2016,Ehlmann+2018}, Na-carbonate absorptions localized in bright spots \cite<3.5 $\mu$m and 4.0 $\mu$m,>{DeSanctis+2016,Stein+2017,Carrozzo+2018}, and absorptions indicating aliphatic organic materials localized in Ernutet crater \cite<3.4 $\mu$m,>{DeSanctis+2017,Pieters+2017}. In addition, data acquired by the Gamma Ray and Neutron Detector (GRaND) suggests that the abundance of H is higher than in many CCs, whereas the Fe content is lower \cite{Prettyman+2017}. These interpreted differences in mineralogy and composition between CCs and Ceres suggest that Ceres' bulk (or at least near-surface) composition and specific alteration pathways could be different.

The inferred volatile-rich composition, and especially the presence of ammoniated phases on Ceres' surface are key to understanding the early evolution of the Solar System. Because the condensation line of NH$_3$ ice is located further out from the Sun than the current orbit of Ceres, the presence of ammoniated phases on Ceres could result from an inward migration of the dwarf planet after a more distant formation \cite{McKinnon2012,DeSanctis+2015,Vokrouhlicky+2016}. Such migration might have been induced by interactions between, and the migration of, the giant planets during or after planetary formation \cite{Levison+2009,Walsh+2011}. Another possible explanation for NH$_3$ ice is the in-situ accretion of volatile-rich pebbles drifting inward from the outer Solar System \cite{DeSanctis+2015}. \citeA{Mcsween+2017} proposed that ammonia might be released from organics in Ceres' interior at relatively high temperatures ($\sim 300^\circ$C). In either case, a quantitative determination of Ceres' mineral assemblages and elemental abundances would provide new constraints on the dynamic history of, and material transport throughout, the early Solar System.

In addition to the aforementioned infrared absorption bands, Ceres' surface is overall darker than any of the phases mentioned above; therefore, some darkening agent must lower Ceres' albedo, and dark materials need to be incorporated in modeling of composition from infrared spectra. Previous spectral modeling (or unmixing) of Ceres' globally-averaged VIR spectrum showed that, in addition to Mg-phyllosilicates, NH$_4$-bearing phases, and Mg/Ca carbonates, the low albedo \cite<0.09,>{Ciarniello+2017} requires the presence of $\sim$60--80 wt.\% of dark materials, then assumed to be mainly C (amorphous carbon) or Fe (magnetite) \cite{DeSanctis+2015}. However, the derived elemental abundances of C or Fe from these darkening agents significantly exceed those typically encountered in CCs, of a few wt.\% C and $\sim$20 wt.\% Fe \cite{Lodders+1998}. The abundances and spatial distributions of H, C, K, Fe, and constraints on average atomic mass were determined from GRaND data throughout Ceres' regolith, to depths of a few decimeters \cite{Prettyman+2017,Prettyman+2018a,Lawrence+2018}. Though GRaND has spatial resolution much larger than that of VIR ($\sim$400 km vs. $\sim$1 km, respectively), the informed globally-averaged abundances of C ($8$--$14$ wt.\%) and Fe ($15$--$17$ wt.\%) are higher than the aforementioned values estimated from VIR data. The elemental composition implies that Ceres is rich in C and poor in Fe relative to CCs, possibly due to differentiation and extensive alteration \cite{Prettyman+2017,Prettyman+2018a}. Because the Ceres surface is well-mixed by impacts and relatively homogeneous in composition (e.g. crater ejecta vs. other surface materials), the discrepancies between the estimates from the two instruments likely arise from the choice of darkening agents in previous spectral unmixing. 

Recently, \citeA{Marchi+2018} performed spectral unmixing by using CCs themselves as spectral endmembers along with amorphous carbon and magnetite as candidate darkening agents to solve the discrepancy. Their spectral unmixing model adopted an artificial multiplicative constant to control the absolute level of reflectance. The stated reason for adding this factor was to compensate for uncertainties in the absolute radiometric accuracy of the data and to compensate for parameters of the model not well characterized, such as the porosity. However, the factor could lead to bias on estimating the abundance of dark materials, and thus we revisit the problem without the multiplicative constant.

Here, we revisit the derivation of mineral and elemental compositions from VIR data \cite{DeSanctis+2015,Marchi+2018} to reconcile the differences between VIR-derived and GRaND-derived compositions, and provide statistically rigorous uncertainties by finding a full set of possible mineral assemblages that can reasonably explain both the chemical and mineralogical datasets using a Bayesian method for spectral unmixing \cite{Lapotre+2017a} (Section \ref{sec:method}). As a control experiment, we first derive Ceres mineral abundances from a VIR spectrum under the same assumptions and using the same mineral endmembers with amorphous carbon \cite{Zubko+1996} or magnetite as darkening agents as in \citeA{DeSanctis+2015}, or meteorite-derived insoluble organic matter (IOM) (Section \ref{sec:unmixing1}). Next, we apply our model to CCs whose mineral and elemental compositions have been independently determined and discuss the contribution of multiple darkening agents such as carbon, magnetite, sulfides, tochilinite, and other oxides to their spectral properties (Section \ref{sec:chondrites}). These oxides are opaque and known to be present in CCs \cite{Howard+2009,Garenne+2016}. Finally, we perform spectral unmixing by using CCs as spectral endmembers without the multiplicative constant introduced by \citeA{Marchi+2018}, following \citeA{Kurokawa+2018} (Section \ref{sec:unmixing2}). The latter approach ensures that multiple darkening agents, which are actually present in meteorite bodies, are included in the modeling. CCs considered in this study include CM- and CI-type meteorites - Murchison (CM2), Cold Bokkeveld (CM2), and Ivuna (CI1) - as well as Tagish Lake (C2 ungrouped) because it is known to be rich in smectite and organics \cite{Blinova+2014}. We discuss the mineral assemblages which satisfy both VIR spectrum and GRaND elemental concentrations and implications for the origin and evolution of the dwarf planet (Section \ref{sec:discussion}).

\section{Methods}
\label{sec:method}

\begin{landscape}
\begin{table}
    \centering
    \footnotesize
    \begin{tabular}{lllllll}
    \hline
    Mineral & Condition & ID or Reference & refractive index n & Grain size ($\mu$m) & Density (g/cc) & Used in previous studies? \\
    \hline
    antigorite (heated at 500$^\circ$C) & Room temp., ambient-dry atm. & AT-TXH-007 (LAAT07) & 1.565$^{\rm d}$ & 62.5$^{\rm b}$ & 2.54$^{\rm d}$ & \citeA{DeSanctis+2015} \\
    magnesite & Room temp., ambient-dry atm. & JB-JLB-946 (BKR1JB946A) & 1.615$^{\rm d}$ & 22.5$^{\rm b}$ & 2.97$^{\rm d}$ & \citeA{DeSanctis+2015} \\
    magnetite (refractive indices) & -- & \citeA{Querry1985} & -- & -- & 5.15$^{\rm d}$  & -- \\
    NH$_4$-montmorillonite & Room temp., ambient-dry atm. & JB-JLB-189 (397F189) & 1.43$^{\rm c}$ & 62.5$^{\rm a}$ &  2.35$^{\rm e}$ & \citeA{DeSanctis+2015} \\
    NH$_4$-saponite & -100$^\circ$C, under vacuum & \citeA{Ehlmann+2018} & 1.56$^{\rm d}$ & 75$^{\rm b}$ & 2.3$^{\rm d}$ & -- \\
    brucite & Room temp., ambient-dry atm. & JB-JLB-944-A (BKR1JB944A) & 1.58$^{\rm d}$ & 22.5$^{\rm b}$ & 2.39$^{\rm d}$ & \citeA{DeSanctis+2015} \\
    amorphous carbon & Room temp., vacuum & \citeA{Zubko+1996} & 1.78--1.84$^{\rm f}$ & -$^{\rm f}$ & 2.0$^{\rm d}$ & \citeA{DeSanctis+2015} \\
    Cold Bokkeveld IOM & Room temp., dry air & \citeA{Kaplan+2018} & 1.6$^{\rm a}$ & 25$^{\rm b}$ & 2.0$^{\rm d}$ & -- \\
    Murchison (heated at 600$^\circ$C) & Room temp., ambient-dry atm. & MT-JMS-190 (BKR1MT190) & 1.6$^{\rm g}$ & 17.5$^{\rm b}$ & 2.71$^{\rm h}$ & -- \\
    Cold Bokkeveld & Room temp., dry air & \begin{tabular}{l}
         \citeA{Ehlmann+2018}, \\
         MT-JMS-186 (BKR1MT186) 
    \end{tabular} & 1.6$^{\rm g}$ & 75$^{\rm b}$ & 2.71$^{\rm h}$ & -- \\
    Tagish Lake & Room temp., ambient-dry atm. & \begin{tabular}{c}
         MT-MEZ-011 (LCMT11)  \\
         MT-MEZ-012 (LCMT12)
    \end{tabular} & 1.6$^{\rm g}$ & 62.5$^{\rm b}$ & 2.6$^{\rm h}$ & -- \\
    Ivuna (heated at 100$^\circ$C) & Room temp., ambient-dry atm. & MP-TXH-018-F (LAMP18F) & 1.6$^{\rm g}$ & 22.5$^{\rm b}$ & 2.27$^{\rm h}$ & \citeA{Marchi+2018} \\
    \hline
    \end{tabular}
    \caption{\raggedright
    Endmember spectra: name, condition of spectral measurement, RELAB sample (spectrum) ID or reference, the real part of refractive index n, grain size assumed in refractive index calculation, density, and the previous study which used the data (if exists).
    a: Assumed a typical value of particulate samples.
    b: $({\rm minimum\ size}+{\rm maximum\ size})/2$.
    c: https://refractiveindex.info/, \citeA{Querry1987}.
    d: http://webmineral.com/.
    e: \citeA{Glotch+2009}.
    f: We used optical constants as a function of wavelength as reported in \citeA{Zubko+1996}.
    g: Assumed a typical value among minerals. 
    h: \citeA{Britt+2003,Ralchenko+2014}. }
    \label{tab:members}
\end{table}
\end{landscape}

\begin{figure}
    \centering
    \includegraphics[width=0.9\linewidth]{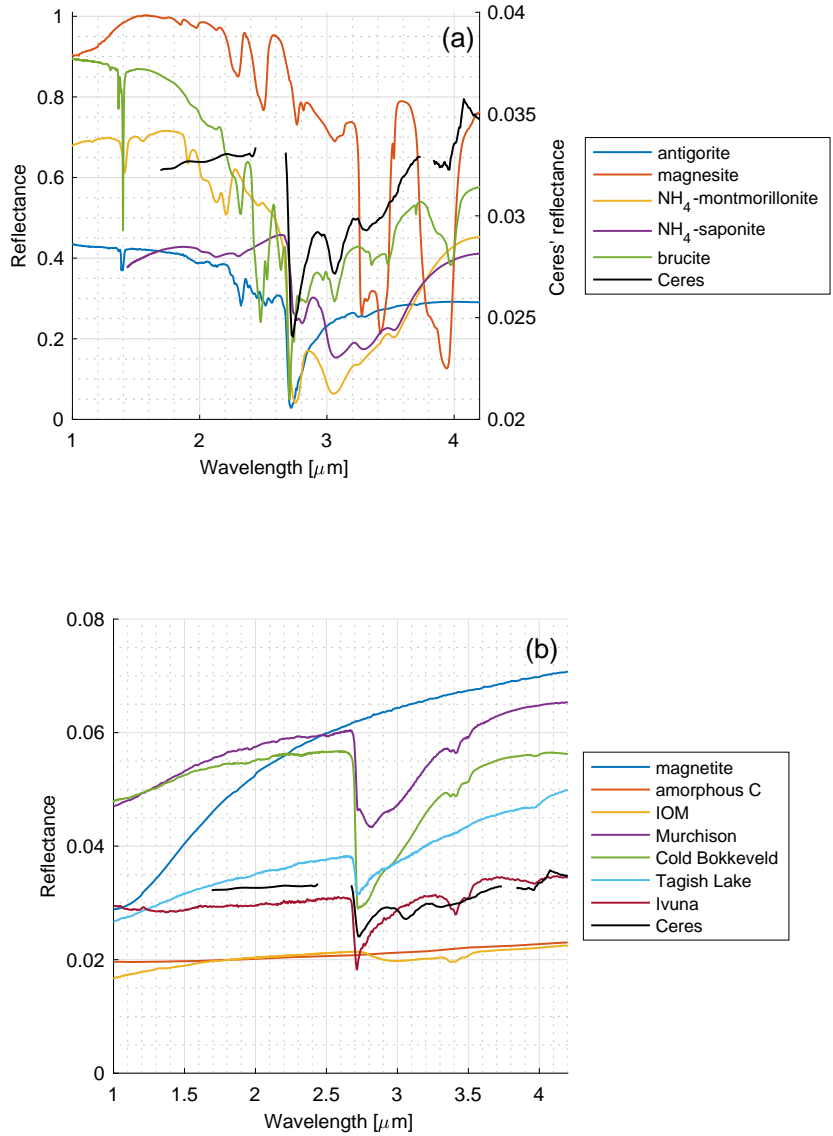}
    \caption{Ceres and endmember spectra used in this study. (a) Minerals with absorption features. Right and left axes denote the reflectance of Ceres and endmember phases, respectively. (b) Dark phases and CCs.}
    \label{fig:endmember_spectra}
\end{figure}

\begin{table}
    \centering
    \begin{tabular}{lllll}
    \hline
    Endmember & H [wt.\%] & C [wt.\%] & K [wt.\%] & Fe [wt.\%] \\
    \hline
    NH$_4$-montmorillonite & $4.04$ $^{\rm a}$ & 0 & $1.43$ $^{\rm b}$ & 0 \\
    NH$_4$-saponite & $2.10$ $^{\rm a}$ & 0 & $0.82$ $^{\rm b}$ & 0 \\
    antigorite & $1.34$ $^{\rm a}$ & 0 & 0 & $14.0$ $^{\rm a}$ \\
    magnesite & 0 & $14.3$ $^{\rm a}$ & 0 & 0 \\
    amorphous carbon & 0 & 100 & 0 & 0 \\
    magnetite & 0 & 0 & 0 & $72.4$ $^{\rm a}$ \\
    IOM & $4.58$ $^{\rm d}$ & $95.4$ $^{\rm d}$ & 0 & 0 \\
    Murchison & $1.07$ $^{\rm e}$ & $2.08$ $^{\rm e}$ & $3.81\times 10^{-2}$ $^{\rm i}$ & $20.5$ $^{\rm i}$ \\
    Cold Bokkeveld & $1.36$ $^{\rm e}$ & $2.45$ $^{\rm e}$ & $4.1\times 10^{-2}$ $^{\rm i}$ & $19.8$ $^{\rm i}$ \\
    Tagish Lake & $1.5$ $^{\rm g}$ & $3.6$ $^{\rm g}$ & $6.5\times 10^{-2}$ $^{\rm g}$ & $19.3$ $^{\rm g}$ \\
    Ivuna & $1.52$ $^{\rm e}$ & $3.50$ $^{\rm e}$ & $4.3\times 10^{-2}$ $^{\rm k}$ & $18.6$ $^{\rm j}$ \\
    \hline
    \end{tabular}
    \caption{\raggedright
    Elemental abundances of endmembers.
    a: http://webmineral.com/
    b: Maximum value assuming that all Na was replaced with K.
    c: \citeA{Bishop+2002}.
    d: \citeA{Kaplan+2018}.
    e: \citeA{Alexander+2012}.
    f: \citeA{Kerridge1985}.
    g: \citeA{Brown+2000}.
    h: \citeA{Grady+2002}.
    i: \citeA{Rubin+2007}.
    j: \citeA{Wolf+Palme2001}.
    k: \citeA{Barrat+2012}.
    }
    \label{tab:elements}
\end{table}

\begin{landscape}
\begin{table}
    \centering
    \begin{tabular}{ll}
    \hline
    Assemblage name & Endmembers \\
    \hline
    A1 & NH$_4$-montmorillonite, antigorite, magnesite, amorphous carbon \\
    A2 & NH$_4$-montmorillonite, antigorite, magnesite, magnetite \\
    A3 & NH$_4$-montmorillonite, antigorite, magnesite, IOM \\
    A4 & NH$_4$-saponite, antigorite, magnesite, IOM \\
    A5 & brucite, antigorite, magnesite, IOM \\
    A6 & NH$_4$-montmorillonite, antigorite, magnesite, magnetite, IOM \\
    B1 & NH$_4$-montmorillonite, antigorite, magnesite, IOM, Murchison \\
    B2 & NH$_4$-montmorillonite, antigorite, magnesite, IOM, Cold Bokkeveld \\
    B3 & NH$_4$-montmorillonite, antigorite, magnesite, IOM, Tagish Lake \\
    B4 & NH$_4$-montmorillonite, antigorite, magnesite, IOM, Ivuna \\
    B5 & NH$_4$-montmorillonite, antigorite, magnesite, magnetite, IOM, Murchison \\
    B6 & NH$_4$-montmorillonite, antigorite, magnesite, magnetite, IOM, Cold Bokkeveld \\
    B7 & NH$_4$-montmorillonite, antigorite, magnesite, magnetite, IOM, Tagish Lake \\
    B8 & NH$_4$-montmorillonite, antigorite, magnesite, magnetite, IOM, Ivuna \\
    B9 & NH$_4$-montmorillonite, antigorite, magnesite, magnetite, amorphous carbon, Murchison \\
    B10 & NH$_4$-montmorillonite, antigorite, magnesite, magnetite, amorphous carbon, Cold Bokkeveld \\
    B11 & NH$_4$-montmorillonite, antigorite, magnesite, magnetite, amorphous carbon, Tagish Lake \\
    B12 & NH$_4$-montmorillonite, antigorite, magnesite, magnetite, amorphous carbon, Ivuna \\
    \hline
    \end{tabular}
    \caption{
    Mixing assemblages used in this study.
    }
    \label{tab:models}
\end{table}
\end{landscape}

\subsection{Modeling and fits to data}
\label{subsec:methods1}

We first derive abundances and grain sizes of endmember phases on Ceres' surface from the globally averaged reflectance by using a Markov Chain Monte Carlo (MCMC) implementation of the radiative transfer model of \citeA{Hapke2012}, developed by \citeA{Lapotre+2017a}. Ceres' VIR spectrum was first reported by \citeA{DeSanctis+2015}, and then recalibrated by \citeA{Carrozzo+2016}. Photometric parameters were then derived from the recalibrated global average spectra by \citeA{Ciarniello+2017}. Finally, \citeA{Marchi+2018} utilized an average spectrum as derived from the photometric parameters of \citeA{Ciarniello+2017}, which we herein use. With calculated mineral abundances on hand, we next compute their corresponding elemental abundances of H, C, K, and Fe and compare them to GRaND-derived abundances \cite{Prettyman+2017,Prettyman+2018a,Prettyman+2018b}.

Here we briefly summarize the mathematical formulation of the Bayesian algorithm, which is fully described in \citeA{Lapotre+2017a}. This method was tested on laboratory mixtures by \citeA{Lapotre+2017a} and applied to real remotely sensed planetary data \cite{Lapotre+2017b,Rampe+2018}. Inputs to the algorithm are the orbiter-based reflectance spectrum, $R(\lambda)$, and the optical indices of endmember phases thought to be on Ceres' surface, $n(\lambda)$ and $k(\lambda)$, where $n$ and $k$ are the real and imaginary indices of refraction, and are functions of wavelength, $\lambda$. At each step, the code chooses a set of grain size $D_i$ (10--1000 $\mu$m) and mass abundance $m_i$ ($<$100 wt.\%) of each endmember phase $i$. The upper bound of grain sizes was chosen to encompass those derived by previous studies \cite{DeSanctis+2015,Marchi+2018}, whereas the lower bound is impoased by limitation of Hapke's model, which assumes the grain size is larger than the photon wavelength \cite{Hapke2012}. For the given set of parameters, we compute $I/F$ (the radiance factor, hereafter simply referred to as the reflectance), defined by
\begin{equation}
    I/F = \frac{\omega}{4}\frac{\mu_0}{\mu+\mu_0}[ (1+B(g))p(g) + H(\omega,\mu) H(\omega,\mu_0) -1 ],
\end{equation}
where $\mu$ and $\mu_0$ are the cosines of the incidence and emission angles. The function $H(\omega,\mu)$ is the Chandrasekhar integral function associated with the observation geometry \cite{Hapke2012}. The average spectrum is defined in standard viewing geometry (the incidence angle $i = 30^\circ$, the emission angle $e = 0^\circ$, and the phase angle $g = 30^\circ$). We assume $B = 0$ (no opposition effect contribution because this effect is only significant for phase angles $g < 2^\circ$, whereas the VIR observation phase angles were much greater than $2^\circ$) and the single-particle phase function (SPPF) $p = 1$ (isotropic scatters, which assumption is further discussed below). The single scattering albedo of a mixture of grains, $\omega$ is computed from a linear combination of the single scattering albedos of endmembers, $\omega_i$, calculated from the refractive indices and grain size, such that,
\begin{equation}
    \omega = \sum_i \frac{m_i/\rho_i D_i}{(\sum_i m_i/\rho_i D_i)} \omega_i,
\end{equation}
where $\rho_i$ is the density of endmember phase $i$ (Table \ref{tab:members}).
The computed spectrum is compared to VIR $I/F$ data and the chosen parameter set is either accepted or rejected with the probability determined by goodness of the fit between data and model spectrum \cite<for details, see>{Lapotre+2017a}. We use a covariance value of $3\times 10^{-6}$, which empirically we found to appropriately account for the magnitude of noise in VIR averaged reflectance. We specify a MCMC chain lengths of $10^6$ for models with four endmembers and $2 \times 10^6$ for those that include more endmember phases. Modeling outputs are the probability distribution functions (PDFs) of grain size $D_i$ (10--1000 $\mu$m) and mass abundance $m_i$ ($<$100 wt.\%) of each endmember phase $i$, as determined by the models accepted by the spectral comparison routine (hereafter called accepted models). We note that our term \textquotedblleft accepted model" follows the terminology of the MCMC method, and it does not mean that any accepted model should be treated equally. Models with poor spectral fit can be accepted with low probability. We put additional selection criteria when necessary (see below in this section). A useful descriptor evaluated from PDFs is the maximum a posteriori probability (MAP) model, which corresponds to the most sampled area of the parameter space, i.e., the most probable assemblage in the MCMC calculation. Because acceptance probability is determined by goodness of the spectral fit, the MAP model corresponds to the best fit explored by the model.

Prior to the MCMC spectral mixing, the imaginary refractive index $k(\lambda)$ is calculated from the reflectance data of endmembers given in the form of the reflectance factor, 
\begin{equation}
    {\rm REFF} = \frac{\omega}{4}\frac{1}{\mu+\mu_0}[ p(g) + H(\omega,\mu) H(\omega,\mu_0) -1 ],
\end{equation}
and an assumed constant $n$ value (a reasonable assumption at the wavelengths herein considered) following the methodology of \citeA{Lucey1998}, \citeA{Roush+1990}, \citeA{Roush2003}, and \citeA{Lapotre+2017a} (Table \ref{tab:members}). In reality, $n$ varies slightly over the wavelengths of interest. For instance, serpentine and other hydrous minerals show a few \% variation in $n$ values in the 2.5--3.0 wavelength range \cite{Mooney+Knacke1985,Dalton+Pitman2012}. However, a sensitivity analysis (not shown) revealed that 10\% variation in $n$ does not influence our modeling as variations in $n$ are readily compensated through derived $k$ values. Because bulk $n$ average values of CCs are unknown, a typical value of minerals found in CCs ($n = 1.6$) is applied.

We assumed isotropic scatter ($p=1$) for all phases considered. We note that the assumption of isotropic scatter may not be correct for some phases such as magnetite \cite{Mustard+Pieters1989}. Because anisotropy depends on the physical properties of particles, evaluating the SPPF requires reflectance measurements with various phase angles for each sample, which are not always available. Thus, we assumed $p=1$ in the absence of further constraints. We note that analysis of VIR data obtained for different phase angles showed that the SPPF of Ceres' averaged surface is fairly symmetric in IR \cite{Ciarniello+2017,Ciarniello+2020}.

With PDFs of mineral abundances $m_i$ on hand, we compute the corresponding PDFs of elemental abundances of H, C, K, and Fe for all accepted models, and compare them to GRaND-derived abundances \cite{Prettyman+2017,Prettyman+2018a,Prettyman+2018b}. Assumed elemental abundances for each endmember phase are listed in Table \ref{tab:elements}. Reported K abundances are upper bounds, as all Na in montmorillonite and saponite is assumed to have been substituted by K in our calculations. Although VIR and GRaND probe different thicknesses of materials, we assume that the upper half meter (probed by GRaND) is well mixed with no difference to the optical surface (probed by VIR) because the surface of Ceres has been subjected to billions of years of continual impact bombardment. This assumption is also consistent with the general lack of evidence for space weathering on Ceres \cite<e.g.,>{Pieters+Noble2016}. An exception to this assumption is H, whose abundance was found to increase from the equator to the poles, consistent with the stability of water ice \cite{Prettyman+2017}. In order to exclude the effect of spatially variable water-ice content, we used the equatorial average reported by \citeA{Prettyman+2017}.

Our MCMC radiative transfer model equally treats all the spectral range where reliable VIR data are available to compute accepted models.
However, when we finally show the models which satisfy both spectral and elemental constraints (Section \ref{sec:unmixing2}), we additionally filter the results to show those that have, not only an acceptable $\chi^2$ over the entire spectral range, but also that have the depths of the absorption features at $2.7$, $3.1$, $3.4$, and $4.0$ $\mu$m being at least 50\% of those in VIR data.
The absorption depth $d$ is defined as,
\begin{equation}
    d = \frac{R_c-R_\lambda}{R_c},
\end{equation}
where $R_c$ is the continuum reflectance given by linear interpolation of two local maximums across the absorption feature, and $R_\lambda$ is the reflectance at the peak of absorption.

\subsection{Spectral endmembers}

The investigated sets of endmember phases are listed in Table \ref{tab:models}, and their corresponding model parameters in Tables \ref{tab:members} and \ref{tab:elements}. Whereas assemblages A1--A6 use single phases as endmembers, assemblages B1--B12 use CC assemblages. Endmember reflectance spectra are compared to that of Ceres' surface in Figure \ref{fig:endmember_spectra}.

Our choice of endmembers was guided by (i) phases that have been positively identified on Ceres by examination of absorptions in infrared spectra \cite<e.g.,>{DeSanctis+2015,DeSanctis+2016,DeSanctis+2017,Ammannito+2016}, (ii) knowledge of phases present in carbonaceous chondrite meteorites (presumed to be similar to those in hydrous asteroids), and (iii) knowledge of GRaND-derived elemental concentrations for Ceres' equatorial ice-free regolith in comparison to carbonaceous chondrites \cite{Prettyman+2017,Prettyman+2018a,Prettyman+2018b}. 
Where possible and reasonable given the radiative transfer model’s sensitivity to endmember albedo and slope, we used the same endmembers as \citeA{DeSanctis+2015,DeSanctis+2016}. 

In addition, spectra of endmember phases that have not been previously considered for Ceres were selected from available spectral libraries (e.g., RELAB, USGS) and other published studies. Criteria for selection of a given spectrum were that (i) it did not suffer from uncorrected measurement artifacts, (ii) it was typical relative to all library spectra of the same sample or sample type, and, when possible, (iii) it was acquired from samples with grain sizes (and preparation) no larger than a few tens of $\mu {\rm m}$. 
In all cases, spectra of particulate samples were used, and either dry air purged or low temperature, vacuum-measured spectra were considered for the shortwave infrared wavelength range. 
Measurements used biconical reflectance measurements in VNIR to set the overall reflectance level, approximating bidirectional geometry by using acquisitions with small angles \cite<e.g.,>{Schaepman-Strub+2006}. 
In order to obtain the reflectance of wavelengths from $1.7\ {\rm \mu m}$ to $4.2\ {\rm \mu m}$, FTIR measurements were spliced and scaled to match the VNIR.

\subsubsection{Antigorite (Mg-serpentine)}

The phyllosilicate-related absorption band in Ceres' spectrum is centered at $2.72\ {\rm \mu m}$ (Figure \ref{fig:endmember_spectra}), consistent with Mg-serpentine rather than Fe-serpentine \cite{Takir+2013}, although we cannot determine the precise Mg-/Fe-content based on band position alone \cite{Schafer+2018}.
We use the same Mg-serpentine (antigorite) as \citeA{DeSanctis+2015} (Figure \ref{fig:endmember_spectra}). 
To dehydrate the sample, it was heated at 500 $^\circ$C for 1 week in a vacuum glass tube, and then sieved to $<$125 ${\rm \mu m}$.
Reflectance data acquired on the sample shows a sharp 2.72-${\rm \mu m}$ absorption from Mg-OH, similar to what is observed on Ceres. 
The serpentine also has weaker absorptions near 1.4, 2.3, and 2.5 ${\rm \mu m}$, which are not readily detected in Ceres' spectra.

\subsubsection{Magnesite (Mg-carbonate)}

The carbonate-related absorption bands in Ceres' spectra are generally broad (e.g., at $\sim 3.5\ {\rm \mu m}$). Though different types of carbonates (magnesite, dolomite, and calcite) show variation in the shape of absorption bands, a previous study found that the limited difference does not influence spectral fitting  \cite{DeSanctis+2015}.
Here, we assume magnesite following \citeA{DeSanctis+2015} and use the same reflectance data (Figure \ref{fig:endmember_spectra}). 
The sample was sieved to $<$45 ${\rm \mu m}$.
Mg-carbonate has 3.4- and 3.96-${\rm \mu m}$ absorptions, as observed on Ceres, as well as weaker absorptions at 2.31 and 2.51 ${\rm \mu m}$. 

\subsubsection{NH$_4$-montmorillonite (Al-smectite)}

Ammoniated smectites were hypothesized by \citeA{King+1992} to be the cause of Ceres' distinctive (relative to other dark asteroids) $\sim$3.05-${\rm \mu m}$ absorption, also corroborated by \citeA{DeSanctis+2015}. 
Montmorillonites, a form of a smectite with octahedral Al, are rarely (if ever) observed in CCs. Nevertheless, we adopt a RELAB-measured NH$_4$-montmorillonite for some runs, following \citeA{DeSanctis+2015} (Figure \ref{fig:endmember_spectra}).
The sample was sieved to $<$125 ${\rm \mu m}$.

\subsubsection{NH$_4$-saponite (Mg-smectite)}

Saponites, or Mg-smectites, are observed in CCs \cite<e.g.,>{Mcsween+2017}. 
We use newly obtained endmember spectra acquired of ammoniated Mg-smectites at $-100^\circ$C in a vacuum by \citeA{Ehlmann+2018}, which was not available when \citeA{DeSanctis+2015} was published (Figure \ref{fig:endmember_spectra}).
The sample was sieved to $<$150 ${\rm \mu m}$.

\subsubsection{Brucite}

The presence of brucite on Ceres was hypothesized by \citeA{Milliken+Rivkin2009} from the presence of a 3.05-${\rm \mu m}$ absorption.
Because brucite has an additional 2.5-${\rm \mu m}$ absorption that precludes good fits to VIR data \cite{DeSanctis+2015}, we include brucite as an endmember for one of our runs to quantify the limits on brucite abundance, using the same spectrum as \citeA{DeSanctis+2015} (Figure \ref{fig:endmember_spectra}).
The sample was sieved to $<$45 ${\rm \mu m}$.

\subsubsection{Insoluble Organic Matter (IOM)}

Most organic matter is dark and relatively featureless over the VNIR wavelength ranges \cite{Kaplan+2018}, and organic matter was hypothesized by \citeA{DeSanctis+2015} to be one possible cause of Ceres’ low VNIR albedo. 
To identify suitable endmembers, we examined spectra of insoluble organic matter from meteorites studied by \citeA{Alexander+2012}. 
We selected a spectrum of $<$50 ${\rm \mu m}$ grains of Cold Bokkeveld IOM (H/C $\sim$58) based on their low reflectance and lack of features, which are suitable properties as a darkening agent. 
The published version of the \citeA{Kaplan+2018} data has an unusually positive spectral slope compared to IOM spectral data in RELAB (e.g., OG-CMA-001 to -004), so we corrected the slope of the data using the overlapping 1.7--2.4 ${\rm \mu m}$ range measured in bidirectional geometry and spliced these VNIR measurements (Kaplan, Milliken, and Hiroi, pers. comm.) with the \citeA{Kaplan+2018} FTIR data to obtain an endmember spectrum (Figure \ref{fig:endmember_spectra}). 
The resultant spectrum shows a C-H stretch at 3.4 ${\rm \mu m}$ and a broad 3-${\rm \mu m}$ feature. 

\subsubsection{Amorphous carbon}

Amorphous carbon is dark and featureless, and may form from organic matter via exposure by UV/ionizing radiation \cite{Hendrix+2016,Kaplan+2018}.
We used the optical constants for amorphous grains produced under arc discharge between amorphous carbon electrodes in an H$_2$ atmosphere (named ACH2 sample) \cite{Zubko+1996}, which is the same sample discussed and proposed as a possible dark phase on Ceres' surface by \citeA{DeSanctis+2015} (Figure \ref{fig:endmember_spectra}).
We note that carbon may exist in the form of graphite on Ceres as suggested from UV spectra \cite{Hendrix+2016}, which is also dark and featureless in VIR wavelengths.

\subsubsection{Magnetite (Fe-oxide)}

Magnetite as well as other Fe oxides and sulfides are relatively featureless, with a weak and broad 1-${\rm \mu m}$ absorption; they often occur in the matrix of CCs, and were thus hypothesized by \citeA{DeSanctis+2015} to be a possible cause of Ceres’ low VNIR albedo. 
Whereas \citeA{DeSanctis+2015} used refractive indices computed from the reflectance data, we use the refractive index data from \citeA{Querry1985} because these optical constants are assumed to be more accurate. We show the reflectance spectrum computed from the refractive indices in Figure \ref{fig:endmember_spectra}.

\subsubsection{Murchison}

As a fairly typical CM2 chondrite, the Murchison meteorite was chosen as an endmember. 
At VNIR and SWIR wavelengths, available Murchison reflectance spectral values at 2 ${\rm \mu m}$ range from 0.025--0.10. 
The absorption strength at 3 $\mu$m varies from 30--45$\%$. 
All examined reflectance spectra of Murchison samples had a 2.80-${\rm \mu m}$ absorption, typical of both structurally bound H$_2$O and Fe serpentines \cite<e.g.,>{Bishop+1994,King+Clark1989}. 
We chose the high SNR spectrum MT-JMS-190, measured under dry air in RELAB over $<$35-${\rm \mu m}$ grains, which is typical of Murchison and has been fully characterized in prior studies \cite{Bland+2004,Howard+2009,Howard+2011,McAdam+2015}.
This sample shows absorption bands centered around 2.72 ${\rm \mu m}$ (weak) and 2.80 ${\rm \mu m}$ (strong) (Figure \ref{fig:endmember_spectra}), which are thought to indicate small amounts of Mg-serpentine and more Fe-serpentine \cite{Takir+2013}, possibly with some contribution from structural or bound water \cite{Bishop+1994} .

\subsubsection{Cold Bokkeveld}

Cold Bokkeveld was chosen as an example of a more altered CM2 chondrite, relatively enriched in serpentine (specifically, Mg-serpentine rather than Fe-serpentine) \cite<e.g.,>{Howard+2009}. 
Particulate samples of Cold Bokkeveld have reflectance values at 2 ${\rm \mu m}$ ranging from 0.03 to 0.07. 
Absorption strength near 3 ${\rm \mu m}$ varies from 17--50$\%$, depending on the specific sample/spectrum. 
All spectra show a minimum at 2.72 ${\rm \mu m}$, characteristic of Mg-OH. 
Select samples, however, have a stronger minimum at 2.78 ${\rm \mu m}$ , which we interpret as possibly due to bound H$_2$O from exposure to terrestrial relative humidity \cite<Table 3b in>{Bishop+1994}. 
Interestingly, a spectrum of a Cold Bokkeveld sample measured after heating and under vacuum \cite{Takir+2013} also shows a 2.78-${\rm \mu m}$ absorption, so it is possible that compositional heterogeneity in the meteorite alone leads to a band shift, with some samples having a 2.72-${\rm \mu m}$ and others a 2.78-${\rm \mu m}$ absorption. 
We choose a spectrum with a 2.72 ${\rm \mu m}$ absorption and very weak additional 3-$\mu$m absorptions, which is a better match to Ceres’s average VIR spectrum.

In the VNIR, we averaged the spectra of different Cold Bokkeveld samples, two $<$35-${\rm \mu m}$-sample spectra measured by \citeA{McAdam+2015} (RELAB MT-JMS-186) and the two $<$150 ${\rm \mu m}$-sample spectra of \citeA{Ehlmann+2018}. 
We used the VNIR reflectance to set the albedo at 2.15 ${\rm \mu m}$, and scaled the FTIR spectrum to form a continuous spectrum with the VNIR. 
\citeA{McAdam+2015}'s MT-JMS-186 was acquired under dry-air purge in RELAB. It has a sharp 2.72-${\rm \mu m}$ absorption and almost no 2.78 ${\rm \mu m}$ absorption. In contrast, \citeA{Ehlmann+2018}'s ammonium-treated Cold Bokkeveld sample was acquired at 173 K under vacuum, and has a sharp 2.72-${\rm \mu m}$ and a weak 3.1-${\rm \mu m}$ absorptions.
The FTIR data of \citeA{McAdam+2015} was used to form our full VIR endmember spectrum (Figure \ref{fig:endmember_spectra}).

\subsubsection{Tagish Lake}

Tagish Lake was chosen because it is a C2 carbonaceous chondrite with a high degree of alteration that in many (but not all) subsamples includes Mg-smectite as well as Mg-serpentine \cite{Zolensky+2002,Izawa+2010,Blinova+2014}. 
There are few published spectra of Tagish Lake. Reflectance values at 2 ${\rm \mu m}$ range from 0.02 to 0.045 for particulate samples. Band strength at 3-${\rm \mu m}$ is 15--25$\%$ for most samples. 
\citeA{Ehlmann+2018} data were acquired at Ceres temperatures under vacuum, but display somewhat unusual shape and relative strengths around the 2.7-${\rm \mu m}$ and 3.1-${\rm \mu m}$ absorptions relative to ambient data and other meteorites. 
This warrants further investigation, and we chose to use an ambient, dry air purge spectrum that appear to be more similar to Ceres spectra. 
In addition, the MT-MEZ-011 and -012 samples of \citeA{Hiroi+2001} are also used. Those with K* and L* filenames were acquired in 2001 and 2002, whereas those with B* filenames correspond to the same samples but acquired 10 years later. The B* spectra show clear evidence for more water (an additional 2.76-${\rm \mu m}$ absorption and possible a 2.8-${\rm \mu m}$ absorption), raising the question of whether Tagish Lake sample may have reacted or hydrated over time on Earth. 
We thus used an average of the RELAB spectra acquired in years 2001 and 2002 only (Figure \ref{fig:endmember_spectra}).
The sample was sieved to $<$125 ${\rm \mu m}$.

\subsubsection{Ivuna}

Ivuna was chosen because it is a CI1 carbonaceous chondrite with a high degree of alteration that is still primitive in composition and has a comparatively high carbonate content \cite{Alexander+2012}, as is thought to be the case of Ceres. Few spectra of the CI1 or CM1 exist in public databases, and most of those that do are also affected by post-fall terrestrial weathering \cite<e.g., Orgueil,>{Bland+2006}. After examination of all available Ivuna spectra and other CI1 and CM1 candidates, we chose to use the VNIR reflectance of RELAB's MP-TXH-018-F (LAMP18F), a sample of Ivuna that had been heated to 100$^\circ$C and that showed by loss of a 1.9-${\rm \mu m}$ absorption that water had been driven off relative to spectra at ambient. FTIR spectrum from \citeA{Takir+2013} of an Ivuna sample that was first heated and then measured under low temperature in vacuum similar to the condition of Ceres' surface was scaled and joined to the RELAB spectrum at 2.15 ${\rm \mu m}$. The reconstructed endmember spectrum has a sharp 2.72-${\rm \mu m}$ absorption, like Ceres (Figure \ref{fig:endmember_spectra}). The sample was sieved to $<$45 ${\rm \mu m}$.

\section{Results: Spectral modeling of Ceres' surface}
\label{sec:unmixing1}

\begin{figure}
    \centering
    \includegraphics[width=0.8\linewidth]{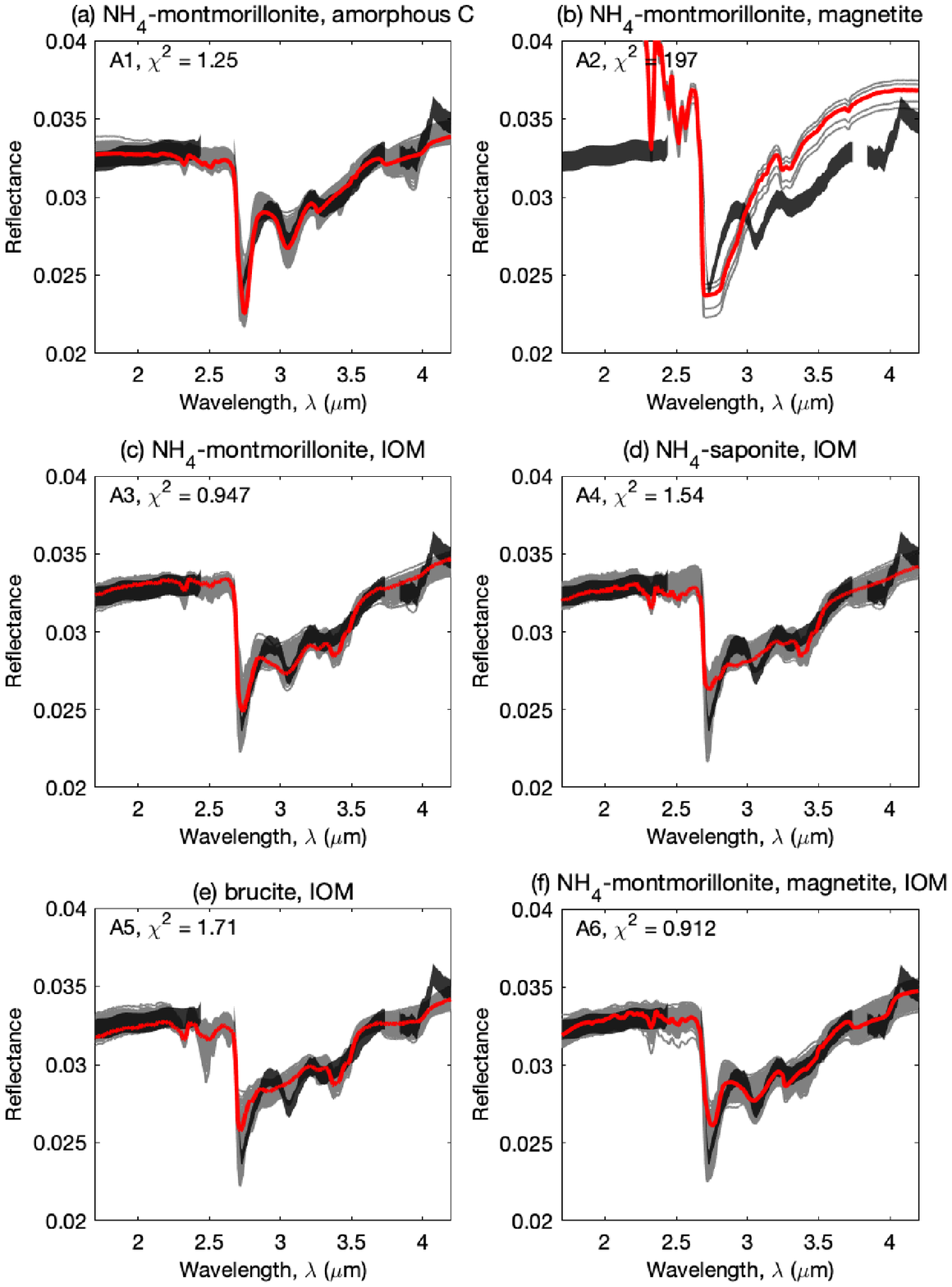}
    \caption{A comparison of modeled spectra with antigorite, magnesite, and the additional phyllosilicate phases and darkening agents indicated in the panel title (assemblages A1--A6; Table \ref{tab:models}) to Ceres' average spectrum \cite{Ciarniello+2017,Marchi+2018} derived from VIR data. The VIR-derived spectrum in cluding 1$\sigma$ error bars as in \citeA{DeSanctis+2015} (black), a set of 1000 accepted and randomly selected models (gray), and the MAP (red) are shown.}
    \label{fig:A_spectra}
\end{figure}

\begin{figure}
    \centering
    \includegraphics[width=\linewidth]{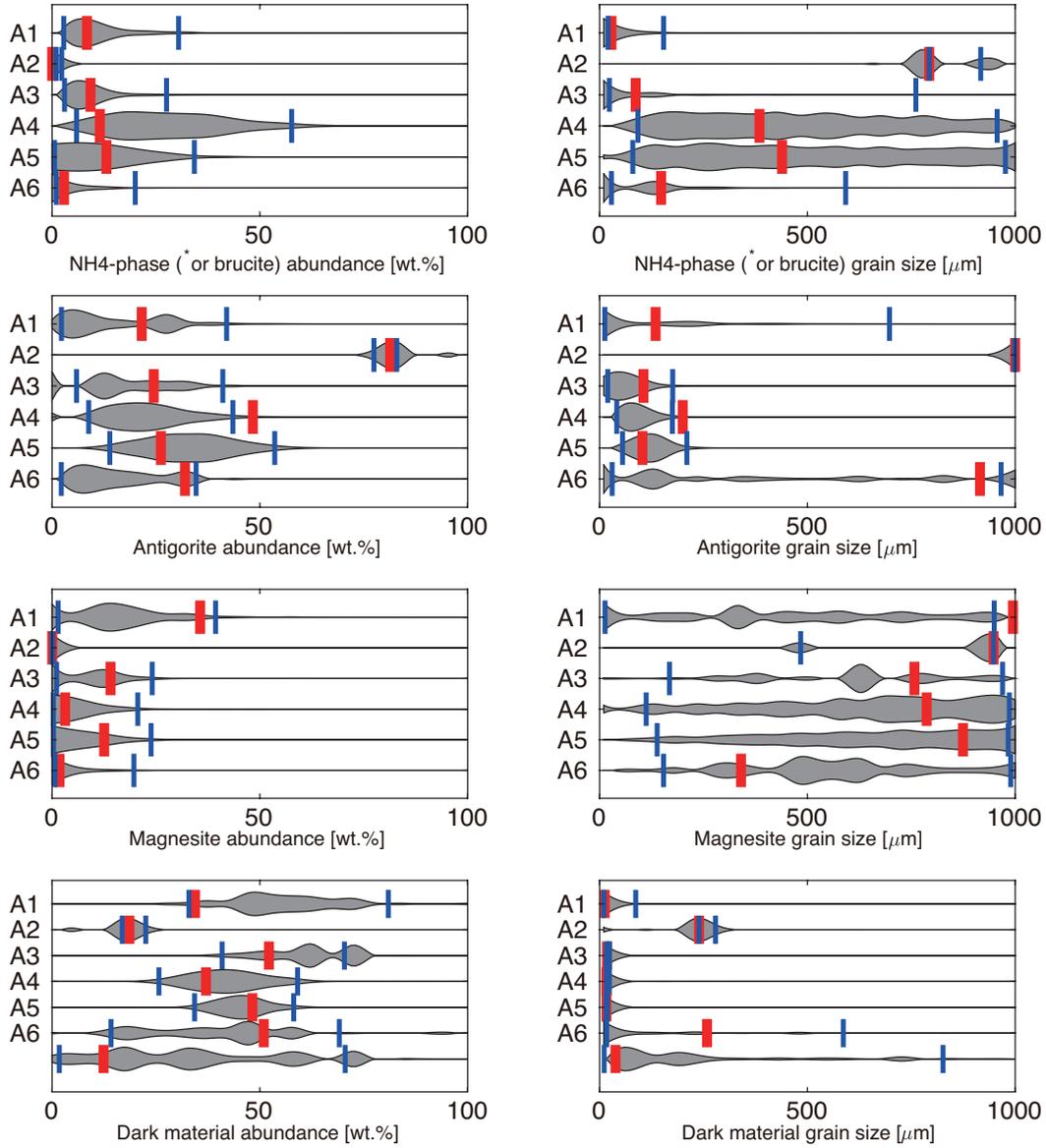}
    \caption{ PDFs of mineral abundances and grain sizes for assemblages A1--A6 (Table \ref{tab:models}). Accepted models (gray), the MAP (red), and 95\%-confidence intervals (the range between two blue lines) are shown. The PDFs are smoothed by spline interpolation of binned data (bin number = 20). The two distributions of dark materials in A6 correspond to IOM (upper) and magnetite (lower).}
    \label{fig:A_minerals_A1-6}
\end{figure}

\begin{figure}
    \centering
    \includegraphics[width=0.8\linewidth]{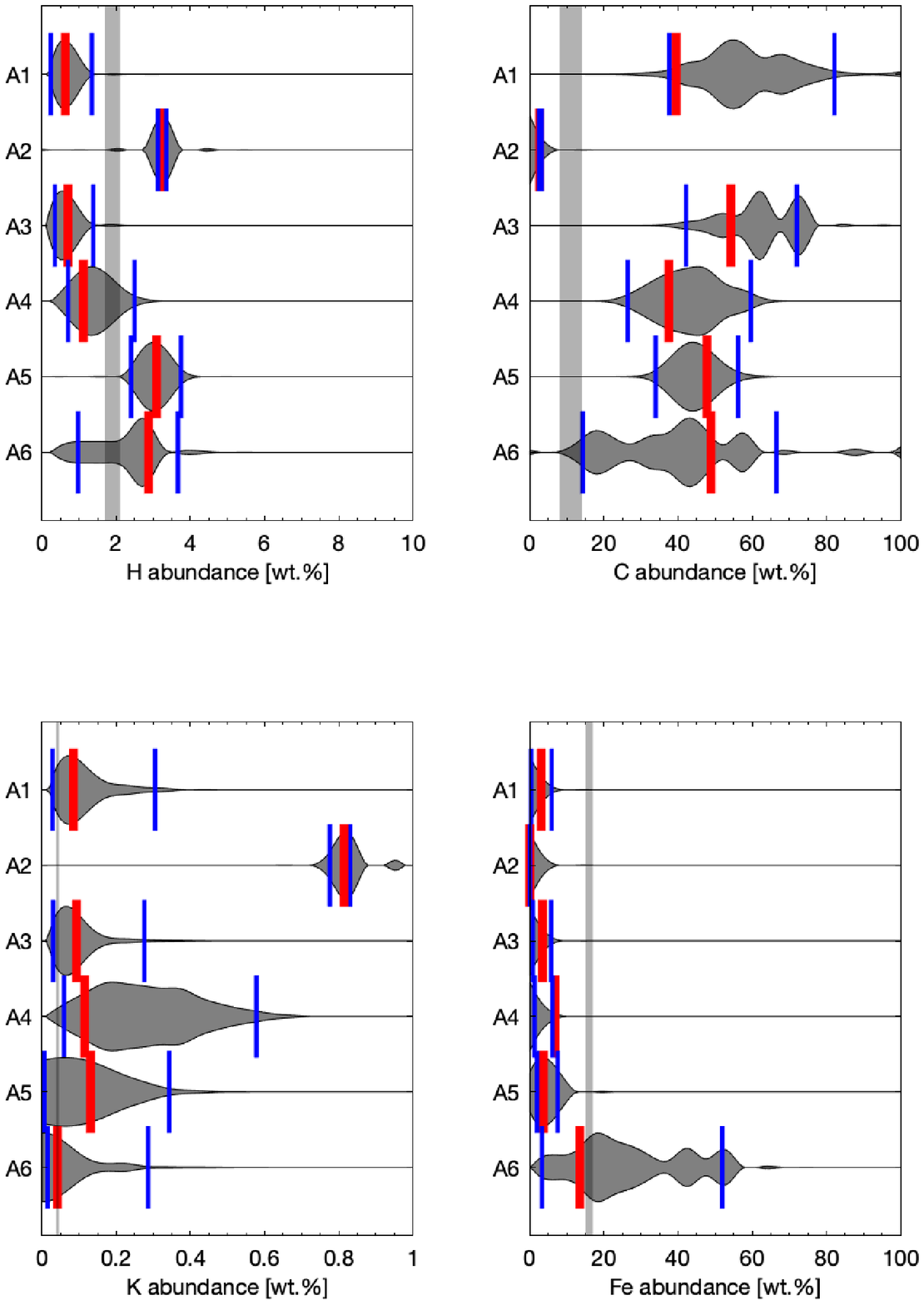}
    \caption{ PDFs of elemental abundances for assemblages A1--A6 (Table \ref{tab:models}). Accepted models (gray), the MAP (red), and 95\%-confidence intervals (the range between two blue lines) are shown. PDFs are smoothed by spline interpolation of binned data (bin number = 20). GRaND measurements with 1$\sigma$ uncertainties \cite{Prettyman+2017,Prettyman+2018a,Prettyman+2018b} are shown for comparison (gray areas). We note that K abundance is an upper bound as it is assumed that all Na in montmorillonite or saponite was substituted by K (Section \ref{sec:method}).}
    \label{fig:A_elements}
\end{figure}

\begin{figure}
    \centering
    \includegraphics[width=\linewidth]{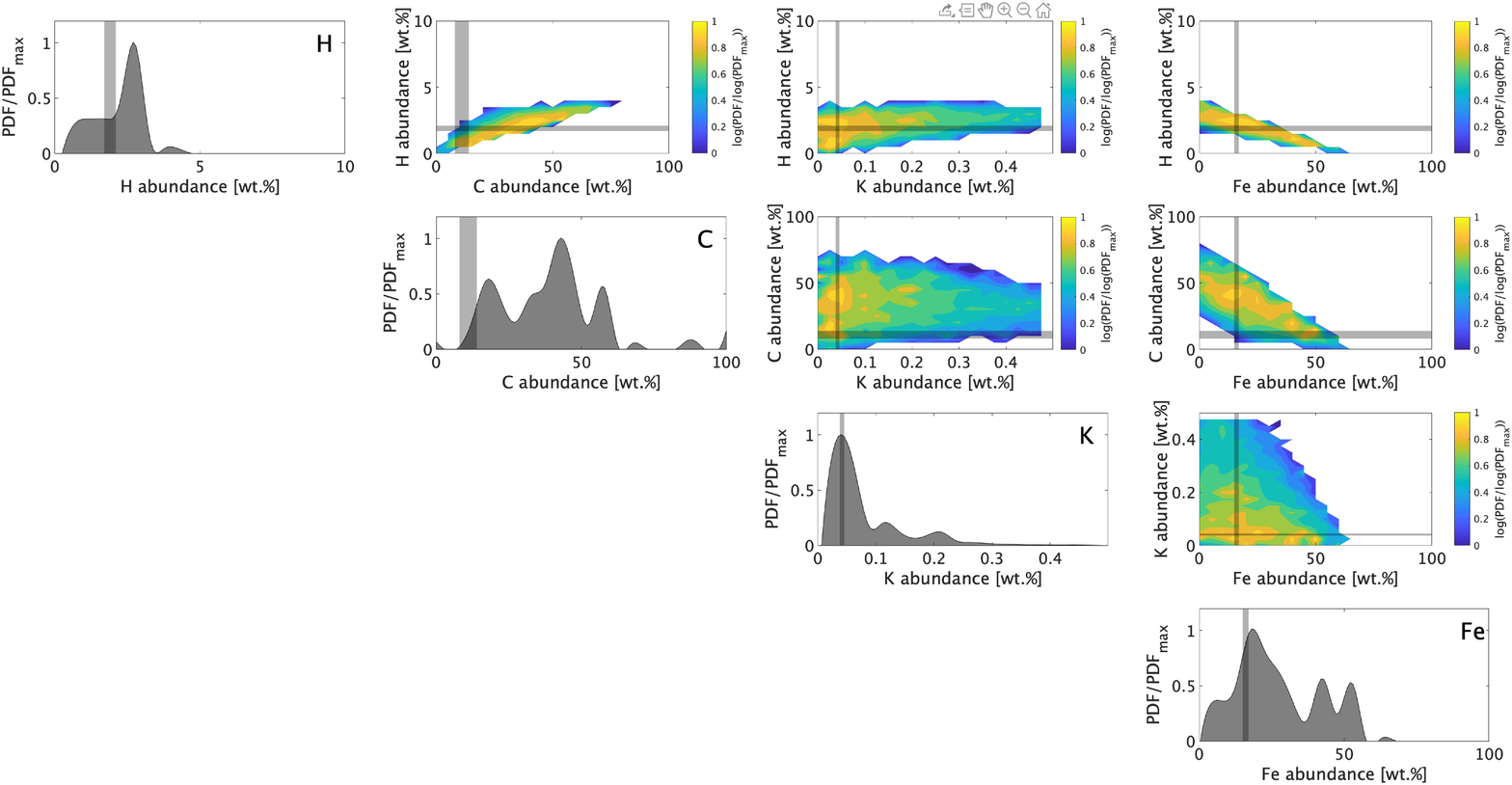}
    \caption{Correlation plot of elemental abundances for assemblage A6. The color contours show the PDF scaled by the peak value in a logarithmic scale (log(PDF)/log(PDF${\rm _{max}}$)). GRaND measurements \cite{Prettyman+2017,Prettyman+2018a,Prettyman+2018b} are shown for comparison (gray areas).}
    \label{fig:correlation_A6}
\end{figure}

First, we model the VIR spectrum with endmember phases following the approach of \citeA{DeSanctis+2015}, but using the Bayesian algorithm of \citeA{Lapotre+2017a} in order to provide statistically rigorous uncertainties on abundances and grain sizes.
Modeled spectra, PDFs of mineral abundances and grain sizes, and PDFs of elemental abundances are shown in Figures \ref{fig:A_spectra}--\ref{fig:A_elements}, respectively.

Modeled reflectance spectra reproduce the general properties of Ceres' VIR spectrum (Figure \ref{fig:A_spectra}): the overall low reflectance and absorption features of phyllosilicates (2.7 ${\rm \mu m}$), ammonia-bearing phases (3.05 ${\rm \mu m}$), and carbonates (3.5 ${\rm \mu m}$ and 4.0 ${\rm \mu m}$).

However, there are differences between assemblages when we focus on MAP models.
The results show that dark phases and phases with positive spectral slopes are important to reproduce Ceres' average spectrum. The MAP model of amorphous-carbon-containing assemblage (A1, Figure \ref{fig:A_spectra}a) reproduces the absorption features and the absolute level of reflectance, but its flat spectral slope at $<$2.5 ${\rm \mu m}$ is inconsistent with the VIR data showing a positive slope. Magnetite shows a positive slope (Figure \ref{fig:endmember_spectra}), but the magnetite-containing assemblage (A2, Figure \ref{fig:A_spectra}b) failed to reproduce Ceres' average spectrum. This is likely because magnetite's reflectance calculated from our specific optical constants is not as dark as Ceres (Figure \ref{fig:endmember_spectra}), as opposed to \citeA{DeSanctis+2015} who utilized darker magnetite reflectance data. The MAP model of assemblage A3 (with NH$_4$-montmorillonite and IOM, Figure \ref{fig:A_spectra}c) shows an overall better fit to the data than the others, thanks to IOM being redder than amorphous carbon (A1) and darker than magnetite (A2) (Figure \ref{fig:endmember_spectra}). Mixing magnetite and IOM improves the fitting only slightly (A6, Figure \ref{fig:A_spectra}f).

Ammoniated-saponite and brucite show worse fits to Ceres' average spectrum than NH$_4$-montmorillonite. Ammoniated-saponite-bearing assemblage (A4, Figure \ref{fig:A_spectra}d) has a higher MAP $\chi^2$ value than that of A3 (Figure \ref{fig:A_spectra}c) with shallower 3.05 ${\rm \mu m}$, deeper 3.4 ${\rm \mu m}$ absorption features, and underfit at 4.0 ${\rm \mu m}$, even though NH$_4$-saponite has been considered to the best candidate for the origin of Ceres' 3.05 ${\rm \mu m}$ absorption. The mismatch is possibly due to the specific NH$_4$-saponite endmember's different broad spectral continuum shape 3--4 ${\rm \mu m}$ relative to Ceres (Figure \ref{fig:endmember_spectra}), which introduces fit problems for that wavelength range. The brucite-bearing assemblage (A5, Figure \ref{fig:A_spectra}e) shows an unsatisfactory fit, as previously reported \cite{DeSanctis+2015,DeSanctis+2018,DeAngelis+2016}. 

Though the MAP models do not always show a clear carbonate absorption (3.5 and 4 ${\rm \mu m}$), some of the accepted models (Figure \ref{fig:A_spectra}, gray lines) do generate spectra displaying these absorption features. This is because the MCMC algorithm accepts or rejects modeled spectra using the goodness of fit over the entire wavelength range rather than using absorption features only. In later steps, we thus filter for absorption band depth (see Methods Section \ref{subsec:methods1}).

As previously reported by \citeA{Lapotre+2017a} and \citeA{Lapotre+2017b}, posterior PDFs show that the unmixing inversion does not always result in unique, well-constrained solutions for mineral abundances and grain sizes (Figure \ref{fig:A_minerals_A1-6}), which highlights the importance of considering rigorous uncertainties. Grain-size PDFs tend to be broad and/or multi-modal because grain sizes easily trade off between most endmembers in setting the albedo and band depths. In contrast, dark endmembers (IOM, amorphous carbon, and magnetite), which exert a primary control on overall albedo, typically have a unimodal grain-size PDF as it controls the absolute level of reflectance. In A6, IOM has a unimodal grain-size PDF but magnetite does not, because the overall albedo is chiefly controlled by IOM in this case. PDFs of mineral abundances are narrower than those of grain sizes, but their 95$\%$-confidence intervals (2$\sigma$) are typically a few tens of percent.

Though assemblages A1--A3 contain all the same minerals except for the dark phase, the resulting abundance and grain-size PDFs significantly differ, highlighting the important role played by the darkening phase in setting the goodness of the spectral fit.
Similarly, assemblages A3 and A4 differ only in their NH$_4$-bearing phases, and yet, the PDFs of all endmember phases are affected by this difference. 
The continuum shapes of specific endmembers, in addition to absorptions, significantly influences the quantitative results, as demonstrated by the higher abundance of NH$_4$-saponite in assemblage A4 relative to NH$_4$-montmorillonite for assemblages A1--A3.

Though unmixing of the near-IR spectrum of Ceres can rule out some sets of mineral assemblages (A2 in Figure \ref{fig:A_spectra}b), it does not alone provide definitive constraints on mineral abundances. Therefore, combining spectral unmixing with additional available constraints is critical to further constraining our understanding of Ceres' surface composition.
We next compare the PDFs of elemental abundances corresponding to inverted mineral assemblages from VIR with GRaND data (Figure \ref{fig:A_elements}).

We did not find pure mineral mixing models that simultaneously satisfy the constraints on elemental abundances from GRaND and the spectral properties from VIR (both acceptable $\chi^2$ and the absorption band depths comparable to Ceres, Section \ref{sec:method}) for assemblages A1--A6. Mimicking the PDFs of mineral phases and grain sizes (Figure \ref{fig:A_minerals_A1-6}), the PDFs of elemental abundances (Figure \ref{fig:A_elements}) are not always sharp. 
The modeled 95\% confidence intervals of either C (in A1, A2, A4, and A5) or Fe (in A3) abundances in assemblages assuming single dark phase are overall higher than estimates from GRaND data. Assemblages with both IOM and magnetite (A6) display the lowest discrepancy in C and Fe abundances with GRaND data. There are some accepted models in A6 which satisfy GRaND constraints, but we found that those models do not meet the requirement on absorption band depths (at least 50\% of those in Ceres' average spectrum). The C and Fe abundances in A6 accepted models show an anti-correlation (Figure \ref{fig:correlation_A6}), and both phases are overpredicted from VIR relative to GRaND measurements for the majority of accepted models. Moreover, many models failed to match the H abundance, either in excess or in depletion. Because we only considered an upper bound for K abundance in montmorillonite and saponite by assuming that K substituted for all Na, accepted models could still be consistent and reconciled with GRaND estimates, if montmorillonite or saponite on Ceres had the appropriate interlayer cation K abundance. Collectively, these results suggest that the models so far did not include all the necessary endmembers, and additional darkening agents containing less C and Fe should be included. In Section \ref{sec:unmixing2}, we show that these discrepancies can be reduced when carbonaceous chondrites are included as endmembers.

\section{Spectral modeling of carbonaceous chondrites}
\label{sec:chondrites}

\subsection{Why are carbonaceous meteorites dark?}

\begin{landscape}
\begin{table}
    \centering
    \begin{tabular}{lllll}
    \hline
    Phase & RELAB sample (spectrum) ID & Refractive index n$^{\rm a}$ & Grain size (${\rm \mu m}$) $^{\rm b}$ & Abundance [wt.\%] \\
    \hline
    olivine (Fo$_{100}$) & SR-JFM-043 (BKR1SR043) & 1.635$^{\rm c}$ & 85.5 & 7.4$^{\rm e}$ \\
    olivine (Fo$_{80}$) & DD-MDD-087 (BKR1DD087) & 1.673$^{\rm c}$ & 22.5 & 4.14$^{\rm e}$ \\
    olivine (Fo$_{40}$) & DD-MDD-094 (BKR1DD094) & 1.750$^{\rm c}$ & 22.5 & 4.04$^{\rm e}$ \\
    enstatite & JB-JLB-238 (397F238) & 1.65 & 22.5 & 8.39$^{\rm e}$ \\
    calcite & CA-EAC-010 (LACA10) & 1.62 & 22.5 & 1.03$^{\rm e}$ \\
    magnetite & \citeA{Querry1985} & -- & -- & 1.79$^{\rm e}$ \\
    pyrrhotite & PY-LXM-001 (LAPY01) & 1.8 & 47.5 & 1.44$^{\rm e}$ \\
    cronstedtite & CR-EAC-021 (LACR21) & 1.76 & 22.5 & 51.7$^{\rm e}$ \\
    chrysotile & CR-TXH-006 (LACR06) & 1.565 & 22.5$^{\rm d}$ & 17.9$^{\rm e}$ \\
    IOM in Murchison & OG-CMA-002 (BKR1OG002) & 1.8 & 13.75 & 2.08$^{\rm e}$ \\
    \hline
    \end{tabular}
    \caption{
    Endmember data for Murchison modeling: Sample IDs, n values and grains sizes for refractive indices calculations, and abundances in spectral mixing calculations.
    a: http://webmineral.com/
    b: $({\rm minimum\ size}+{\rm maximum\ size})/2$
    c: \citeA{Lucey1998}.
    d: Assumed.
    e: \citeA{Howard+2009,Alexander+2012}
    }
    \label{tab:Murchison}
\end{table}
\end{landscape}

\begin{figure}
    \centering
    \includegraphics[width=0.49\linewidth]{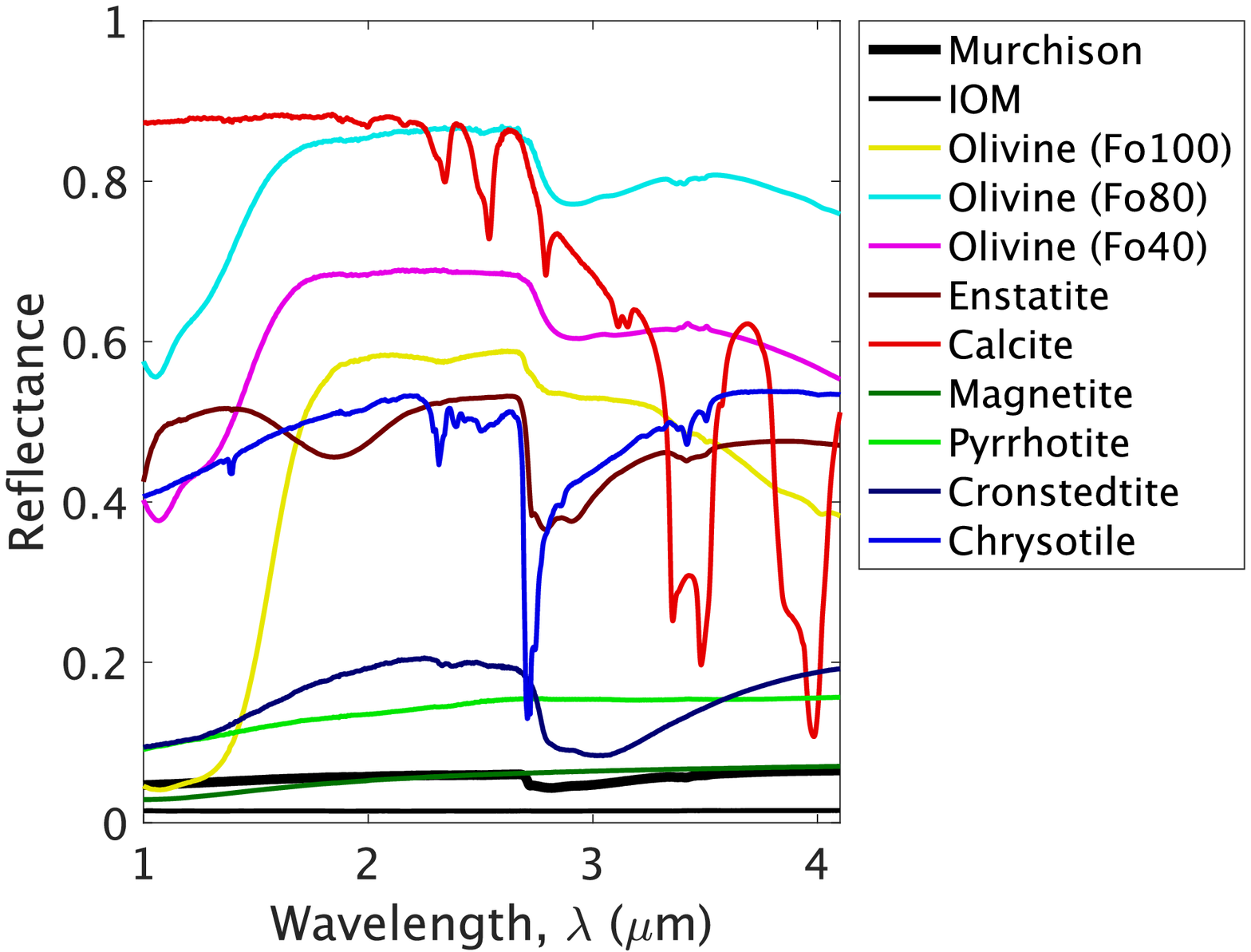}
    \includegraphics[width=0.49\linewidth]{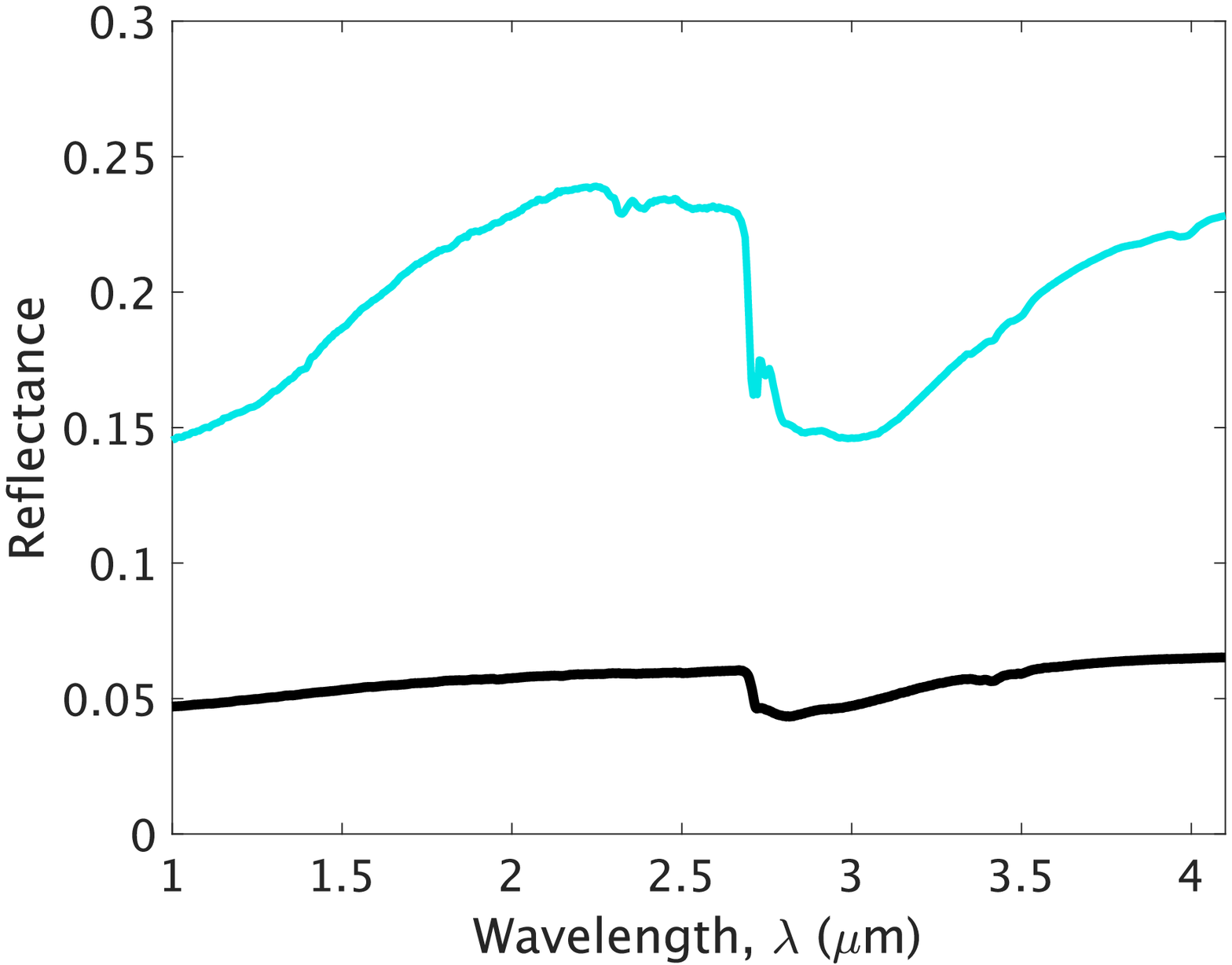}
    \caption{(a) Reflectance spectra of Murchison and minerals found in Murchison. (b) A comparison of a model reflectance (blue) computed by the linear-SSA-mixing model to Murchison's reflectance (black).}
    \label{fig:ModelingMurchison}
\end{figure}

The discrepancy between the inferred large quantities of darkening agents from VIR unmixing models and those allowed by C and Fe abundances derived from GRaND data raises the question of what phases are responsible for Ceres' low albedo and why (e.g., the composition(s) of darkening agent(s) versus their physical textures and spatial relationships with other phases).

Carbon is a main candidate as the reflectances of CCs have been shown to correlate with their carbon contents \cite{Cloutis+2012}. Carbon in CCs chiefly exists as IOM \cite{Kerridge1985,Alexander+2012}, which is dark and IOM reflectance decreases as IOM matures \cite<i.e., as the H/C ratio decreases,>{Moroz+1998,Quirico+2016,Kaplan+2018}. However, a comparison between primitive and metamorphosed CM chondrites does not show a clear difference in their albedos at $2\ {\rm \mu m}$, though metamorphosed ones are inferred to contain less C from elemental analysis \cite{Garenne+2016,Beck+2018}. \citeA{Garenne+2016} and \citeA{Beck+2018} argued that the albedo of CCs is primarily controlled by their matrix-to-chondrule ratio.

In addition, sulfides and iron oxides in the matrix also likely contribute to lowering the albedo. Iron sulfides (e.g., troilite and pyrrhotite) and Fe-Ni alloys have been identified in carbonaceous chondrites and as opaque phases mixed with refractory polyaromatic carbon on the 67P/Churyumov-Gerasimenko comet observed by the Rosetta spacecraft \cite{Quirico+2016,Rousseau+2018}.

In addition to mineralogy, the mixing structure may also affect albedo. For example, 
small dark grains effectively decrease the reflectance when mixed with large bright grains (also called the “coating effect”, e.g.,  \citeA{Cloutis+1990,Pommerol+Schmitt2008,Rousseau+2018}). Moreover, laboratory measurements on a pure opaque phase (iron sulfides) showed that finer grains provide the lowest albedo in sub-micron sizes, with a behavior basically opposite to \textquotedblleft classical silicates" \cite{Rousseau+2018}. Darkening from small grains is sometimes more effective than predicted by linear single-scattering-albedo mixing models \cite{Clark1983}.
In particular, submicroscopic metallic Fe produced by space weathering in lunar regolith is a potent darkening agent, even in limited abundances \cite{Shkuratov+1999,Hapke2001,Li+Li2011}. Modeling the effect of such inclusions on reflectance requires more complex radiative transfer models \cite{Mustard+Hays1997}. If dark phases are coatings, their effect on spectral properties and modeled abundances would be higher than their true abundances.

To quantitatively examine the effect of these structural and compositional phenomena in CCs, we first compute a reflectance spectrum using endmembers and abundances as measured in Murchison by independent techniques \cite{Howard+2009}, assuming a grain size of $31.5\ {\rm \mu m}$, and compare it to an actual reflectance spectrum of Murchison (Figure \ref{fig:ModelingMurchison}). Measured Murchison composition used to model a synthetic reflectance spectrum is listed in Table \ref{tab:Murchison}.

We find that our synthetic Murchison spectrum has an overall higher albedo than the true Murchison spectrum, suggesting that some darkening phases or mechanisms are missing in the spectral modeling and that pure-endmember geometric optics radiative transfer models are not appropriate for dark asteroids.
Next, we explore how using CCs as phase endmembers in spectral modeling may affect spectral fit and compare with GRaND data.

\section{Spectral modeling of Ceres using meteorites as endmembers}
\label{sec:unmixing2}

\newpage

\begin{figure}
    \centering
    \includegraphics[width=1.0\linewidth]{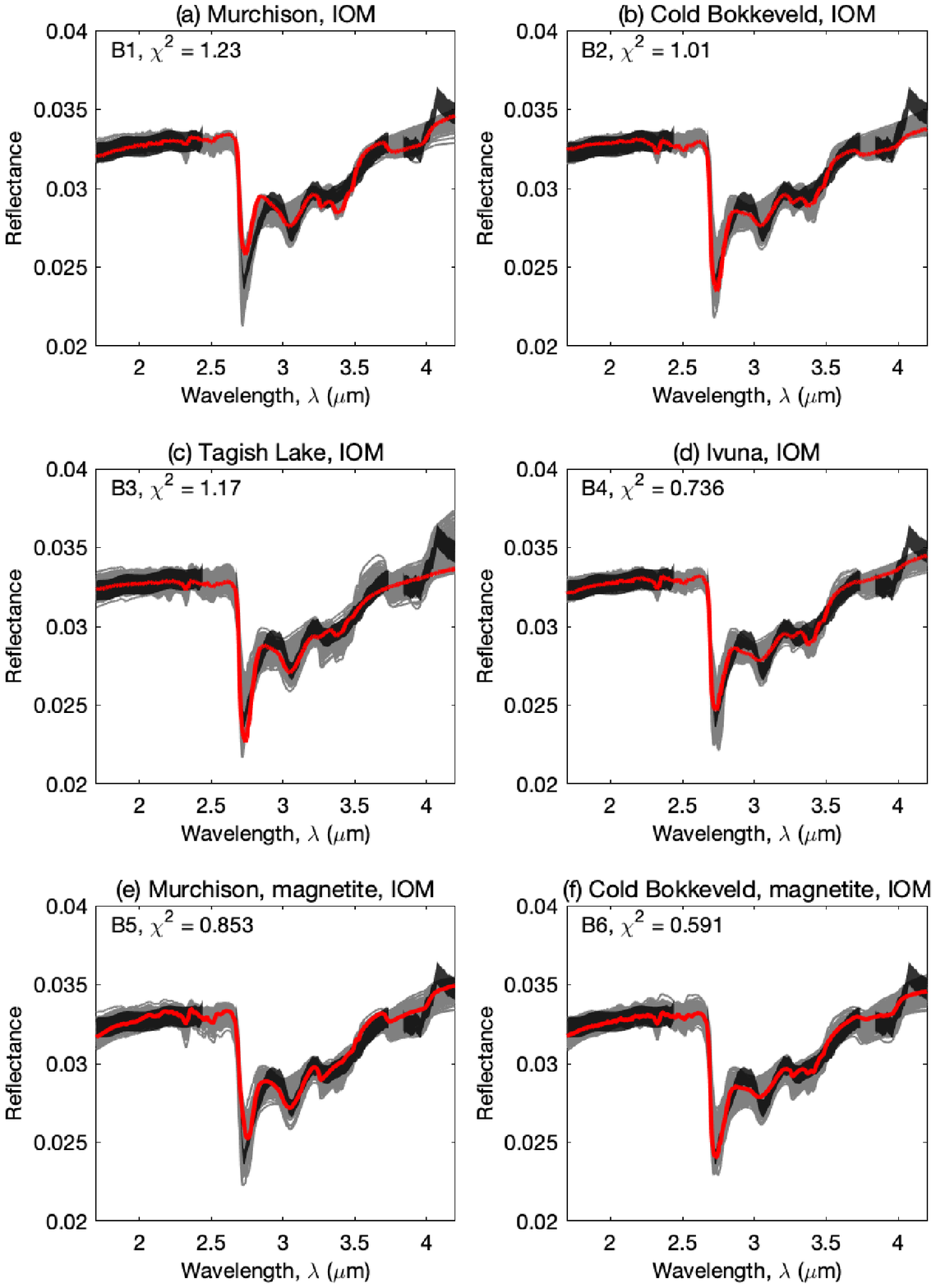}
\end{figure}

\newpage

\begin{figure}
    \centering
    \includegraphics[width=1.0\linewidth]{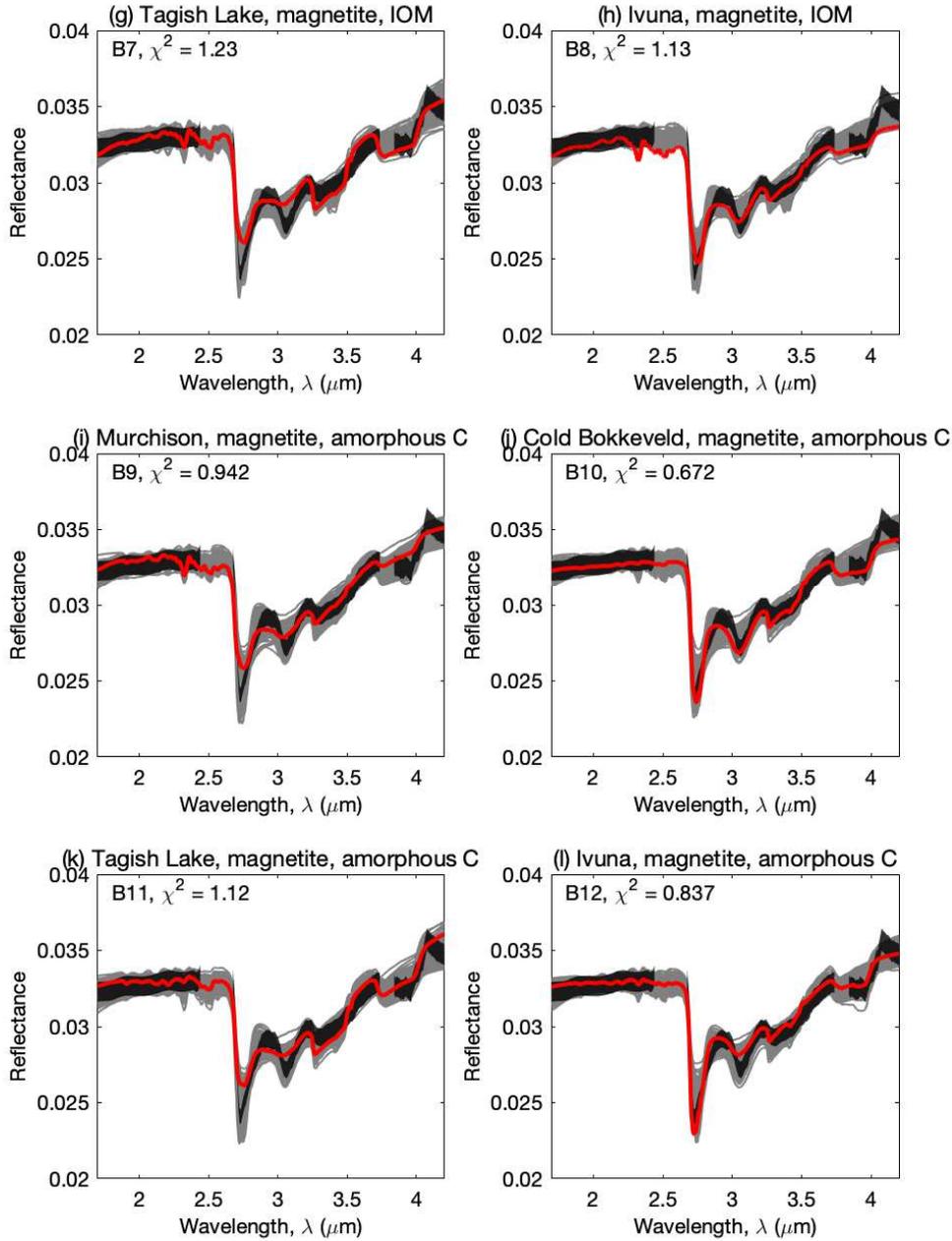}
    \caption{A comparison of modeled spectra with NH$_4$-montmorillonite, antigorite, magnesite, and the darkening agents indicated in the panel title (assemblages B1--B12; Table \ref{tab:models}) to Ceres' average spectrum \cite{Ciarniello+2017,Marchi+2018} derived from VIR data. The VIR-derived spectrum in cluding 1$\sigma$ error bars as in \citeA{DeSanctis+2015} (black), a set of 1000 accepted and randomly selected models (gray), and the MAP (red) are shown.}
    \label{fig:B1-12_spectra}
\end{figure}

\newpage

\begin{figure}
    \centering
    \includegraphics[width=0.8\linewidth]{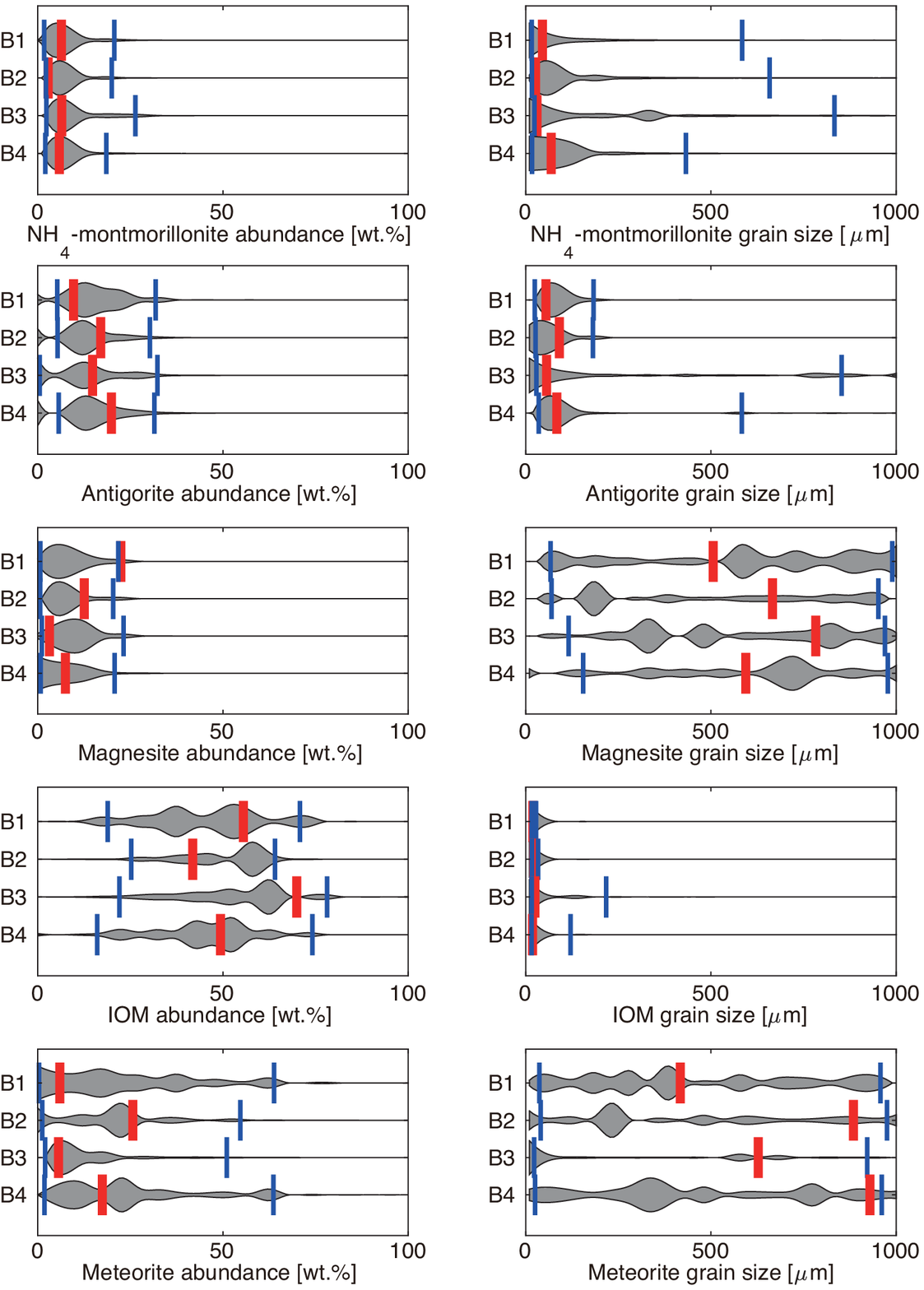}
    \caption{PDFs of phase abundances and grain sizes for assemblages B1--B4 (Table \ref{tab:models}). All accepted models (gray), MAP (red), and 95\%-confidence intervals (the range between two blue lines) are shown. PDFs are smoothed by spline interpolation of binned data (bin number = 20).}
    \label{fig:B1-4_minerals}
\end{figure}

\newpage

\begin{figure}
    \centering
    \includegraphics[width=0.8\linewidth]{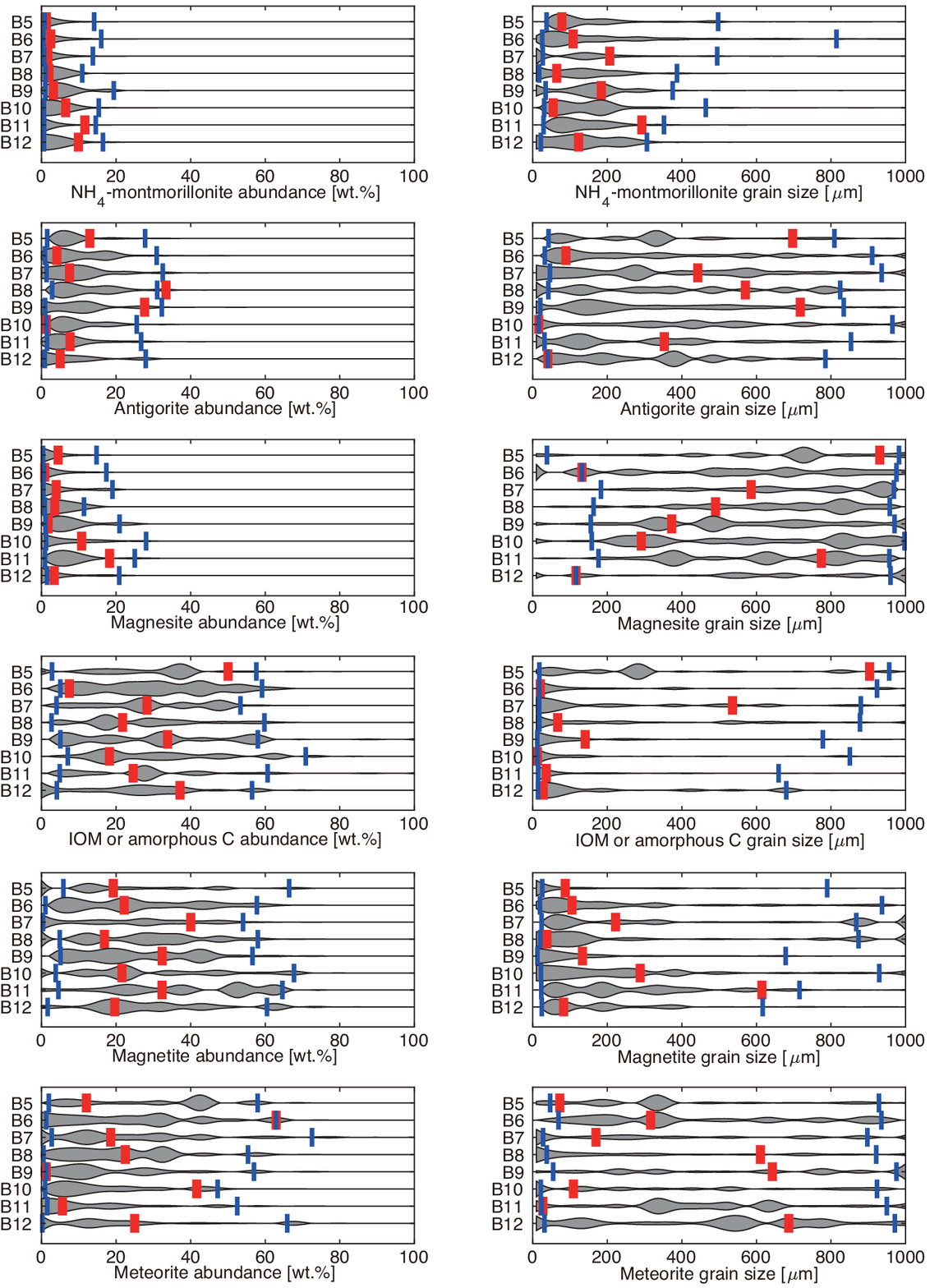}
    \caption{PDFs of phase abundances and grain sizes for assemblages B5--B12 (Table \ref{tab:models}). All accepted models (gray), MAP (red), and 95\%-confidence intervals (range between two blue lines) are shown. PDFs are smoothed black by spline interpolation of binned data (bin number = 20).}
    \label{fig:B5-12_minerals}
\end{figure}

\newpage

\begin{figure}
    \centering
    \includegraphics[width=0.8\linewidth]{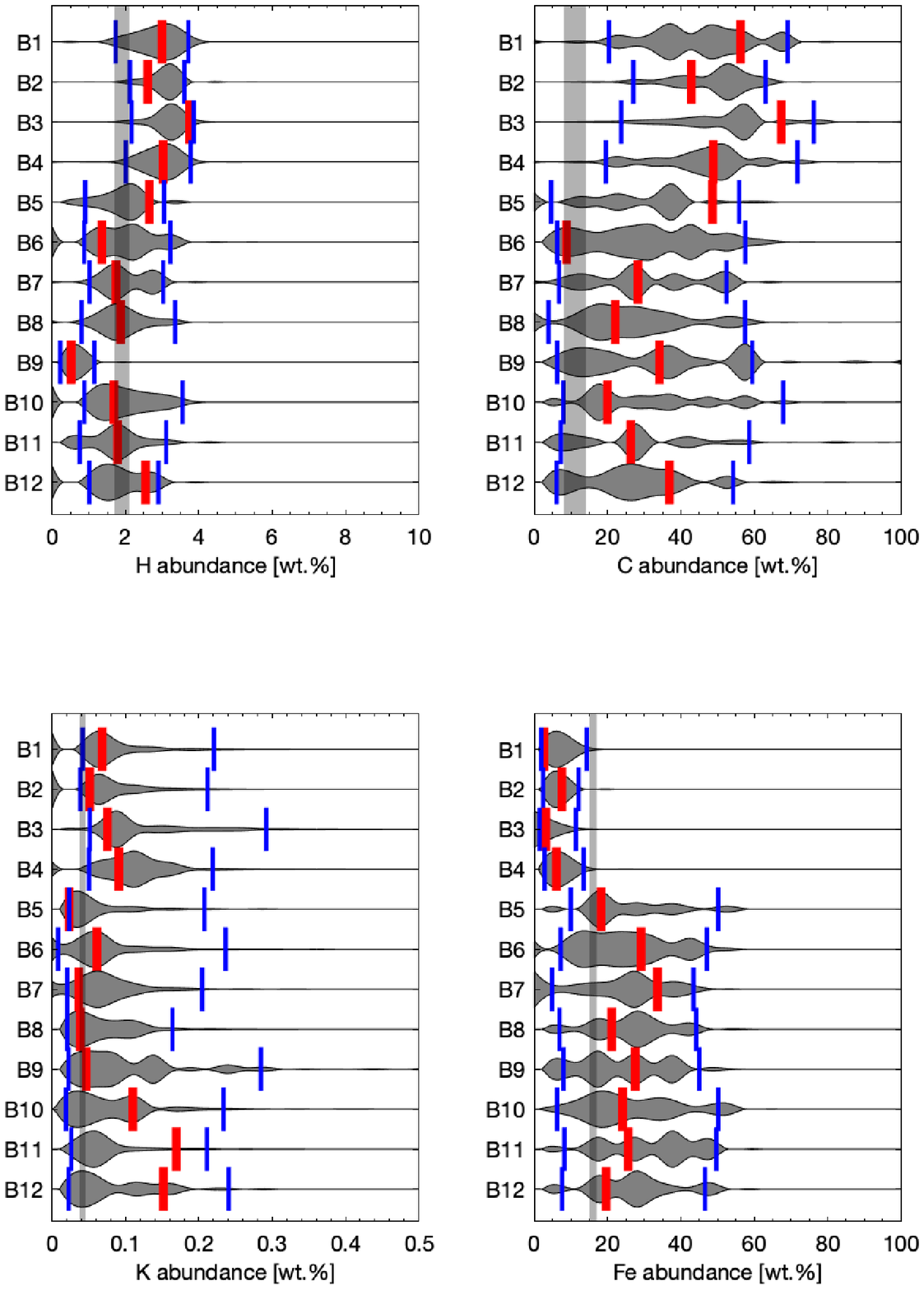}
    \caption{ PDFs of elemental abundances for assemblages B1--B12 (Table \ref{tab:models}). All accepted models (gray), MAP (red), and 95\%-confidence intervals (range between two blue lines) are shown. PDFs are smoothed by spline interpolation of binned data (bin number = 20). GRaND measurements \cite{Prettyman+2017,Prettyman+2018a,Prettyman+2018b} are shown for comparison (gray areas). K abundance is an upper bound, as all Na in montmorillonite and saponite is assumed to have been substituted by K (Section \ref{sec:method}).}
    \label{fig:B1-12_elements}
\end{figure}

\newpage

\begin{figure}
    \centering
    \includegraphics[width=\linewidth]{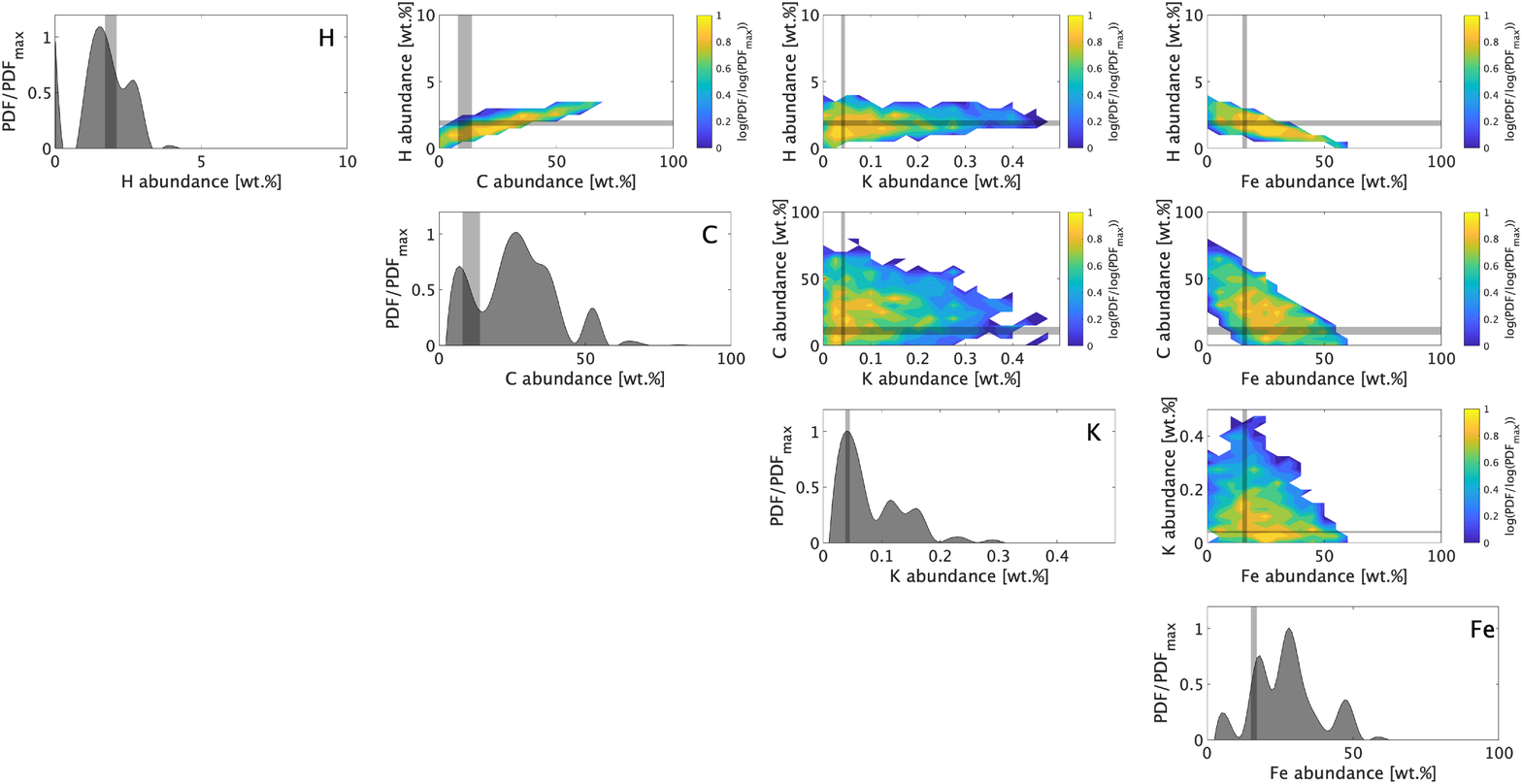}
    \caption{Correlation plot of elemental abundances for assemblage B12. The color contours show the PDF scaled by the peak value in a logarithmic scale (log(PDF)/log(PDF${\rm _{max}}$)). GRaND measurements \cite{Prettyman+2017,Prettyman+2018a,Prettyman+2018b} are shown for comparison (gray areas). Assemblages B5--B12 show similar results, except for B9 which resulted in less H (Figure \ref{fig:B1-12_elements}).}
    \label{fig:correlation_B12}
\end{figure}

\newpage

\begin{figure}
    \centering
    \includegraphics[width=0.49\linewidth]{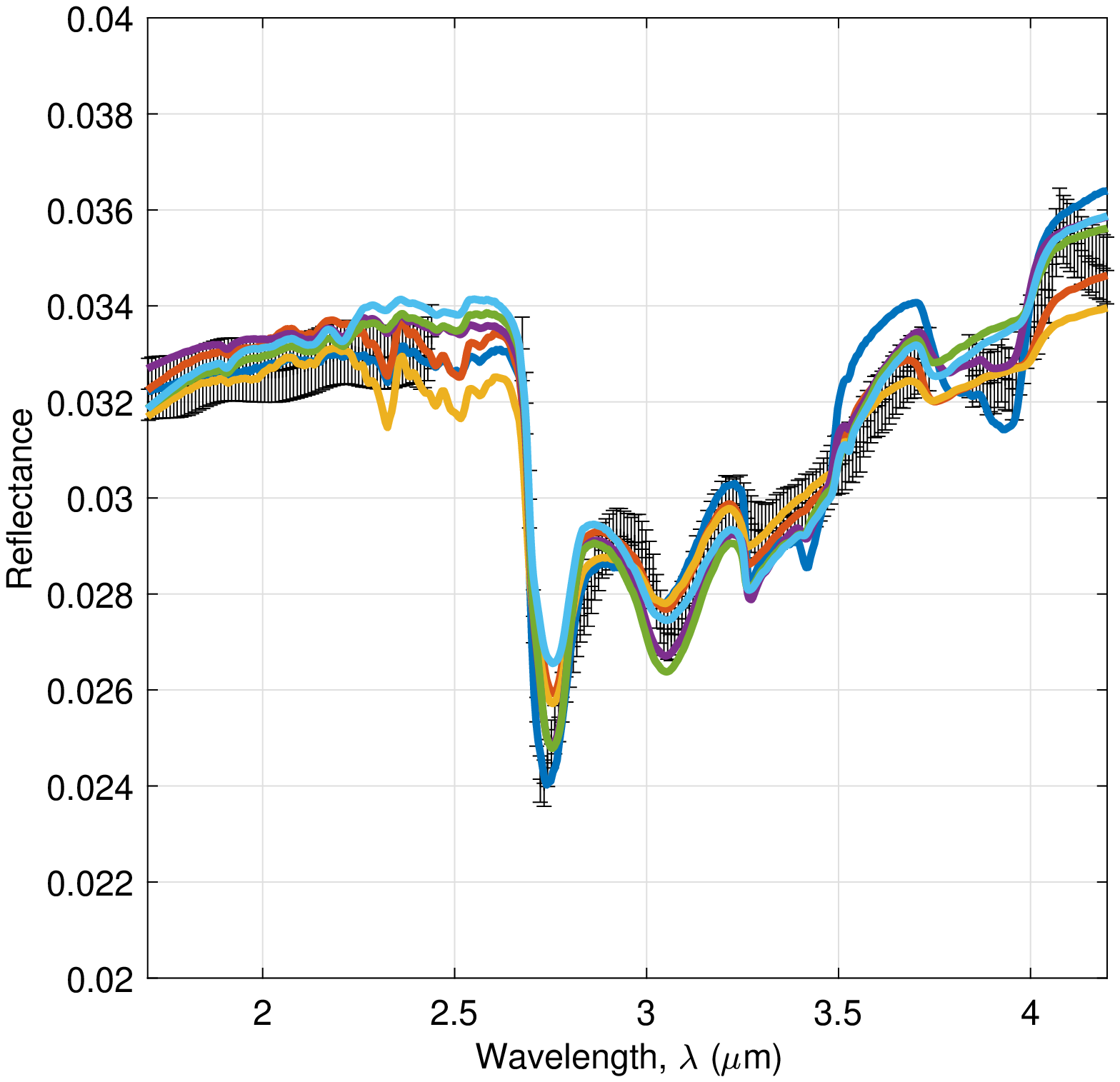}
    \includegraphics[width=0.49\linewidth]{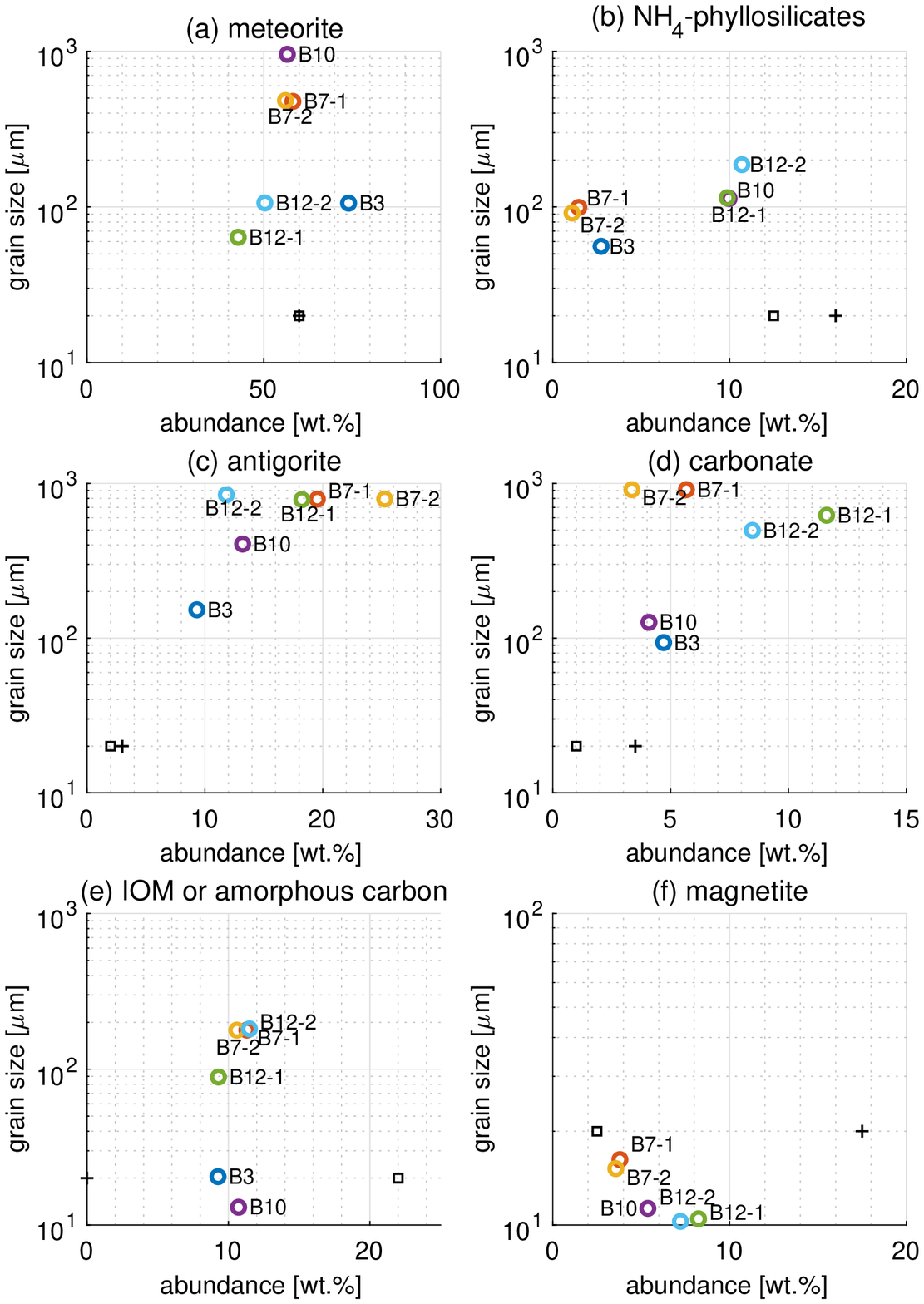}
    \caption{Models which satisfy both the spectral (VIR) and elemental (GRaND) constraints (Section 2.1). Left: Modeled spectra of assemblages B3 (blue), B7 (orange and yellow), B10 (purple), and B12 (green and cyan). Ceres' average spectrum is shown for comparison (black). Light: Modeled phase abundances and grain sizes (Table \ref{tab:VIR+GRaND}, circles). \citeA{Marchi+2018}'s models, their Case A (cross) and Case B (square), are shown for comparison. We note that the cross and squire overlap with each other in panel (a).}
    \label{fig:VIR+GRaND}
\end{figure}

\newpage

\begin{table}
    \centering
    \begin{tabular}{lrrrrrr}
          & Meteorite & NH$_4$-montmorillonite & Antigorite & Magnesite & IOM or amorphous C & Magnetite \\ \hline
          B3 & 74 & 2.7 & 9.3 & 4.7 & 9.3 & 0 \\ 
          B7-1 & 58 & 1.5 & 20 & 5.7 & 11 & 3.8 \\ 
          B7-2 & 56 & 1.1 & 25 & 3.3 & 11 & 3.6 \\
          B10 & 57 & 10 & 13 & 4.1 & 11 & 5.4 \\ 
          B12-1 & 43 & 9.9 & 18 & 12 & 9.3 & 8.2 \\ 
          B12-2 & 50 & 11 & 12 & 8.5 & 11 & 7.2 \\ 
    \end{tabular}
    \caption{Phase abundances (wt.\%) in models which satisfy both the spectral (VIR) and elemental (GRaND) constraints.}
    \label{tab:VIR+GRaND}
\end{table}

From darkest to brightest, we investigate the reflectance spectra of Ivuna, Tagish Lake, Cold Bokkeveld, and Murchison (Figure \ref{fig:endmember_spectra}).
CI chondrites are rich in iron sulfides compared to CM chondrites \cite{Howard+2009,Howard+2011,King+2015}, possibly explaining why Ivuna has the lowest albedo. However, there is no simple correlation between albedo and C or Fe abundances (Table \ref{tab:elements}), which is consistent with the inference that multiple phases contribute to darkening.

Modeled spectra are compared with Ceres' VIR spectrum in Figure \ref{fig:B1-12_spectra}. NH$_4$-montmorillonite, antigorite, magnesite are assumed for all mixtures and the endmembers for carbonaceous chondrite and darkening agent endmember vary. Assemblages B1--B4 (Figures \ref{fig:B1-12_spectra}a--\ref{fig:B1-12_spectra}d) differ in the choice of CCs used as an endmember (Table \ref{tab:models}). All of their MAP spectra provide a good fit to Ceres, though they are outside error bars in some wavelengths, e.g., shallower $\sim2.7\ {\rm \mu m}$ absorption in B1, B7, B9, and B11, and  $\sim3.9\ {\rm \mu m}$ absorption in B3 and B4.
Specifically, both CM and CI types appear to provide good analogs for Ceres' dark material, when IOM is added.
Addition of magnetite or substitution of amorphous carbon for IOM in assemblages B5--B12 do not always improve the spectral fit significantly (Figures \ref{fig:B1-12_spectra}e--\ref{fig:B1-12_spectra}l).

Minerals having characteristic absorption features (NH$_4$-montmorillonite, antigorite, and magnesite) are relatively well constrained with estimated abundances $<$20\% and typically with modes and MAPs closer to $\sim$10\% (Figures \ref{fig:B1-4_minerals} and \ref{fig:B5-12_minerals}).
The abundances of dark phases (IOM, amorphous carbon, magnetite, and meteorite) display broad and/or multi-modal PDFs, a likely result of trade-off relations between the abundances of various dark phases in matching the albedo of Ceres.
Notably, the presence of CCs typically reduces the amount of IOM required (Figures \ref{fig:B1-4_minerals} and \ref{fig:B5-12_minerals} vs. Figure \ref{fig:A_minerals_A1-6}).

Elemental abundances in CC-bearing assemblages (B1--B12; Figure \ref{fig:B1-12_elements}) are more consistent with GRaND data than those in assemblages without CCs (A1--A6; Figure \ref{fig:A_elements}).
For nearly all of the investigated mixtures, there exists a model solution for which the modeled elemental abundances match GRaND data.
Overall, the mismatch between VIR and GRaND-derived H and C abundances was reduced by adding CCs as endmembers in assemblages B1--B4.
Further adding magnetite (B5--B8) and substituting amorphous carbon for IOM (B9--B12) makes the H and C abundances comparable to the estimate from GRaND data, although this improvement does not result from more H contributed by magnetite or IOM, but because less of the dark phases is required to match the albedo, leading to higher abundances of hydrated minerals and CCs.
K abundances are not significantly affected by the specific assemblages, and our upper bounds remain consistent with GRaND data.
Whereas Fe abundances for assemblages without magnetite (B1--B4) are smaller than GRaND estimates, assemblages with magnetite (B5--12) are comparable within 2$\sigma$ in PDFs.
Carbon abundances remain larger than GRaND constraints in most models; however, many models are compatible. 
Compared to B1--B4, adding magnetite (B5--B8) or substituting amorphous carbon for IOM (B9--B12) reduces the C abundance.
Finally, there is a clear anti-correlation between the abundances of C and Fe for assemblages B5--B12 (Figure \ref{fig:correlation_B12}), but contrary to assemblages A1--A6 (Figure \ref{fig:correlation_A6}), several accepted models now satisfy the GRaND constraints.

Including various CCs as an endmember in spectral modeling of VIR data not only produces good spectral fits, but it resolves the discrepancy between VIR- and GRaND-derived C and Fe abundances. We showed the spectra and mineral abundances of select models (out of $2\times 10^6$ models) that satisfy both spectral (VIR) and elemental (GRaND) constraints (Figure \ref{fig:VIR+GRaND} and Table \ref{tab:VIR+GRaND}). We additionally filtered the results in Figure \ref{fig:VIR+GRaND} and Table \ref{tab:VIR+GRaND} to show those that not only have an acceptable $\chi^2$ over the entire spectral range, but also have absorption depths at $2.7$, $3.1$, $3.4$, and $4.0$ $\mu$m at least 50\% of those in VIR data (Section \ref{sec:method}). Only 6 accepted models from assemblages B3, B7, B10, and B12 fully reconcile VIR and GRaND data. We note that their modeled spectra do not match Ceres' average spectrum for the entire wavelength range because of tread-offs to satisfy the constraints on elemental abundances. These models suggest that Ceres is composed of CI/CM-chondrites-like materials ($40$--$70$ wt.\%) with excess amounts of darkening phases such as carbon and magnetite ($10$--$20$ wt.\% in total), as well as hydrous minerals ($10$--$25$ wt.\%), carbonates ($4$--$12$ wt.\%), and NH$_4$-bearing phyllosilicates ($1$--$11$ wt.\%). 

\section{Discussion}
\label{sec:discussion}

\subsection{Comparisons with previous studies}

Our spectral unmixing highlights the importance of identifying all spectral endmembers in the mixture and not including those not in fact present. Any model that includes artificial scaling or synthetic endmembers cannot provide absolute abundances. If a model contains all the endmembers that are actually present, and accounts adequately for their radiative contribution, either with a mathematical formulation or with compound endmembers as we did in this study, the best fit (or the MAP model in our MCMC method) will correspond to the true mineralogical - and elemental - abundances.

Here, we improved on the initial results of \citeA{DeSanctis+2015}, who reported a single best-fit estimate of mineral abundances to Ceres’ VIR data. Specifically, we derived full probability distributions of possible phase assemblages using the same radiative transfer model, but combining it with an MCMC algorithm (Figure \ref{fig:A_minerals_A1-6}, \ref{fig:B1-4_minerals}, and \ref{fig:B5-12_minerals}).
For all assemblages, we found that the 2$\sigma$ intervals of phase abundances are typically 5--20\% for phases that are responsible for absorption features (NH$_4$-bearing phases, antigorite, and magnesite). In contrast, 2$\sigma$ intervals are wider for featureless dark materials, especially when multiple dark phases are included in the mixture.
The range reflects noise in VIR observations as well as the uncertainty in grain sizes.
Considering the variety of possible mineral assemblages (assemblages A1--A6 and B1--B12) makes the uncertainty even larger.
Nevertheless, our results quantitatively corroborate those of \citeA{DeSanctis+2015} that Ceres is enriched in NH$_4$-phyllosilicate, carbonate, and serpentine relative to CCs.

We updated the optical constants for magnetite from those used in \citeA{DeSanctis+2015} (Section \ref{sec:method}). The reflectance of magnetite computed from the exact optical constant was brighter than Ceres' average reflectance, which made magnetite no longer most suitable as the darkening agent (Figure \ref{fig:A_spectra}b), in contrast to the results of \citeA{DeSanctis+2015} but consistent with Fe contents lower than CCs that have been reported by GRaND \cite{Prettyman+2018b}. Assumption of isotropic scattering in our model may underestimate the darkening effect of magnetite due to its forward scattering \cite{Mustard+Pieters1989}. However, analysis of Ceres' photometric properties \cite{Ciarniello+2017,Ciarniello+2020} showed back scattering or symmetrical scattering, depending on different assumptions, which suggests that strong forward scattering is unlikely to be a major cause of Ceres' low albedo.

Our work shows that mimicking CC properties, in particular composition and/or texture of their diverse opaque phases is important to fitting Ceres’ spectral properties with modest additional amounts of IOM or amorphous carbon ($\sim$10 wt.\%) and smaller amounts of magnetite (3--8 wt.\%) that are fully consistent with both VIR and GRaND data. \citeA{Marchi+2018} also modeled Ceres' spectra with CC endmembers. They reconciled the discrepancy between VIR-derived and GRaND-derived elemental compositions, with a model where the absolute value of reflectance is a free parameter. In contrast, we succeeded to reconcile VIR and GRaND constraints without using the multiplicative factor free-parameter on absolute reflectance. Compared to \citeA{Marchi+2018}, our model led to larger grain sizes because of its control on the absolute level of reflectance (Figure \ref{fig:VIR+GRaND}). Given the recently observed properties of Ryugu \cite{Watanabe+2019,Sugita+2019,Jaumann+2019} and Bennu \cite{Lauretta+2019,Dellagiustina+2019,Walsh+2019}, also hydrous asteroids and with surfaces including large cobble to boulder-sized clasts, a Ceres regolith of exclusively fine particles may be unlikely. The different model assumptions resulted in the differences in all phase abundances as well: our model results have more serpentine (antigorite) and carbonates but less NH$_4$-phyllosilicates than those in \citeA{Marchi+2018}. Our model results show that dark phases only require $\sim$10 wt.\% carbon (versus the up to 20 wt.\% in \citeA{Marchi+2018}) and $<$10 wt.\% magnetite (versus the up to 18 wt.\% in \citeA{Marchi+2018}).

The crustal density reflects the mineral composition, but we do not consider it as a constraint on the phase abundance in our model. The density of Ceres' 30--40 km kilometer thick crust is 1,200--1,400 ${\rm kg\ m^{-3}}$, as inferred from gravity measurements, and is thus significantly lower than any endmember phase because ice is inferred to represent many 10's wt.\% in the Ceres bulk crust \cite{Ermakov+2017}. With impact gardening \cite{Stein+2017}, the surface is likely very porous and so the bulk density does not necessarily equals the solid density. Additionally, as discussed by \citeA{Marchi+2018}, the VIR measurements probe a dehydrated near-surface layer where the density is not constrained by gravity data.

\subsection{Implications for formation and evolution of Ceres}
\label{subsec:TagishLake}

Whereas multiple mixtures are permitted, all accepted models suggest that Ceres experienced an alteration process similar to, but more thorough than, CCs \cite{Mcsween+2017}.
Alternatively, alteration on Ceres may have involved systems richer in volatile species (H, C, N) than is typical for CCs, possibly due to accretion of materials located outside NH$_3$ snowline \cite{DeSanctis+2015} and/or ice-rock differentiation \cite{Park+2016,Ermakov+2017}. 
Such a distant origin has been discussed for Tagish Lake parent body from the high abundance of $^{13}$C-rich carbonates \cite{fujiya2019migration}.
Our model showed that Ceres can be modeled by mixing of Tagish Lake-like materials with amorphous carbon, IOM and other hydrous minerals (B3, Figure \ref{fig:VIR+GRaND}). The position of 2.7 ${\rm \mu m}$ absorption in assemblage B3 matches that of Ceres. These results are consistent with the distant origin of Ceres (and Tagish Lake parent body) beyond the NH$_3$ and CO$_2$ snow lines.
The lack of ammoniated-phyllosilicates in Tagish Lake may suggest the different alteration pathways between the two bodies, as a result of their different sizes or Tagish Lake's source depth.
Models of the accretion, differentiation, and alteration of asteroids that account for the differences in surface bulk composition are required to determine whether an outer-Solar-System origin needs to be invoked.
 
\section{Conclusions}
\label{sec:conclusions}

We derived Ceres' surface phase abundances, with full uncertainties, for a set of possible phase assemblages from data acquired by Dawn's VIR spectrometer. We find that, within 2$\sigma$, abundances of NH$_4$-bearing phases, serpentine, and carbonate are of about 5--20\%. Dark phases have larger uncertainties. Comparing the resulting elemental abundances of H, C, K, and Fe with those derived from GRaND data demonstrates that only including pure endmember phases in the models leads to overestimated abundances of C and Fe (through over-modeling of IOM, carbonaceous materials, and magnetite). Relative to the true reflectance of CCs, the spectral unmixing method using independently measured CC abundances underestimates the darkening effect of the dark materials. The unmixing of Ceres' reflectance spectra using CCs as an endmember resolves the discrepancy from elemental abundances of C and Fe measured by GRaND. Ceres' average reflectance spectrum (VIR) and elemental abundances (GRaND) are reproduced by CC-like composition, e.g., like that of including Tagish Lake with modest amounts of additional carbon and minor additions of Fe oxides like magnetite. Ceres is composed of CI/CM-chondrite-like materials ($40$--$70$ wt.\%), with additional carbon like IOM or amorphous carbon ($10$ wt.\%), magnetite ($3$--$8$ wt.\%), serpentine ($10$--$25$ wt.\%), carbonates ($4$--$12$ wt.\%), and NH$_4$-bearing phyllosilicates ($1$--$11$ wt.\%). The best match in the position of 2.7 ${\rm \mu m}$ absorption in Assemblage B3 is consistent with the hypothetical similarity in the formation and evolution history with Tagish Lake parent body. Our results are consistent with the scenario that Ceres accreted materials which originate outside NH$_3$ and CO$_2$ snow lines and that its surface materials experienced greater alteration than CCs.

\acknowledgments
We thank the editor and two anonymous reviewers for comments which improved this manuscript. We thank Driss Takir and Hannah Kaplan for sharing meteorite and organics reflectance data.
We also thank to Alexis Templeton for discussing her early analyses of iron speciation and sulfidization in the serpentinized Oman Drilling Program ophiolite core. This work was supported by the JSPS Core-to-Core Program "International Network of Planetary Sciences".
H.K. was supported by JSPS KAKENHI Grant number 17H01175, 17H06457, 18K13602, 19H01960, and 19H05072. B.L.E thanks the Dawn team for welcoming her as a Science Affiliate to collaborate on data analysis during the mission's Ceres phase. 
All data from this work can be downloaded at  https://doi.org/ 10.22002/D1.1628 \cite{https://doi.org/10.22002/d1.1628}.
MATLAB scripts for MCMC computation are available at http://resolver.caltech.edu/CaltechAUTHORS:20170302-115016869.
Laboratory spectra presented in this paper are available from the RELAB spectral library and the authors of corresponding publications.

\bibliography{references.bib}

\begin{thebibliography}{}

\bibitem [\protect \citeauthoryear {%
Alexander%
\ \protect \BOthers {.}}{%
Alexander%
\ \protect \BOthers {.}}{%
{\protect \APACyear {2012}}%
}]{%
Alexander+2012}
\APACinsertmetastar {%
Alexander+2012}%
\begin{APACrefauthors}%
Alexander, C\BPBI O.%
, Bowden, R.%
, Fogel, M.%
, Howard, K.%
, Herd, C.%
\BCBL {}\ \BBA {} Nittler, L.%
\end{APACrefauthors}%
\unskip\
\newblock
\APACrefYearMonthDay{2012}{}{}.
\newblock
{\BBOQ}\APACrefatitle {The provenances of asteroids, and their contributions to
  the volatile inventories of the terrestrial planets} {The provenances of
  asteroids, and their contributions to the volatile inventories of the
  terrestrial planets}.{\BBCQ}
\newblock
\APACjournalVolNumPages{Science}{337}{6095}{721--723}.
\PrintBackRefs{\CurrentBib}

\bibitem [\protect \citeauthoryear {%
Ammannito%
\ \protect \BOthers {.}}{%
Ammannito%
\ \protect \BOthers {.}}{%
{\protect \APACyear {2016}}%
}]{%
Ammannito+2016}
\APACinsertmetastar {%
Ammannito+2016}%
\begin{APACrefauthors}%
Ammannito, E.%
, DeSanctis, M.%
, Ciarniello, M.%
, Frigeri, A.%
, Carrozzo, F.%
, Combe, J\BHBI P.%
\BDBL {}others%
\end{APACrefauthors}%
\unskip\
\newblock
\APACrefYearMonthDay{2016}{}{}.
\newblock
{\BBOQ}\APACrefatitle {Distribution of phyllosilicates on the surface of
  {C}eres} {Distribution of phyllosilicates on the surface of {C}eres}.{\BBCQ}
\newblock
\APACjournalVolNumPages{Science}{353}{6303}{aaf4279}.
\PrintBackRefs{\CurrentBib}

\bibitem [\protect \citeauthoryear {%
Barrat%
\ \protect \BOthers {.}}{%
Barrat%
\ \protect \BOthers {.}}{%
{\protect \APACyear {2012}}%
}]{%
Barrat+2012}
\APACinsertmetastar {%
Barrat+2012}%
\begin{APACrefauthors}%
Barrat, J\BHBI A.%
, Zanda, B.%
, Moynier, F.%
, Bollinger, C.%
, Liorzou, C.%
\BCBL {}\ \BBA {} Bayon, G.%
\end{APACrefauthors}%
\unskip\
\newblock
\APACrefYearMonthDay{2012}{}{}.
\newblock
{\BBOQ}\APACrefatitle {Geochemistry of {CI} chondrites: Major and trace
  elements, and {C}u and {Z}n isotopes} {Geochemistry of {CI} chondrites: Major
  and trace elements, and {C}u and {Z}n isotopes}.{\BBCQ}
\newblock
\APACjournalVolNumPages{Geochimica et Cosmochimica Acta}{83}{}{79--92}.
\PrintBackRefs{\CurrentBib}

\bibitem [\protect \citeauthoryear {%
Beck%
\ \protect \BOthers {.}}{%
Beck%
\ \protect \BOthers {.}}{%
{\protect \APACyear {2018}}%
}]{%
Beck+2018}
\APACinsertmetastar {%
Beck+2018}%
\begin{APACrefauthors}%
Beck, P.%
, Maturilli, A.%
, Garenne, A.%
, Vernazza, P.%
, Helbert, J.%
, Quirico, E.%
\BCBL {}\ \BBA {} Schmitt, B.%
\end{APACrefauthors}%
\unskip\
\newblock
\APACrefYearMonthDay{2018}{}{}.
\newblock
{\BBOQ}\APACrefatitle {What is controlling the reflectance spectra (0.35--150
  $\mu$m) of hydrated (and dehydrated) carbonaceous chondrites?} {What is
  controlling the reflectance spectra (0.35--150 $\mu$m) of hydrated (and
  dehydrated) carbonaceous chondrites?}{\BBCQ}
\newblock
\APACjournalVolNumPages{Icarus}{313}{}{124--138}.
\PrintBackRefs{\CurrentBib}

\bibitem [\protect \citeauthoryear {%
J.~Bishop%
, Banin%
, Mancinelli%
\BCBL {}\ \BBA {} Klovstad%
}{%
J.~Bishop%
\ \protect \BOthers {.}}{%
{\protect \APACyear {2002}}%
}]{%
Bishop+2002}
\APACinsertmetastar {%
Bishop+2002}%
\begin{APACrefauthors}%
Bishop, J.%
, Banin, A.%
, Mancinelli, R.%
\BCBL {}\ \BBA {} Klovstad, M.%
\end{APACrefauthors}%
\unskip\
\newblock
\APACrefYearMonthDay{2002}{}{}.
\newblock
{\BBOQ}\APACrefatitle {Detection of soluble and fixed {NH}$_4^+$ in clay
  minerals by {DTA} and {IR} reflectance spectroscopy: a potential tool for
  planetary surface exploration} {Detection of soluble and fixed {NH}$_4^+$ in
  clay minerals by {DTA} and {IR} reflectance spectroscopy: a potential tool
  for planetary surface exploration}.{\BBCQ}
\newblock
\APACjournalVolNumPages{Planetary and Space Science}{50}{1}{11--19}.
\PrintBackRefs{\CurrentBib}

\bibitem [\protect \citeauthoryear {%
J\BPBI L.~Bishop%
, Pieters%
\BCBL {}\ \BBA {} Edwards%
}{%
J\BPBI L.~Bishop%
\ \protect \BOthers {.}}{%
{\protect \APACyear {1994}}%
}]{%
Bishop+1994}
\APACinsertmetastar {%
Bishop+1994}%
\begin{APACrefauthors}%
Bishop, J\BPBI L.%
, Pieters, C\BPBI M.%
\BCBL {}\ \BBA {} Edwards, J\BPBI O.%
\end{APACrefauthors}%
\unskip\
\newblock
\APACrefYearMonthDay{1994}{}{}.
\newblock
{\BBOQ}\APACrefatitle {Infrared spectroscopic analyses on the nature of water
  in montmorillonite} {Infrared spectroscopic analyses on the nature of water
  in montmorillonite}.{\BBCQ}
\newblock
\BIn{} \APACrefbtitle {Clays and clay minerals.} {Clays and clay minerals.}
\PrintBackRefs{\CurrentBib}

\bibitem [\protect \citeauthoryear {%
P.~Bland%
, Zolensky%
, Benedix%
\BCBL {}\ \BBA {} Sephton%
}{%
P.~Bland%
\ \protect \BOthers {.}}{%
{\protect \APACyear {2006}}%
}]{%
Bland+2006}
\APACinsertmetastar {%
Bland+2006}%
\begin{APACrefauthors}%
Bland, P.%
, Zolensky, M.%
, Benedix, G.%
\BCBL {}\ \BBA {} Sephton, M.%
\end{APACrefauthors}%
\unskip\
\newblock
\APACrefYearMonthDay{2006}{}{}.
\newblock
{\BBOQ}\APACrefatitle {Weathering of chondritic meteorites} {Weathering of
  chondritic meteorites}.{\BBCQ}
\newblock
\APACjournalVolNumPages{Meteorites and the early solar system
  II}{1}{}{853--867}.
\PrintBackRefs{\CurrentBib}

\bibitem [\protect \citeauthoryear {%
P\BPBI A.~Bland%
, Cressey%
\BCBL {}\ \BBA {} Menzies%
}{%
P\BPBI A.~Bland%
\ \protect \BOthers {.}}{%
{\protect \APACyear {2004}}%
}]{%
Bland+2004}
\APACinsertmetastar {%
Bland+2004}%
\begin{APACrefauthors}%
Bland, P\BPBI A.%
, Cressey, G.%
\BCBL {}\ \BBA {} Menzies, O\BPBI N.%
\end{APACrefauthors}%
\unskip\
\newblock
\APACrefYearMonthDay{2004}{}{}.
\newblock
{\BBOQ}\APACrefatitle {Modal mineralogy of carbonaceous chondrites by {X}-ray
  diffraction and {M}{\"o}ssbauer spectroscopy} {Modal mineralogy of
  carbonaceous chondrites by {X}-ray diffraction and {M}{\"o}ssbauer
  spectroscopy}.{\BBCQ}
\newblock
\APACjournalVolNumPages{Meteoritics \& Planetary Science}{39}{1}{3--16}.
\PrintBackRefs{\CurrentBib}

\bibitem [\protect \citeauthoryear {%
Blinova%
, Zega%
, Herd%
\BCBL {}\ \BBA {} Stroud%
}{%
Blinova%
\ \protect \BOthers {.}}{%
{\protect \APACyear {2014}}%
}]{%
Blinova+2014}
\APACinsertmetastar {%
Blinova+2014}%
\begin{APACrefauthors}%
Blinova, A\BPBI I.%
, Zega, T\BPBI J.%
, Herd, C\BPBI D.%
\BCBL {}\ \BBA {} Stroud, R\BPBI M.%
\end{APACrefauthors}%
\unskip\
\newblock
\APACrefYearMonthDay{2014}{}{}.
\newblock
{\BBOQ}\APACrefatitle {Testing variations within the {T}agish {L}ake meteorite
  -- {I}: Mineralogy and petrology of pristine samples} {Testing variations
  within the {T}agish {L}ake meteorite -- {I}: Mineralogy and petrology of
  pristine samples}.{\BBCQ}
\newblock
\APACjournalVolNumPages{Meteoritics \& Planetary Science}{49}{4}{473--502}.
\PrintBackRefs{\CurrentBib}

\bibitem [\protect \citeauthoryear {%
Britt%
, Yeomans%
, Housen%
\BCBL {}\ \BBA {} Consolmagno%
}{%
Britt%
\ \protect \BOthers {.}}{%
{\protect \APACyear {2003}}%
}]{%
Britt+2003}
\APACinsertmetastar {%
Britt+2003}%
\begin{APACrefauthors}%
Britt, D\BPBI T.%
, Yeomans, D.%
, Housen, K.%
\BCBL {}\ \BBA {} Consolmagno, G.%
\end{APACrefauthors}%
\unskip\
\newblock
\APACrefYearMonthDay{2003}{}{}.
\newblock
{\BBOQ}\APACrefatitle {Asteroid density, porosity, and structure} {Asteroid
  density, porosity, and structure}.{\BBCQ}
\newblock

\PrintBackRefs{\CurrentBib}

\bibitem [\protect \citeauthoryear {%
Brown%
\ \protect \BOthers {.}}{%
Brown%
\ \protect \BOthers {.}}{%
{\protect \APACyear {2000}}%
}]{%
Brown+2000}
\APACinsertmetastar {%
Brown+2000}%
\begin{APACrefauthors}%
Brown, P\BPBI G.%
, Hildebrand, A\BPBI R.%
, Zolensky, M\BPBI E.%
, Grady, M.%
, Clayton, R\BPBI N.%
, Mayeda, T\BPBI K.%
\BDBL {}others%
\end{APACrefauthors}%
\unskip\
\newblock
\APACrefYearMonthDay{2000}{}{}.
\newblock
{\BBOQ}\APACrefatitle {The fall, recovery, orbit, and composition of the
  {T}agish {L}ake meteorite: A new type of carbonaceous chondrite} {The fall,
  recovery, orbit, and composition of the {T}agish {L}ake meteorite: A new type
  of carbonaceous chondrite}.{\BBCQ}
\newblock
\APACjournalVolNumPages{Science}{290}{5490}{320--325}.
\PrintBackRefs{\CurrentBib}

\bibitem [\protect \citeauthoryear {%
Carrozzo%
\ \protect \BOthers {.}}{%
Carrozzo%
\ \protect \BOthers {.}}{%
{\protect \APACyear {2018}}%
}]{%
Carrozzo+2018}
\APACinsertmetastar {%
Carrozzo+2018}%
\begin{APACrefauthors}%
Carrozzo, F\BPBI G.%
, De~Sanctis, M\BPBI C.%
, Raponi, A.%
, Ammannito, E.%
, Castillo-Rogez, J.%
, Ehlmann, B\BPBI L.%
\BDBL {}others%
\end{APACrefauthors}%
\unskip\
\newblock
\APACrefYearMonthDay{2018}{}{}.
\newblock
{\BBOQ}\APACrefatitle {Nature, formation, and distribution of carbonates on
  {C}eres} {Nature, formation, and distribution of carbonates on
  {C}eres}.{\BBCQ}
\newblock
\APACjournalVolNumPages{Science advances}{4}{3}{e1701645}.
\PrintBackRefs{\CurrentBib}

\bibitem [\protect \citeauthoryear {%
Carrozzo%
\ \protect \BOthers {.}}{%
Carrozzo%
\ \protect \BOthers {.}}{%
{\protect \APACyear {2016}}%
}]{%
Carrozzo+2016}
\APACinsertmetastar {%
Carrozzo+2016}%
\begin{APACrefauthors}%
Carrozzo, F\BPBI G.%
, Raponi, A.%
, De~Sanctis, M.%
, Ammannito, E.%
, Giardino, M.%
, D’Aversa, E.%
\BDBL {}Tosi, F.%
\end{APACrefauthors}%
\unskip\
\newblock
\APACrefYearMonthDay{2016}{}{}.
\newblock
{\BBOQ}\APACrefatitle {Artifacts reduction in VIR/Dawn data} {Artifacts
  reduction in vir/dawn data}.{\BBCQ}
\newblock
\APACjournalVolNumPages{Review of Scientific Instruments}{87}{12}{124501}.
\PrintBackRefs{\CurrentBib}

\bibitem [\protect \citeauthoryear {%
Ciarniello%
\ \protect \BOthers {.}}{%
Ciarniello%
\ \protect \BOthers {.}}{%
{\protect \APACyear {2017}}%
}]{%
Ciarniello+2017}
\APACinsertmetastar {%
Ciarniello+2017}%
\begin{APACrefauthors}%
Ciarniello, M.%
, De~Sanctis, M.%
, Ammannito, E.%
, Raponi, A.%
, Longobardo, A.%
, Palomba, E.%
\BDBL {}others%
\end{APACrefauthors}%
\unskip\
\newblock
\APACrefYearMonthDay{2017}{}{}.
\newblock
{\BBOQ}\APACrefatitle {Spectrophotometric properties of dwarf planet {C}eres
  from the {VIR} spectrometer on board the {D}awn mission} {Spectrophotometric
  properties of dwarf planet {C}eres from the {VIR} spectrometer on board the
  {D}awn mission}.{\BBCQ}
\newblock
\APACjournalVolNumPages{Astronomy \& Astrophysics}{598}{}{A130}.
\PrintBackRefs{\CurrentBib}

\bibitem [\protect \citeauthoryear {%
Ciarniello%
\ \protect \BOthers {.}}{%
Ciarniello%
\ \protect \BOthers {.}}{%
{\protect \APACyear {2020}}%
}]{%
Ciarniello+2020}
\APACinsertmetastar {%
Ciarniello+2020}%
\begin{APACrefauthors}%
Ciarniello, M.%
, De~Sanctis, M.%
, Raponi, A.%
, Rousseau, B.%
, Longobardo, A.%
, Li, J\BHBI Y.%
\BDBL {}others%
\end{APACrefauthors}%
\unskip\
\newblock
\APACrefYearMonthDay{2020}{}{}.
\newblock
{\BBOQ}\APACrefatitle {Ceres observed at low phase angles by {VIR}-{D}awn}
  {Ceres observed at low phase angles by {VIR}-{D}awn}.{\BBCQ}
\newblock
\APACjournalVolNumPages{Astronomy \& Astrophysics}{634}{}{A39}.
\PrintBackRefs{\CurrentBib}

\bibitem [\protect \citeauthoryear {%
Clark%
}{%
Clark%
}{%
{\protect \APACyear {1983}}%
}]{%
Clark1983}
\APACinsertmetastar {%
Clark1983}%
\begin{APACrefauthors}%
Clark, R\BPBI N.%
\end{APACrefauthors}%
\unskip\
\newblock
\APACrefYearMonthDay{1983}{}{}.
\newblock
{\BBOQ}\APACrefatitle {Spectral properties of mixtures of montmorillonite and
  dark carbon grains: Implications for remote sensing minerals containing
  chemically and physically adsorbed water} {Spectral properties of mixtures of
  montmorillonite and dark carbon grains: Implications for remote sensing
  minerals containing chemically and physically adsorbed water}.{\BBCQ}
\newblock
\APACjournalVolNumPages{Journal of Geophysical Research: Solid
  Earth}{88}{B12}{10635--10644}.
\PrintBackRefs{\CurrentBib}

\bibitem [\protect \citeauthoryear {%
E.~Cloutis%
, Hudon%
, Hiroi%
\BCBL {}\ \BBA {} Gaffey%
}{%
E.~Cloutis%
\ \protect \BOthers {.}}{%
{\protect \APACyear {2012}}%
}]{%
Cloutis+2012}
\APACinsertmetastar {%
Cloutis+2012}%
\begin{APACrefauthors}%
Cloutis, E.%
, Hudon, P.%
, Hiroi, T.%
\BCBL {}\ \BBA {} Gaffey, M.%
\end{APACrefauthors}%
\unskip\
\newblock
\APACrefYearMonthDay{2012}{}{}.
\newblock
{\BBOQ}\APACrefatitle {Spectral reflectance properties of carbonaceous
  chondrites 4: Aqueously altered and thermally metamorphosed meteorites}
  {Spectral reflectance properties of carbonaceous chondrites 4: Aqueously
  altered and thermally metamorphosed meteorites}.{\BBCQ}
\newblock
\APACjournalVolNumPages{Icarus}{220}{2}{586--617}.
\PrintBackRefs{\CurrentBib}

\bibitem [\protect \citeauthoryear {%
E\BPBI A.~Cloutis%
, Gaffey%
, Smith%
\BCBL {}\ \BBA {} Lambert%
}{%
E\BPBI A.~Cloutis%
\ \protect \BOthers {.}}{%
{\protect \APACyear {1990}}%
}]{%
Cloutis+1990}
\APACinsertmetastar {%
Cloutis+1990}%
\begin{APACrefauthors}%
Cloutis, E\BPBI A.%
, Gaffey, M\BPBI J.%
, Smith, D\BPBI G.%
\BCBL {}\ \BBA {} Lambert, R\BPBI S\BPBI J.%
\end{APACrefauthors}%
\unskip\
\newblock
\APACrefYearMonthDay{1990}{}{}.
\newblock
{\BBOQ}\APACrefatitle {Reflectance spectra of mafic silicate-opaque assemblages
  with applications to meteorite spectra} {Reflectance spectra of mafic
  silicate-opaque assemblages with applications to meteorite spectra}.{\BBCQ}
\newblock
\APACjournalVolNumPages{Icarus}{84}{2}{315--333}.
\PrintBackRefs{\CurrentBib}

\bibitem [\protect \citeauthoryear {%
Dalton~III%
\ \BBA {} Pitman%
}{%
Dalton~III%
\ \BBA {} Pitman%
}{%
{\protect \APACyear {2012}}%
}]{%
Dalton+Pitman2012}
\APACinsertmetastar {%
Dalton+Pitman2012}%
\begin{APACrefauthors}%
Dalton~III, J.%
\BCBT {}\ \BBA {} Pitman, K.%
\end{APACrefauthors}%
\unskip\
\newblock
\APACrefYearMonthDay{2012}{}{}.
\newblock
{\BBOQ}\APACrefatitle {Low temperature optical constants of some hydrated
  sulfates relevant to planetary surfaces} {Low temperature optical constants
  of some hydrated sulfates relevant to planetary surfaces}.{\BBCQ}
\newblock
\APACjournalVolNumPages{Journal of Geophysical Research: Planets}{117}{E9}{}.
\PrintBackRefs{\CurrentBib}

\bibitem [\protect \citeauthoryear {%
De~Angelis%
, Manzari%
, De~Sanctis%
, Ammannito%
\BCBL {}\ \BBA {} Di~Iorio%
}{%
De~Angelis%
\ \protect \BOthers {.}}{%
{\protect \APACyear {2016}}%
}]{%
DeAngelis+2016}
\APACinsertmetastar {%
DeAngelis+2016}%
\begin{APACrefauthors}%
De~Angelis, S.%
, Manzari, P.%
, De~Sanctis, M.%
, Ammannito, E.%
\BCBL {}\ \BBA {} Di~Iorio, T.%
\end{APACrefauthors}%
\unskip\
\newblock
\APACrefYearMonthDay{2016}{}{}.
\newblock
{\BBOQ}\APACrefatitle {{VIS-IR} study of brucite--clay--carbonate mixtures:
  Implications for Ceres surface composition} {{VIS-IR} study of
  brucite--clay--carbonate mixtures: Implications for ceres surface
  composition}.{\BBCQ}
\newblock
\APACjournalVolNumPages{Icarus}{280}{}{315--327}.
\PrintBackRefs{\CurrentBib}

\bibitem [\protect \citeauthoryear {%
DellaGiustina%
\ \protect \BOthers {.}}{%
DellaGiustina%
\ \protect \BOthers {.}}{%
{\protect \APACyear {2019}}%
}]{%
Dellagiustina+2019}
\APACinsertmetastar {%
Dellagiustina+2019}%
\begin{APACrefauthors}%
DellaGiustina, D.%
, Emery, J.%
, Golish, D.%
, Rozitis, B.%
, Bennett, C.%
, Burke, K.%
\BDBL {}others%
\end{APACrefauthors}%
\unskip\
\newblock
\APACrefYearMonthDay{2019}{}{}.
\newblock
{\BBOQ}\APACrefatitle {Properties of rubble-pile asteroid (101955) {B}ennu from
  {OSIRIS-RE}x imaging and thermal analysis} {Properties of rubble-pile
  asteroid (101955) {B}ennu from {OSIRIS-RE}x imaging and thermal
  analysis}.{\BBCQ}
\newblock
\APACjournalVolNumPages{Nature Astronomy}{3}{4}{341--351}.
\PrintBackRefs{\CurrentBib}

\bibitem [\protect \citeauthoryear {%
De~Sanctis%
\ \protect \BOthers {.}}{%
De~Sanctis%
\ \protect \BOthers {.}}{%
{\protect \APACyear {2018}}%
}]{%
DeSanctis+2018}
\APACinsertmetastar {%
DeSanctis+2018}%
\begin{APACrefauthors}%
De~Sanctis, M.%
, Ammannito, E.%
, Carrozzo, F.%
, Ciarniello, M.%
, Giardino, M.%
, Frigeri, A.%
\BDBL {}others%
\end{APACrefauthors}%
\unskip\
\newblock
\APACrefYearMonthDay{2018}{}{}.
\newblock
{\BBOQ}\APACrefatitle {Ceres's global and localized mineralogical composition
  determined by {D}awn's {V}isible and {I}nfrared {S}pectrometer ({VIR})}
  {Ceres's global and localized mineralogical composition determined by
  {D}awn's {V}isible and {I}nfrared {S}pectrometer ({VIR})}.{\BBCQ}
\newblock
\APACjournalVolNumPages{Meteoritics \& Planetary Science}{53}{9}{1844--1865}.
\PrintBackRefs{\CurrentBib}

\bibitem [\protect \citeauthoryear {%
De~Sanctis%
\ \protect \BOthers {.}}{%
De~Sanctis%
\ \protect \BOthers {.}}{%
{\protect \APACyear {2017}}%
}]{%
DeSanctis+2017}
\APACinsertmetastar {%
DeSanctis+2017}%
\begin{APACrefauthors}%
De~Sanctis, M.%
, Ammannito, E.%
, McSween, H.%
, Raponi, A.%
, Marchi, S.%
, Capaccioni, F.%
\BDBL {}others%
\end{APACrefauthors}%
\unskip\
\newblock
\APACrefYearMonthDay{2017}{}{}.
\newblock
{\BBOQ}\APACrefatitle {Localized aliphatic organic material on the surface of
  {C}eres} {Localized aliphatic organic material on the surface of
  {C}eres}.{\BBCQ}
\newblock
\APACjournalVolNumPages{Science}{355}{6326}{719--722}.
\PrintBackRefs{\CurrentBib}

\bibitem [\protect \citeauthoryear {%
De~Sanctis%
\ \protect \BOthers {.}}{%
De~Sanctis%
\ \protect \BOthers {.}}{%
{\protect \APACyear {2015}}%
}]{%
DeSanctis+2015}
\APACinsertmetastar {%
DeSanctis+2015}%
\begin{APACrefauthors}%
De~Sanctis, M.%
, Ammannito, E.%
, Raponi, A.%
, Marchi, S.%
, McCord, T\BPBI B.%
, McSween, H.%
\BDBL {}others%
\end{APACrefauthors}%
\unskip\
\newblock
\APACrefYearMonthDay{2015}{}{}.
\newblock
{\BBOQ}\APACrefatitle {Ammoniated phyllosilicates with a likely outer solar
  system origin on (1) {C}eres} {Ammoniated phyllosilicates with a likely outer
  solar system origin on (1) {C}eres}.{\BBCQ}
\newblock
\APACjournalVolNumPages{Nature}{528}{7581}{241}.
\PrintBackRefs{\CurrentBib}

\bibitem [\protect \citeauthoryear {%
De~Sanctis%
\ \protect \BOthers {.}}{%
De~Sanctis%
\ \protect \BOthers {.}}{%
{\protect \APACyear {2011}}%
}]{%
DeSanctis+2010}
\APACinsertmetastar {%
DeSanctis+2010}%
\begin{APACrefauthors}%
De~Sanctis, M.%
, Coradini, A.%
, Ammannito, E.%
, Filacchione, G.%
, Capria, M.%
, Fonte, S.%
\BDBL {}others%
\end{APACrefauthors}%
\unskip\
\newblock
\APACrefYearMonthDay{2011}{}{}.
\newblock
{\BBOQ}\APACrefatitle {The {VIR} spectrometer} {The {VIR} spectrometer}.{\BBCQ}
\newblock
\BIn{} \APACrefbtitle {The {D}awn mission to minor planets 4 {V}esta and 1
  {C}eres} {The {D}awn mission to minor planets 4 {V}esta and 1 {C}eres}\
  (\BPGS\ 329--369).
\newblock
\APACaddressPublisher{}{Springer}.
\PrintBackRefs{\CurrentBib}

\bibitem [\protect \citeauthoryear {%
De~Sanctis%
\ \protect \BOthers {.}}{%
De~Sanctis%
\ \protect \BOthers {.}}{%
{\protect \APACyear {2016}}%
}]{%
DeSanctis+2016}
\APACinsertmetastar {%
DeSanctis+2016}%
\begin{APACrefauthors}%
De~Sanctis, M.%
, Raponi, A.%
, Ammannito, E.%
, Ciarniello, M.%
, Toplis, M.%
, McSween, H.%
\BDBL {}others%
\end{APACrefauthors}%
\unskip\
\newblock
\APACrefYearMonthDay{2016}{}{}.
\newblock
{\BBOQ}\APACrefatitle {Bright carbonate deposits as evidence of aqueous
  alteration on (1) {C}eres} {Bright carbonate deposits as evidence of aqueous
  alteration on (1) {C}eres}.{\BBCQ}
\newblock
\APACjournalVolNumPages{Nature}{536}{7614}{54}.
\PrintBackRefs{\CurrentBib}

\bibitem [\protect \citeauthoryear {%
Ehlmann%
\ \protect \BOthers {.}}{%
Ehlmann%
\ \protect \BOthers {.}}{%
{\protect \APACyear {2018}}%
}]{%
Ehlmann+2018}
\APACinsertmetastar {%
Ehlmann+2018}%
\begin{APACrefauthors}%
Ehlmann, B\BPBI L.%
, Hodyss, R.%
, Bristow, T\BPBI F.%
, Rossman, G\BPBI R.%
, Ammannito, E.%
, De~Sanctis, M\BPBI C.%
\BCBL {}\ \BBA {} Raymond, C\BPBI A.%
\end{APACrefauthors}%
\unskip\
\newblock
\APACrefYearMonthDay{2018}{}{}.
\newblock
{\BBOQ}\APACrefatitle {Ambient and cold-temperature infrared spectra and {XRD}
  patterns of ammoniated phyllosilicates and carbonaceous chondrite meteorites
  relevant to {C}eres and other solar system bodies} {Ambient and
  cold-temperature infrared spectra and {XRD} patterns of ammoniated
  phyllosilicates and carbonaceous chondrite meteorites relevant to {C}eres and
  other solar system bodies}.{\BBCQ}
\newblock
\APACjournalVolNumPages{Meteoritics \& Planetary Science}{}{}{}.
\PrintBackRefs{\CurrentBib}

\bibitem [\protect \citeauthoryear {%
Ermakov%
\ \protect \BOthers {.}}{%
Ermakov%
\ \protect \BOthers {.}}{%
{\protect \APACyear {2017}}%
}]{%
Ermakov+2017}
\APACinsertmetastar {%
Ermakov+2017}%
\begin{APACrefauthors}%
Ermakov, A.%
, Fu, R.%
, Castillo-Rogez, J.%
, Raymond, C.%
, Park, R.%
, Preusker, F.%
\BDBL {}Zuber, M.%
\end{APACrefauthors}%
\unskip\
\newblock
\APACrefYearMonthDay{2017}{}{}.
\newblock
{\BBOQ}\APACrefatitle {Constraints on {C}eres' internal structure and evolution
  from its shape and gravity measured by the {D}awn spacecraft} {Constraints on
  {C}eres' internal structure and evolution from its shape and gravity measured
  by the {D}awn spacecraft}.{\BBCQ}
\newblock
\APACjournalVolNumPages{Journal of Geophysical Research:
  Planets}{122}{11}{2267--2293}.
\PrintBackRefs{\CurrentBib}

\bibitem [\protect \citeauthoryear {%
Feierberg%
, Lebofsky%
\BCBL {}\ \BBA {} Tholen%
}{%
Feierberg%
\ \protect \BOthers {.}}{%
{\protect \APACyear {1985}}%
}]{%
Feierberg+1985}
\APACinsertmetastar {%
Feierberg+1985}%
\begin{APACrefauthors}%
Feierberg, M\BPBI A.%
, Lebofsky, L\BPBI A.%
\BCBL {}\ \BBA {} Tholen, D\BPBI J.%
\end{APACrefauthors}%
\unskip\
\newblock
\APACrefYearMonthDay{1985}{}{}.
\newblock
{\BBOQ}\APACrefatitle {The nature of {C}-class asteroids from 3-$\mu$m
  spectrophotometry} {The nature of {C}-class asteroids from 3-$\mu$m
  spectrophotometry}.{\BBCQ}
\newblock
\APACjournalVolNumPages{Icarus}{63}{2}{183--191}.
\PrintBackRefs{\CurrentBib}

\bibitem [\protect \citeauthoryear {%
Fujiya%
\ \protect \BOthers {.}}{%
Fujiya%
\ \protect \BOthers {.}}{%
{\protect \APACyear {2019}}%
}]{%
fujiya2019migration}
\APACinsertmetastar {%
fujiya2019migration}%
\begin{APACrefauthors}%
Fujiya, W.%
, Hoppe, P.%
, Ushikubo, T.%
, Fukuda, K.%
, Lindgren, P.%
, Lee, M\BPBI R.%
\BDBL {}Sano, Y.%
\end{APACrefauthors}%
\unskip\
\newblock
\APACrefYearMonthDay{2019}{}{}.
\newblock
{\BBOQ}\APACrefatitle {Migration of {D}-type asteroids from the outer {S}olar
  {S}ystem inferred from carbonate in meteorites} {Migration of {D}-type
  asteroids from the outer {S}olar {S}ystem inferred from carbonate in
  meteorites}.{\BBCQ}
\newblock
\APACjournalVolNumPages{Nature Astronomy}{3}{10}{910--915}.
\PrintBackRefs{\CurrentBib}

\bibitem [\protect \citeauthoryear {%
Garenne%
\ \protect \BOthers {.}}{%
Garenne%
\ \protect \BOthers {.}}{%
{\protect \APACyear {2016}}%
}]{%
Garenne+2016}
\APACinsertmetastar {%
Garenne+2016}%
\begin{APACrefauthors}%
Garenne, A.%
, Beck, P.%
, Montes-Hernandez, G.%
, Brissaud, O.%
, Schmitt, B.%
, Quirico, E.%
\BDBL {}Howard, K.%
\end{APACrefauthors}%
\unskip\
\newblock
\APACrefYearMonthDay{2016}{}{}.
\newblock
{\BBOQ}\APACrefatitle {Bidirectional reflectance spectroscopy of carbonaceous
  chondrites: Implications for water quantification and primary composition}
  {Bidirectional reflectance spectroscopy of carbonaceous chondrites:
  Implications for water quantification and primary composition}.{\BBCQ}
\newblock
\APACjournalVolNumPages{Icarus}{264}{}{172--183}.
\PrintBackRefs{\CurrentBib}

\bibitem [\protect \citeauthoryear {%
Glotch%
\ \BBA {} Rossman%
}{%
Glotch%
\ \BBA {} Rossman%
}{%
{\protect \APACyear {2009}}%
}]{%
Glotch+2009}
\APACinsertmetastar {%
Glotch+2009}%
\begin{APACrefauthors}%
Glotch, T\BPBI D.%
\BCBT {}\ \BBA {} Rossman, G\BPBI R.%
\end{APACrefauthors}%
\unskip\
\newblock
\APACrefYearMonthDay{2009}{}{}.
\newblock
{\BBOQ}\APACrefatitle {Mid-infrared reflectance spectra and optical constants
  of six iron oxide/oxyhydroxide phases} {Mid-infrared reflectance spectra and
  optical constants of six iron oxide/oxyhydroxide phases}.{\BBCQ}
\newblock
\APACjournalVolNumPages{Icarus}{204}{2}{663--671}.
\PrintBackRefs{\CurrentBib}

\bibitem [\protect \citeauthoryear {%
Grady%
, Verchovsky%
, Franchi%
, Wright%
\BCBL {}\ \BBA {} Pillinger%
}{%
Grady%
\ \protect \BOthers {.}}{%
{\protect \APACyear {2002}}%
}]{%
Grady+2002}
\APACinsertmetastar {%
Grady+2002}%
\begin{APACrefauthors}%
Grady, M\BPBI M.%
, Verchovsky, A.%
, Franchi, I.%
, Wright, I.%
\BCBL {}\ \BBA {} Pillinger, C.%
\end{APACrefauthors}%
\unskip\
\newblock
\APACrefYearMonthDay{2002}{}{}.
\newblock
{\BBOQ}\APACrefatitle {Light dement geochemistry of the {T}agish {L}ake {CI}2
  chondrite: Comparison with {CI}1 and {CM}2 meteorites} {Light dement
  geochemistry of the {T}agish {L}ake {CI}2 chondrite: Comparison with {CI}1
  and {CM}2 meteorites}.{\BBCQ}
\newblock
\APACjournalVolNumPages{Meteoritics \& Planetary Science}{37}{5}{713--735}.
\PrintBackRefs{\CurrentBib}

\bibitem [\protect \citeauthoryear {%
Hapke%
}{%
Hapke%
}{%
{\protect \APACyear {2001}}%
}]{%
Hapke2001}
\APACinsertmetastar {%
Hapke2001}%
\begin{APACrefauthors}%
Hapke, B.%
\end{APACrefauthors}%
\unskip\
\newblock
\APACrefYearMonthDay{2001}{}{}.
\newblock
{\BBOQ}\APACrefatitle {Space weathering from {M}ercury to the asteroid belt}
  {Space weathering from {M}ercury to the asteroid belt}.{\BBCQ}
\newblock
\APACjournalVolNumPages{Journal of Geophysical Research:
  Planets}{106}{E5}{10039--10073}.
\PrintBackRefs{\CurrentBib}

\bibitem [\protect \citeauthoryear {%
Hapke%
}{%
Hapke%
}{%
{\protect \APACyear {2012}}%
}]{%
Hapke2012}
\APACinsertmetastar {%
Hapke2012}%
\begin{APACrefauthors}%
Hapke, B.%
\end{APACrefauthors}%
\unskip\
\newblock
\APACrefYear{2012}.
\newblock
\APACrefbtitle {Theory of reflectance and emittance spectroscopy} {Theory of
  reflectance and emittance spectroscopy}.
\newblock
\APACaddressPublisher{}{Cambridge university press}.
\PrintBackRefs{\CurrentBib}

\bibitem [\protect \citeauthoryear {%
Hendrix%
, Vilas%
\BCBL {}\ \BBA {} Li%
}{%
Hendrix%
\ \protect \BOthers {.}}{%
{\protect \APACyear {2016}}%
}]{%
Hendrix+2016}
\APACinsertmetastar {%
Hendrix+2016}%
\begin{APACrefauthors}%
Hendrix, A\BPBI R.%
, Vilas, F.%
\BCBL {}\ \BBA {} Li, J\BHBI Y.%
\end{APACrefauthors}%
\unskip\
\newblock
\APACrefYearMonthDay{2016}{}{}.
\newblock
{\BBOQ}\APACrefatitle {Ceres: Sulfur deposits and graphitized carbon} {Ceres:
  Sulfur deposits and graphitized carbon}.{\BBCQ}
\newblock
\APACjournalVolNumPages{Geophysical Research Letters}{43}{17}{8920--8927}.
\PrintBackRefs{\CurrentBib}

\bibitem [\protect \citeauthoryear {%
Hiroi%
, Zolensky%
\BCBL {}\ \BBA {} Pieters%
}{%
Hiroi%
\ \protect \BOthers {.}}{%
{\protect \APACyear {2001}}%
}]{%
Hiroi+2001}
\APACinsertmetastar {%
Hiroi+2001}%
\begin{APACrefauthors}%
Hiroi, T.%
, Zolensky, M\BPBI E.%
\BCBL {}\ \BBA {} Pieters, C\BPBI M.%
\end{APACrefauthors}%
\unskip\
\newblock
\APACrefYearMonthDay{2001}{}{}.
\newblock
{\BBOQ}\APACrefatitle {The {T}agish {L}ake meteorite: A possible sample from a
  {D}-type asteroid} {The {T}agish {L}ake meteorite: A possible sample from a
  {D}-type asteroid}.{\BBCQ}
\newblock
\APACjournalVolNumPages{Science}{293}{5538}{2234--2236}.
\PrintBackRefs{\CurrentBib}

\bibitem [\protect \citeauthoryear {%
Howard%
, Benedix%
, Bland%
\BCBL {}\ \BBA {} Cressey%
}{%
Howard%
\ \protect \BOthers {.}}{%
{\protect \APACyear {2009}}%
}]{%
Howard+2009}
\APACinsertmetastar {%
Howard+2009}%
\begin{APACrefauthors}%
Howard, K.%
, Benedix, G.%
, Bland, P.%
\BCBL {}\ \BBA {} Cressey, G.%
\end{APACrefauthors}%
\unskip\
\newblock
\APACrefYearMonthDay{2009}{}{}.
\newblock
{\BBOQ}\APACrefatitle {Modal mineralogy of {CM}2 chondrites by {X}-ray
  diffraction ({PSD}-{XRD}). {P}art 1: Total phyllosilicate abundance and the
  degree of aqueous alteration} {Modal mineralogy of {CM}2 chondrites by
  {X}-ray diffraction ({PSD}-{XRD}). {P}art 1: Total phyllosilicate abundance
  and the degree of aqueous alteration}.{\BBCQ}
\newblock
\APACjournalVolNumPages{Geochimica et Cosmochimica Acta}{73}{15}{4576--4589}.
\PrintBackRefs{\CurrentBib}

\bibitem [\protect \citeauthoryear {%
Howard%
, Benedix%
, Bland%
\BCBL {}\ \BBA {} Cressey%
}{%
Howard%
\ \protect \BOthers {.}}{%
{\protect \APACyear {2011}}%
}]{%
Howard+2011}
\APACinsertmetastar {%
Howard+2011}%
\begin{APACrefauthors}%
Howard, K.%
, Benedix, G.%
, Bland, P.%
\BCBL {}\ \BBA {} Cressey, G.%
\end{APACrefauthors}%
\unskip\
\newblock
\APACrefYearMonthDay{2011}{}{}.
\newblock
{\BBOQ}\APACrefatitle {Modal mineralogy of {CM} chondrites by {X}-ray
  diffraction ({PSD}-{XRD}): Part 2. Degree, nature and settings of aqueous
  alteration} {Modal mineralogy of {CM} chondrites by {X}-ray diffraction
  ({PSD}-{XRD}): Part 2. degree, nature and settings of aqueous
  alteration}.{\BBCQ}
\newblock
\APACjournalVolNumPages{Geochimica et Cosmochimica Acta}{75}{10}{2735--2751}.
\PrintBackRefs{\CurrentBib}

\bibitem [\protect \citeauthoryear {%
Izawa%
, Flemming%
, King%
, Peterson%
\BCBL {}\ \BBA {} McCAUSLAND%
}{%
Izawa%
\ \protect \BOthers {.}}{%
{\protect \APACyear {2010}}%
}]{%
Izawa+2010}
\APACinsertmetastar {%
Izawa+2010}%
\begin{APACrefauthors}%
Izawa, M\BPBI R.%
, Flemming, R\BPBI L.%
, King, P\BPBI L.%
, Peterson, R\BPBI C.%
\BCBL {}\ \BBA {} McCAUSLAND, P\BPBI J.%
\end{APACrefauthors}%
\unskip\
\newblock
\APACrefYearMonthDay{2010}{}{}.
\newblock
{\BBOQ}\APACrefatitle {Mineralogical and spectroscopic investigation of the
  {T}agish {L}ake carbonaceous chondrite by {X}-ray diffraction and infrared
  reflectance spectroscopy} {Mineralogical and spectroscopic investigation of
  the {T}agish {L}ake carbonaceous chondrite by {X}-ray diffraction and
  infrared reflectance spectroscopy}.{\BBCQ}
\newblock
\APACjournalVolNumPages{Meteoritics \& Planetary Science}{45}{4}{675--698}.
\PrintBackRefs{\CurrentBib}

\bibitem [\protect \citeauthoryear {%
Jaumann%
\ \protect \BOthers {.}}{%
Jaumann%
\ \protect \BOthers {.}}{%
{\protect \APACyear {2019}}%
}]{%
Jaumann+2019}
\APACinsertmetastar {%
Jaumann+2019}%
\begin{APACrefauthors}%
Jaumann, R.%
, Schmitz, N.%
, Ho, T\BHBI M.%
, Schr{\"o}der, S.%
, Otto, K\BPBI A.%
, Stephan, K.%
\BDBL {}others%
\end{APACrefauthors}%
\unskip\
\newblock
\APACrefYearMonthDay{2019}{}{}.
\newblock
{\BBOQ}\APACrefatitle {Images from the surface of asteroid {R}yugu show rocks
  similar to carbonaceous chondrite meteorites} {Images from the surface of
  asteroid {R}yugu show rocks similar to carbonaceous chondrite
  meteorites}.{\BBCQ}
\newblock
\APACjournalVolNumPages{Science}{365}{6455}{817--820}.
\PrintBackRefs{\CurrentBib}

\bibitem [\protect \citeauthoryear {%
Kaplan%
, Milliken%
\BCBL {}\ \BBA {} O'D.~Alexander%
}{%
Kaplan%
\ \protect \BOthers {.}}{%
{\protect \APACyear {2018}}%
}]{%
Kaplan+2018}
\APACinsertmetastar {%
Kaplan+2018}%
\begin{APACrefauthors}%
Kaplan, H\BPBI H.%
, Milliken, R\BPBI E.%
\BCBL {}\ \BBA {} O'D.~Alexander, C\BPBI M.%
\end{APACrefauthors}%
\unskip\
\newblock
\APACrefYearMonthDay{2018}{}{}.
\newblock
{\BBOQ}\APACrefatitle {New constraints on the abundance and composition of
  organic matter on {C}eres} {New constraints on the abundance and composition
  of organic matter on {C}eres}.{\BBCQ}
\newblock
\APACjournalVolNumPages{Geophysical Research Letters}{}{}{}.
\PrintBackRefs{\CurrentBib}

\bibitem [\protect \citeauthoryear {%
Kerridge%
}{%
Kerridge%
}{%
{\protect \APACyear {1985}}%
}]{%
Kerridge1985}
\APACinsertmetastar {%
Kerridge1985}%
\begin{APACrefauthors}%
Kerridge, J\BPBI F.%
\end{APACrefauthors}%
\unskip\
\newblock
\APACrefYearMonthDay{1985}{}{}.
\newblock
{\BBOQ}\APACrefatitle {Carbon, hydrogen and nitrogen in carbonaceous
  chondrites: Abundances and isotopic compositions in bulk samples} {Carbon,
  hydrogen and nitrogen in carbonaceous chondrites: Abundances and isotopic
  compositions in bulk samples}.{\BBCQ}
\newblock
\APACjournalVolNumPages{Geochimica et Cosmochimica Acta}{49}{8}{1707--1714}.
\PrintBackRefs{\CurrentBib}

\bibitem [\protect \citeauthoryear {%
A.~King%
, Schofield%
, Howard%
\BCBL {}\ \BBA {} Russell%
}{%
A.~King%
\ \protect \BOthers {.}}{%
{\protect \APACyear {2015}}%
}]{%
King+2015}
\APACinsertmetastar {%
King+2015}%
\begin{APACrefauthors}%
King, A.%
, Schofield, P.%
, Howard, K.%
\BCBL {}\ \BBA {} Russell, S.%
\end{APACrefauthors}%
\unskip\
\newblock
\APACrefYearMonthDay{2015}{}{}.
\newblock
{\BBOQ}\APACrefatitle {Modal mineralogy of {CI} and {CI}-like chondrites by
  {X}-ray diffraction} {Modal mineralogy of {CI} and {CI}-like chondrites by
  {X}-ray diffraction}.{\BBCQ}
\newblock
\APACjournalVolNumPages{Geochimica et Cosmochimica Acta}{165}{}{148--160}.
\PrintBackRefs{\CurrentBib}

\bibitem [\protect \citeauthoryear {%
T\BPBI V.~King%
, Clark%
, Calvin%
, Sherman%
\BCBL {}\ \BBA {} Brown%
}{%
T\BPBI V.~King%
\ \protect \BOthers {.}}{%
{\protect \APACyear {1992}}%
}]{%
King+1992}
\APACinsertmetastar {%
King+1992}%
\begin{APACrefauthors}%
King, T\BPBI V.%
, Clark, R.%
, Calvin, W.%
, Sherman, D\BPBI M.%
\BCBL {}\ \BBA {} Brown, R.%
\end{APACrefauthors}%
\unskip\
\newblock
\APACrefYearMonthDay{1992}{}{}.
\newblock
{\BBOQ}\APACrefatitle {Evidence for ammonium-bearing minerals on {C}eres}
  {Evidence for ammonium-bearing minerals on {C}eres}.{\BBCQ}
\newblock
\APACjournalVolNumPages{Science}{255}{5051}{1551--1553}.
\PrintBackRefs{\CurrentBib}

\bibitem [\protect \citeauthoryear {%
T\BPBI V.~King%
\ \BBA {} Clark%
}{%
T\BPBI V.~King%
\ \BBA {} Clark%
}{%
{\protect \APACyear {1989}}%
}]{%
King+Clark1989}
\APACinsertmetastar {%
King+Clark1989}%
\begin{APACrefauthors}%
King, T\BPBI V.%
\BCBT {}\ \BBA {} Clark, R\BPBI N.%
\end{APACrefauthors}%
\unskip\
\newblock
\APACrefYearMonthDay{1989}{}{}.
\newblock
{\BBOQ}\APACrefatitle {Spectral characteristics of chlorites and
  {M}g-serpentines using high-resolution reflectance spectroscopy} {Spectral
  characteristics of chlorites and {M}g-serpentines using high-resolution
  reflectance spectroscopy}.{\BBCQ}
\newblock
\APACjournalVolNumPages{Journal of Geophysical Research: Solid
  Earth}{94}{B10}{13997--14008}.
\PrintBackRefs{\CurrentBib}

\bibitem [\protect \citeauthoryear {%
{Kurokawa}%
\ \protect \BOthers {.}}{%
{Kurokawa}%
\ \protect \BOthers {.}}{%
{\protect \APACyear {2018}}%
}]{%
Kurokawa+2018}
\APACinsertmetastar {%
Kurokawa+2018}%
\begin{APACrefauthors}%
{Kurokawa}, H.%
, {Ehlmann}, B\BPBI L.%
, {Ammannito}, E.%
, {De Sanctis}, M\BPBI C.%
, {Lapotre}, M.%
, {Usui}, T.%
\BDBL {}{Ciarniello}, M.%
\end{APACrefauthors}%
\unskip\
\newblock
\APACrefYearMonthDay{2018}{{\APACmonth{03}}}{}.
\newblock
{\BBOQ}\APACrefatitle {{A {B}ayesian approach to deriving {C}eres surface
  composition from {D}awn {VIR} data: Initial quantification of bright spot and
  typical dark material phases with this method}} {{A {B}ayesian approach to
  deriving {C}eres surface composition from {D}awn {VIR} data: Initial
  quantification of bright spot and typical dark material phases with this
  method}}.{\BBCQ}
\newblock
\BIn{} \APACrefbtitle {Lunar and Planetary Science Conference} {Lunar and
  planetary science conference}\ (\BVOL~49, \BPG~1908).
\PrintBackRefs{\CurrentBib}

\bibitem [\protect \citeauthoryear {%
Kurokawa%
\ \protect \BOthers {.}}{%
Kurokawa%
\ \protect \BOthers {.}}{%
{\protect \APACyear {2020}}%
}]{%
https://doi.org/10.22002/d1.1628}
\APACinsertmetastar {%
https://doi.org/10.22002/d1.1628}%
\begin{APACrefauthors}%
Kurokawa, H.%
, Ehlmann, B\BPBI L.%
, {M. C. De Sanctis}%
, { M. G. A. Lapotre}%
, {T. Usui}%
, {N. T. Stein}%
\BDBL {}{M. Ciarniello}%
\end{APACrefauthors}%
\unskip\
\newblock
\APACrefYearMonthDay{2020}{}{}.
\newblock
\APACrefbtitle {Data accompanying "A probabilistic approach to determination of
  Ceres' average surface composition from {D}awn {VIR} and {GR}a{ND} data".}
  {Data accompanying "a probabilistic approach to determination of ceres'
  average surface composition from {D}awn {VIR} and {GR}a{ND} data".}
\newblock
\APACaddressPublisher{}{CaltechDATA}.
\newblock
\begin{APACrefURL} \url{https://data.caltech.edu/records/1628} \end{APACrefURL}
\newblock
\begin{APACrefDOI} \doi{10.22002/D1.1628} \end{APACrefDOI}
\PrintBackRefs{\CurrentBib}

\bibitem [\protect \citeauthoryear {%
Lapotre%
, Ehlmann%
, Minson%
, Arvidson%
\BCBL {}\ \protect \BOthers {.}}{%
Lapotre%
, Ehlmann%
, Minson%
, Arvidson%
\BCBL {}\ \protect \BOthers {.}}{%
{\protect \APACyear {2017}}%
}]{%
Lapotre+2017b}
\APACinsertmetastar {%
Lapotre+2017b}%
\begin{APACrefauthors}%
Lapotre, M\BPBI G.%
, Ehlmann, B.%
, Minson, S\BPBI E.%
, Arvidson, R.%
, Ayoub, F.%
, Fraeman, A.%
\BDBL {}Bridges, N.%
\end{APACrefauthors}%
\unskip\
\newblock
\APACrefYearMonthDay{2017}{}{}.
\newblock
{\BBOQ}\APACrefatitle {Compositional variations in sands of the {B}agnold
  {D}unes, {G}ale crater, {M}ars, from visible-shortwave infrared spectroscopy
  and comparison with ground truth from the {C}uriosity rover} {Compositional
  variations in sands of the {B}agnold {D}unes, {G}ale crater, {M}ars, from
  visible-shortwave infrared spectroscopy and comparison with ground truth from
  the {C}uriosity rover}.{\BBCQ}
\newblock
\APACjournalVolNumPages{Journal of Geophysical Research:
  Planets}{122}{12}{2489--2509}.
\PrintBackRefs{\CurrentBib}

\bibitem [\protect \citeauthoryear {%
Lapotre%
, Ehlmann%
\BCBL {}\ \BBA {} Minson%
}{%
Lapotre%
, Ehlmann%
\BCBL {}\ \BBA {} Minson%
}{%
{\protect \APACyear {2017}}%
}]{%
Lapotre+2017a}
\APACinsertmetastar {%
Lapotre+2017a}%
\begin{APACrefauthors}%
Lapotre, M\BPBI G.%
, Ehlmann, B\BPBI L.%
\BCBL {}\ \BBA {} Minson, S\BPBI E.%
\end{APACrefauthors}%
\unskip\
\newblock
\APACrefYearMonthDay{2017}{}{}.
\newblock
{\BBOQ}\APACrefatitle {A probabilistic approach to remote compositional
  analysis of planetary surfaces} {A probabilistic approach to remote
  compositional analysis of planetary surfaces}.{\BBCQ}
\newblock
\APACjournalVolNumPages{Journal of Geophysical Research:
  Planets}{122}{5}{983--1009}.
\PrintBackRefs{\CurrentBib}

\bibitem [\protect \citeauthoryear {%
Lauretta%
\ \protect \BOthers {.}}{%
Lauretta%
\ \protect \BOthers {.}}{%
{\protect \APACyear {2019}}%
}]{%
Lauretta+2019}
\APACinsertmetastar {%
Lauretta+2019}%
\begin{APACrefauthors}%
Lauretta, D.%
, DellaGiustina, D.%
, Bennett, C.%
, Golish, D.%
, Becker, K.%
, Balram-Knutson, S.%
\BDBL {}others%
\end{APACrefauthors}%
\unskip\
\newblock
\APACrefYearMonthDay{2019}{}{}.
\newblock
{\BBOQ}\APACrefatitle {The unexpected surface of asteroid (101955) {B}ennu}
  {The unexpected surface of asteroid (101955) {B}ennu}.{\BBCQ}
\newblock
\APACjournalVolNumPages{Nature}{568}{7750}{55--60}.
\PrintBackRefs{\CurrentBib}

\bibitem [\protect \citeauthoryear {%
Lawrence%
\ \protect \BOthers {.}}{%
Lawrence%
\ \protect \BOthers {.}}{%
{\protect \APACyear {2018}}%
}]{%
Lawrence+2018}
\APACinsertmetastar {%
Lawrence+2018}%
\begin{APACrefauthors}%
Lawrence, D\BPBI J.%
, Peplowski, P\BPBI N.%
, Beck, A\BPBI W.%
, Feldman, W\BPBI C.%
, Prettyman, T\BPBI H.%
, Russell, C\BPBI T.%
\BDBL {}others%
\end{APACrefauthors}%
\unskip\
\newblock
\APACrefYearMonthDay{2018}{}{}.
\newblock
{\BBOQ}\APACrefatitle {Compositional variability on the surface of 1 {C}eres
  revealed through {GR}a{ND} measurements of high-energy gamma rays}
  {Compositional variability on the surface of 1 {C}eres revealed through
  {GR}a{ND} measurements of high-energy gamma rays}.{\BBCQ}
\newblock
\APACjournalVolNumPages{Meteoritics \& Planetary Science}{}{}{}.
\PrintBackRefs{\CurrentBib}

\bibitem [\protect \citeauthoryear {%
Levison%
\ \protect \BOthers {.}}{%
Levison%
\ \protect \BOthers {.}}{%
{\protect \APACyear {2009}}%
}]{%
Levison+2009}
\APACinsertmetastar {%
Levison+2009}%
\begin{APACrefauthors}%
Levison, H\BPBI F.%
, Bottke, W\BPBI F.%
, Gounelle, M.%
, Morbidelli, A.%
, Nesvorn{\`y}, D.%
\BCBL {}\ \BBA {} Tsiganis, K.%
\end{APACrefauthors}%
\unskip\
\newblock
\APACrefYearMonthDay{2009}{}{}.
\newblock
{\BBOQ}\APACrefatitle {Contamination of the asteroid belt by primordial
  trans-{N}eptunian objects} {Contamination of the asteroid belt by primordial
  trans-{N}eptunian objects}.{\BBCQ}
\newblock
\APACjournalVolNumPages{Nature}{460}{7253}{364}.
\PrintBackRefs{\CurrentBib}

\bibitem [\protect \citeauthoryear {%
Li%
\ \BBA {} Li%
}{%
Li%
\ \BBA {} Li%
}{%
{\protect \APACyear {2011}}%
}]{%
Li+Li2011}
\APACinsertmetastar {%
Li+Li2011}%
\begin{APACrefauthors}%
Li, S.%
\BCBT {}\ \BBA {} Li, L.%
\end{APACrefauthors}%
\unskip\
\newblock
\APACrefYearMonthDay{2011}{}{}.
\newblock
{\BBOQ}\APACrefatitle {Radiative transfer modeling for quantifying lunar
  surface minerals, particle size, and submicroscopic metallic {F}e} {Radiative
  transfer modeling for quantifying lunar surface minerals, particle size, and
  submicroscopic metallic {F}e}.{\BBCQ}
\newblock
\APACjournalVolNumPages{Journal of Geophysical Research: Planets}{116}{E9}{}.
\PrintBackRefs{\CurrentBib}

\bibitem [\protect \citeauthoryear {%
Lodders%
, Fegley%
, Lodders%
\BCBL {}\ \protect \BOthers {.}}{%
Lodders%
\ \protect \BOthers {.}}{%
{\protect \APACyear {1998}}%
}]{%
Lodders+1998}
\APACinsertmetastar {%
Lodders+1998}%
\begin{APACrefauthors}%
Lodders, K.%
, Fegley, B.%
, Lodders, F.%
\BCBL {}\ \BOthersPeriod {.}\end{APACrefauthors}%
\unskip\
\newblock
\APACrefYear{1998}.
\newblock
\APACrefbtitle {The planetary scientist's companion} {The planetary scientist's
  companion}.
\newblock
\APACaddressPublisher{}{Oxford University Press on Demand}.
\PrintBackRefs{\CurrentBib}

\bibitem [\protect \citeauthoryear {%
Lucey%
}{%
Lucey%
}{%
{\protect \APACyear {1998}}%
}]{%
Lucey1998}
\APACinsertmetastar {%
Lucey1998}%
\begin{APACrefauthors}%
Lucey, P\BPBI G.%
\end{APACrefauthors}%
\unskip\
\newblock
\APACrefYearMonthDay{1998}{}{}.
\newblock
{\BBOQ}\APACrefatitle {Model near-infrared optical constants of olivine and
  pyroxene as a function of iron content} {Model near-infrared optical
  constants of olivine and pyroxene as a function of iron content}.{\BBCQ}
\newblock
\APACjournalVolNumPages{Journal of Geophysical Research:
  Planets}{103}{E1}{1703--1713}.
\PrintBackRefs{\CurrentBib}

\bibitem [\protect \citeauthoryear {%
Marchi%
\ \protect \BOthers {.}}{%
Marchi%
\ \protect \BOthers {.}}{%
{\protect \APACyear {2018}}%
}]{%
Marchi+2018}
\APACinsertmetastar {%
Marchi+2018}%
\begin{APACrefauthors}%
Marchi, S.%
, Raponi, A.%
, Prettyman, T.%
, De~Sanctis, M.%
, Castillo-Rogez, J.%
, Raymond, C.%
\BDBL {}others%
\end{APACrefauthors}%
\unskip\
\newblock
\APACrefYearMonthDay{2018}{}{}.
\newblock
{\BBOQ}\APACrefatitle {An aqueously altered carbon-rich {C}eres} {An aqueously
  altered carbon-rich {C}eres}.{\BBCQ}
\newblock
\APACjournalVolNumPages{Nature Astronomy}{}{}{1}.
\PrintBackRefs{\CurrentBib}

\bibitem [\protect \citeauthoryear {%
McAdam%
, Sunshine%
, Howard%
\BCBL {}\ \BBA {} McCoy%
}{%
McAdam%
\ \protect \BOthers {.}}{%
{\protect \APACyear {2015}}%
}]{%
McAdam+2015}
\APACinsertmetastar {%
McAdam+2015}%
\begin{APACrefauthors}%
McAdam, M.%
, Sunshine, J.%
, Howard, K.%
\BCBL {}\ \BBA {} McCoy, T.%
\end{APACrefauthors}%
\unskip\
\newblock
\APACrefYearMonthDay{2015}{}{}.
\newblock
{\BBOQ}\APACrefatitle {Aqueous alteration on asteroids: Linking the mineralogy
  and spectroscopy of {CM} and {CI} chondrites} {Aqueous alteration on
  asteroids: Linking the mineralogy and spectroscopy of {CM} and {CI}
  chondrites}.{\BBCQ}
\newblock
\APACjournalVolNumPages{Icarus}{245}{}{320--332}.
\PrintBackRefs{\CurrentBib}

\bibitem [\protect \citeauthoryear {%
McKinnon%
}{%
McKinnon%
}{%
{\protect \APACyear {2012}}%
}]{%
McKinnon2012}
\APACinsertmetastar {%
McKinnon2012}%
\begin{APACrefauthors}%
McKinnon, W.%
\end{APACrefauthors}%
\unskip\
\newblock
\APACrefYearMonthDay{2012}{}{}.
\newblock
{\BBOQ}\APACrefatitle {Where did {C}eres accrete?} {Where did {C}eres
  accrete?}{\BBCQ}
\newblock
\BIn{} \APACrefbtitle {Asteroids, Comets, Meteors 2012} {Asteroids, comets,
  meteors 2012}\ (\BVOL\ 1667).
\PrintBackRefs{\CurrentBib}

\bibitem [\protect \citeauthoryear {%
McSween%
\ \protect \BOthers {.}}{%
McSween%
\ \protect \BOthers {.}}{%
{\protect \APACyear {2017}}%
}]{%
Mcsween+2017}
\APACinsertmetastar {%
Mcsween+2017}%
\begin{APACrefauthors}%
McSween, H\BPBI Y.%
, Emery, J\BPBI P.%
, Rivkin, A\BPBI S.%
, Toplis, M\BPBI J.%
, C.~Castillo-Rogez, J.%
, Prettyman, T\BPBI H.%
\BDBL {}Russell, C\BPBI T.%
\end{APACrefauthors}%
\unskip\
\newblock
\APACrefYearMonthDay{2017}{}{}.
\newblock
{\BBOQ}\APACrefatitle {Carbonaceous chondrites as analogs for the composition
  and alteration of {C}eres} {Carbonaceous chondrites as analogs for the
  composition and alteration of {C}eres}.{\BBCQ}
\newblock
\APACjournalVolNumPages{Meteoritics \& Planetary Science}{}{}{}.
\PrintBackRefs{\CurrentBib}

\bibitem [\protect \citeauthoryear {%
Milliken%
\ \BBA {} Rivkin%
}{%
Milliken%
\ \BBA {} Rivkin%
}{%
{\protect \APACyear {2009}}%
}]{%
Milliken+Rivkin2009}
\APACinsertmetastar {%
Milliken+Rivkin2009}%
\begin{APACrefauthors}%
Milliken, R\BPBI E.%
\BCBT {}\ \BBA {} Rivkin, A\BPBI S.%
\end{APACrefauthors}%
\unskip\
\newblock
\APACrefYearMonthDay{2009}{}{}.
\newblock
{\BBOQ}\APACrefatitle {Brucite and carbonate assemblages from altered
  olivine-rich materials on {C}eres} {Brucite and carbonate assemblages from
  altered olivine-rich materials on {C}eres}.{\BBCQ}
\newblock
\APACjournalVolNumPages{Nature Geoscience}{2}{4}{258}.
\PrintBackRefs{\CurrentBib}

\bibitem [\protect \citeauthoryear {%
Mooney%
\ \BBA {} Knacke%
}{%
Mooney%
\ \BBA {} Knacke%
}{%
{\protect \APACyear {1985}}%
}]{%
Mooney+Knacke1985}
\APACinsertmetastar {%
Mooney+Knacke1985}%
\begin{APACrefauthors}%
Mooney, T.%
\BCBT {}\ \BBA {} Knacke, R.%
\end{APACrefauthors}%
\unskip\
\newblock
\APACrefYearMonthDay{1985}{}{}.
\newblock
{\BBOQ}\APACrefatitle {Optical constants of chlorite and serpentine between 2.5
  and 50 $\mu$m} {Optical constants of chlorite and serpentine between 2.5 and
  50 $\mu$m}.{\BBCQ}
\newblock
\APACjournalVolNumPages{Icarus}{64}{3}{493--502}.
\PrintBackRefs{\CurrentBib}

\bibitem [\protect \citeauthoryear {%
Moroz%
, Arnold%
, Korochantsev%
\BCBL {}\ \BBA {} W{\"a}sch%
}{%
Moroz%
\ \protect \BOthers {.}}{%
{\protect \APACyear {1998}}%
}]{%
Moroz+1998}
\APACinsertmetastar {%
Moroz+1998}%
\begin{APACrefauthors}%
Moroz, L.%
, Arnold, G.%
, Korochantsev, A.%
\BCBL {}\ \BBA {} W{\"a}sch, R.%
\end{APACrefauthors}%
\unskip\
\newblock
\APACrefYearMonthDay{1998}{}{}.
\newblock
{\BBOQ}\APACrefatitle {Natural solid bitumens as possible analogs for cometary
  and asteroid organics: 1. reflectance spectroscopy of pure bitumens} {Natural
  solid bitumens as possible analogs for cometary and asteroid organics: 1.
  reflectance spectroscopy of pure bitumens}.{\BBCQ}
\newblock
\APACjournalVolNumPages{Icarus}{134}{2}{253--268}.
\PrintBackRefs{\CurrentBib}

\bibitem [\protect \citeauthoryear {%
Mustard%
\ \BBA {} Hays%
}{%
Mustard%
\ \BBA {} Hays%
}{%
{\protect \APACyear {1997}}%
}]{%
Mustard+Hays1997}
\APACinsertmetastar {%
Mustard+Hays1997}%
\begin{APACrefauthors}%
Mustard, J\BPBI F.%
\BCBT {}\ \BBA {} Hays, J\BPBI E.%
\end{APACrefauthors}%
\unskip\
\newblock
\APACrefYearMonthDay{1997}{}{}.
\newblock
{\BBOQ}\APACrefatitle {Effects of hyperfine particles on reflectance spectra
  from 0.3 to 25 $\mu$m} {Effects of hyperfine particles on reflectance spectra
  from 0.3 to 25 $\mu$m}.{\BBCQ}
\newblock
\APACjournalVolNumPages{Icarus}{125}{1}{145--163}.
\PrintBackRefs{\CurrentBib}

\bibitem [\protect \citeauthoryear {%
Mustard%
\ \BBA {} Pieters%
}{%
Mustard%
\ \BBA {} Pieters%
}{%
{\protect \APACyear {1989}}%
}]{%
Mustard+Pieters1989}
\APACinsertmetastar {%
Mustard+Pieters1989}%
\begin{APACrefauthors}%
Mustard, J\BPBI F.%
\BCBT {}\ \BBA {} Pieters, C\BPBI M.%
\end{APACrefauthors}%
\unskip\
\newblock
\APACrefYearMonthDay{1989}{}{}.
\newblock
{\BBOQ}\APACrefatitle {Photometric phase functions of common geologic minerals
  and applications to quantitative analysis of mineral mixture reflectance
  spectra} {Photometric phase functions of common geologic minerals and
  applications to quantitative analysis of mineral mixture reflectance
  spectra}.{\BBCQ}
\newblock
\APACjournalVolNumPages{Journal of Geophysical Research: Solid
  Earth}{94}{B10}{13619--13634}.
\PrintBackRefs{\CurrentBib}

\bibitem [\protect \citeauthoryear {%
Park%
\ \protect \BOthers {.}}{%
Park%
\ \protect \BOthers {.}}{%
{\protect \APACyear {2016}}%
}]{%
Park+2016}
\APACinsertmetastar {%
Park+2016}%
\begin{APACrefauthors}%
Park, R.%
, Konopliv, A.%
, Bills, B.%
, Rambaux, N.%
, Castillo-Rogez, J.%
, Raymond, C.%
\BDBL {}others%
\end{APACrefauthors}%
\unskip\
\newblock
\APACrefYearMonthDay{2016}{}{}.
\newblock
{\BBOQ}\APACrefatitle {A partially differentiated interior for (1) {C}eres
  deduced from its gravity field and shape} {A partially differentiated
  interior for (1) {C}eres deduced from its gravity field and shape}.{\BBCQ}
\newblock
\APACjournalVolNumPages{Nature}{537}{7621}{515}.
\PrintBackRefs{\CurrentBib}

\bibitem [\protect \citeauthoryear {%
C.~Pieters%
\ \protect \BOthers {.}}{%
C.~Pieters%
\ \protect \BOthers {.}}{%
{\protect \APACyear {2017}}%
}]{%
Pieters+2017}
\APACinsertmetastar {%
Pieters+2017}%
\begin{APACrefauthors}%
Pieters, C.%
, Nathues, A.%
, Thangjam, G.%
, Hoffmann, M.%
, Platz, T.%
, De~Sanctis, M.%
\BDBL {}others%
\end{APACrefauthors}%
\unskip\
\newblock
\APACrefYearMonthDay{2017}{}{}.
\newblock
{\BBOQ}\APACrefatitle {Geologic constraints on the origin of red organic-rich
  material on {C}eres} {Geologic constraints on the origin of red organic-rich
  material on {C}eres}.{\BBCQ}
\newblock
\APACjournalVolNumPages{Meteoritics \& Planetary Science}{}{}{}.
\PrintBackRefs{\CurrentBib}

\bibitem [\protect \citeauthoryear {%
C\BPBI M.~Pieters%
\ \BBA {} Noble%
}{%
C\BPBI M.~Pieters%
\ \BBA {} Noble%
}{%
{\protect \APACyear {2016}}%
}]{%
Pieters+Noble2016}
\APACinsertmetastar {%
Pieters+Noble2016}%
\begin{APACrefauthors}%
Pieters, C\BPBI M.%
\BCBT {}\ \BBA {} Noble, S\BPBI K.%
\end{APACrefauthors}%
\unskip\
\newblock
\APACrefYearMonthDay{2016}{}{}.
\newblock
{\BBOQ}\APACrefatitle {Space weathering on airless bodies} {Space weathering on
  airless bodies}.{\BBCQ}
\newblock
\APACjournalVolNumPages{Journal of Geophysical Research:
  Planets}{121}{10}{1865--1884}.
\PrintBackRefs{\CurrentBib}

\bibitem [\protect \citeauthoryear {%
Pommerol%
\ \BBA {} Schmitt%
}{%
Pommerol%
\ \BBA {} Schmitt%
}{%
{\protect \APACyear {2008}}%
}]{%
Pommerol+Schmitt2008}
\APACinsertmetastar {%
Pommerol+Schmitt2008}%
\begin{APACrefauthors}%
Pommerol, A.%
\BCBT {}\ \BBA {} Schmitt, B.%
\end{APACrefauthors}%
\unskip\
\newblock
\APACrefYearMonthDay{2008}{}{}.
\newblock
{\BBOQ}\APACrefatitle {Strength of the {H}$_2${O} near-infrared absorption
  bands in hydrated minerals: Effects of particle size and correlation with
  albedo} {Strength of the {H}$_2${O} near-infrared absorption bands in
  hydrated minerals: Effects of particle size and correlation with
  albedo}.{\BBCQ}
\newblock
\APACjournalVolNumPages{Journal of Geophysical Research: Planets}{113}{E10}{}.
\PrintBackRefs{\CurrentBib}

\bibitem [\protect \citeauthoryear {%
Prettyman%
, Yamashita%
, Ammannito%
, Castillo-Rogez%
\BCBL {}\ \protect \BOthers {.}}{%
Prettyman%
, Yamashita%
, Ammannito%
, Castillo-Rogez%
\BCBL {}\ \protect \BOthers {.}}{%
{\protect \APACyear {2018}}%
}]{%
Prettyman+2018a}
\APACinsertmetastar {%
Prettyman+2018a}%
\begin{APACrefauthors}%
Prettyman, T.%
, Yamashita, N.%
, Ammannito, E.%
, Castillo-Rogez, J.%
, Ehlmann, B.%
, McSween, H.%
\BDBL {}others%
\end{APACrefauthors}%
\unskip\
\newblock
\APACrefYearMonthDay{2018}{}{}.
\newblock
{\BBOQ}\APACrefatitle {Carbon on {C}eres: Implications for Origins and Interior
  Evolution} {Carbon on {C}eres: Implications for origins and interior
  evolution}.{\BBCQ}
\newblock
\BIn{} \APACrefbtitle {Lunar and Planetary Science Conference} {Lunar and
  planetary science conference}\ (\BVOL~49).
\PrintBackRefs{\CurrentBib}

\bibitem [\protect \citeauthoryear {%
Prettyman%
, Yamashita%
, Ammannito%
, Ehlmann%
\BCBL {}\ \protect \BOthers {.}}{%
Prettyman%
, Yamashita%
, Ammannito%
, Ehlmann%
\BCBL {}\ \protect \BOthers {.}}{%
{\protect \APACyear {2018}}%
}]{%
Prettyman+2018b}
\APACinsertmetastar {%
Prettyman+2018b}%
\begin{APACrefauthors}%
Prettyman, T.%
, Yamashita, N.%
, Ammannito, E.%
, Ehlmann, B.%
, McSween, H.%
, Mittlefehldt, D.%
\BDBL {}others%
\end{APACrefauthors}%
\unskip\
\newblock
\APACrefYearMonthDay{2018}{}{}.
\newblock
{\BBOQ}\APACrefatitle {Elemental composition and mineralogy of {V}esta and
  {C}eres: Distribution and origins of hydrogen-bearing species} {Elemental
  composition and mineralogy of {V}esta and {C}eres: Distribution and origins
  of hydrogen-bearing species}.{\BBCQ}
\newblock
\APACjournalVolNumPages{Icarus}{}{}{}.
\PrintBackRefs{\CurrentBib}

\bibitem [\protect \citeauthoryear {%
Prettyman%
\ \protect \BOthers {.}}{%
Prettyman%
\ \protect \BOthers {.}}{%
{\protect \APACyear {2017}}%
}]{%
Prettyman+2017}
\APACinsertmetastar {%
Prettyman+2017}%
\begin{APACrefauthors}%
Prettyman, T.%
, Yamashita, N.%
, Toplis, M.%
, McSween, H.%
, Sch{\"o}rghofer, N.%
, Marchi, S.%
\BDBL {}others%
\end{APACrefauthors}%
\unskip\
\newblock
\APACrefYearMonthDay{2017}{}{}.
\newblock
{\BBOQ}\APACrefatitle {Extensive water ice within {C}eres’ aqueously altered
  regolith: Evidence from nuclear spectroscopy} {Extensive water ice within
  {C}eres’ aqueously altered regolith: Evidence from nuclear
  spectroscopy}.{\BBCQ}
\newblock
\APACjournalVolNumPages{Science}{355}{6320}{55--59}.
\PrintBackRefs{\CurrentBib}

\bibitem [\protect \citeauthoryear {%
Querry%
}{%
Querry%
}{%
{\protect \APACyear {1985}}%
}]{%
Querry1985}
\APACinsertmetastar {%
Querry1985}%
\begin{APACrefauthors}%
Querry, M.%
\end{APACrefauthors}%
\unskip\
\newblock
\APACrefYearMonthDay{1985}{}{}.
\newblock
{\BBOQ}\APACrefatitle {Optical Constants, Contractor Report {CRDC-CR}-85034}
  {Optical constants, contractor report {CRDC-CR}-85034}.{\BBCQ}
\newblock

\PrintBackRefs{\CurrentBib}

\bibitem [\protect \citeauthoryear {%
Querry%
}{%
Querry%
}{%
{\protect \APACyear {1987}}%
}]{%
Querry1987}
\APACinsertmetastar {%
Querry1987}%
\begin{APACrefauthors}%
Querry, M.%
\end{APACrefauthors}%
\unskip\
\newblock
\APACrefYearMonthDay{1987}{}{}.
\newblock
\APACrefbtitle {Optical constants of minerals and other materials from the
  millimeter to the ultraviolet} {Optical constants of minerals and other
  materials from the millimeter to the ultraviolet}\ \APACbVolEdTR{}{\BTR{}}.
\newblock
\APACaddressInstitution{}{CHEMICAL RESEARCH DEVELOPMENT AND ENGINEERING CENTER
  ABERDEEN PROVING GROUNDMD}.
\PrintBackRefs{\CurrentBib}

\bibitem [\protect \citeauthoryear {%
Quirico%
\ \protect \BOthers {.}}{%
Quirico%
\ \protect \BOthers {.}}{%
{\protect \APACyear {2016}}%
}]{%
Quirico+2016}
\APACinsertmetastar {%
Quirico+2016}%
\begin{APACrefauthors}%
Quirico, E.%
, Moroz, L.%
, Schmitt, B.%
, Arnold, G.%
, Faure, M.%
, Beck, P.%
\BDBL {}others%
\end{APACrefauthors}%
\unskip\
\newblock
\APACrefYearMonthDay{2016}{}{}.
\newblock
{\BBOQ}\APACrefatitle {Refractory and semi-volatile organics at the surface of
  comet 67{P}/{C}huryumov-{G}erasimenko: insights from the {VIRTIS}/{R}osetta
  imaging spectrometer} {Refractory and semi-volatile organics at the surface
  of comet 67{P}/{C}huryumov-{G}erasimenko: insights from the
  {VIRTIS}/{R}osetta imaging spectrometer}.{\BBCQ}
\newblock
\APACjournalVolNumPages{Icarus}{272}{}{32--47}.
\PrintBackRefs{\CurrentBib}

\bibitem [\protect \citeauthoryear {%
Ralchenko%
\ \protect \BOthers {.}}{%
Ralchenko%
\ \protect \BOthers {.}}{%
{\protect \APACyear {2014}}%
}]{%
Ralchenko+2014}
\APACinsertmetastar {%
Ralchenko+2014}%
\begin{APACrefauthors}%
Ralchenko, M.%
, Britt, D.%
, Samson, C.%
, Herd, C.%
, Herd, R.%
\BCBL {}\ \BBA {} McCausland, P.%
\end{APACrefauthors}%
\unskip\
\newblock
\APACrefYearMonthDay{2014}{}{}.
\newblock
{\BBOQ}\APACrefatitle {Bulk physical properties of the {T}agish Lake meteorite
  frozen pristine fragments (abstract\# 1021)} {Bulk physical properties of the
  {T}agish lake meteorite frozen pristine fragments (abstract\# 1021)}.{\BBCQ}
\newblock
\BIn{} \APACrefbtitle {14th Lunar and Planetary Science Conference. CD-ROM.}
  {14th lunar and planetary science conference. cd-rom.}
\PrintBackRefs{\CurrentBib}

\bibitem [\protect \citeauthoryear {%
Rampe%
\ \protect \BOthers {.}}{%
Rampe%
\ \protect \BOthers {.}}{%
{\protect \APACyear {2018}}%
}]{%
Rampe+2018}
\APACinsertmetastar {%
Rampe+2018}%
\begin{APACrefauthors}%
Rampe, E.%
, Lapotre, M.%
, Bristow, T.%
, Arvidson, R.%
, Morris, R.%
, Achilles, C.%
\BDBL {}others%
\end{APACrefauthors}%
\unskip\
\newblock
\APACrefYearMonthDay{2018}{}{}.
\newblock
{\BBOQ}\APACrefatitle {Sand mineralogy within the {B}agnold {D}unes, {G}ale
  crater, as observed in situ and from orbit} {Sand mineralogy within the
  {B}agnold {D}unes, {G}ale crater, as observed in situ and from orbit}.{\BBCQ}
\newblock
\APACjournalVolNumPages{Geophysical Research Letters}{45}{18}{9488--9497}.
\PrintBackRefs{\CurrentBib}

\bibitem [\protect \citeauthoryear {%
Rivkin%
, Volquardsen%
\BCBL {}\ \BBA {} Clark%
}{%
Rivkin%
\ \protect \BOthers {.}}{%
{\protect \APACyear {2006}}%
}]{%
Rivkin+2006}
\APACinsertmetastar {%
Rivkin+2006}%
\begin{APACrefauthors}%
Rivkin, A.%
, Volquardsen, E.%
\BCBL {}\ \BBA {} Clark, B.%
\end{APACrefauthors}%
\unskip\
\newblock
\APACrefYearMonthDay{2006}{}{}.
\newblock
{\BBOQ}\APACrefatitle {The surface composition of {C}eres: Discovery of
  carbonates and iron-rich clays} {The surface composition of {C}eres:
  Discovery of carbonates and iron-rich clays}.{\BBCQ}
\newblock
\APACjournalVolNumPages{Icarus}{185}{2}{563--567}.
\PrintBackRefs{\CurrentBib}

\bibitem [\protect \citeauthoryear {%
T.~Roush%
, Pollack%
, Witteborn%
, Bregman%
\BCBL {}\ \BBA {} Simpson%
}{%
T.~Roush%
\ \protect \BOthers {.}}{%
{\protect \APACyear {1990}}%
}]{%
Roush+1990}
\APACinsertmetastar {%
Roush+1990}%
\begin{APACrefauthors}%
Roush, T.%
, Pollack, J.%
, Witteborn, F.%
, Bregman, J.%
\BCBL {}\ \BBA {} Simpson, J.%
\end{APACrefauthors}%
\unskip\
\newblock
\APACrefYearMonthDay{1990}{}{}.
\newblock
{\BBOQ}\APACrefatitle {Ice and minerals on {C}allisto: A reassessment of the
  reflectance spectra} {Ice and minerals on {C}allisto: A reassessment of the
  reflectance spectra}.{\BBCQ}
\newblock
\APACjournalVolNumPages{Icarus}{86}{2}{355--382}.
\PrintBackRefs{\CurrentBib}

\bibitem [\protect \citeauthoryear {%
T\BPBI L.~Roush%
}{%
T\BPBI L.~Roush%
}{%
{\protect \APACyear {2003}}%
}]{%
Roush2003}
\APACinsertmetastar {%
Roush2003}%
\begin{APACrefauthors}%
Roush, T\BPBI L.%
\end{APACrefauthors}%
\unskip\
\newblock
\APACrefYearMonthDay{2003}{}{}.
\newblock
{\BBOQ}\APACrefatitle {Estimated optical constants of the {T}agish {L}ake
  meteorite} {Estimated optical constants of the {T}agish {L}ake
  meteorite}.{\BBCQ}
\newblock
\APACjournalVolNumPages{Meteoritics \& Planetary Science}{38}{3}{419--426}.
\PrintBackRefs{\CurrentBib}

\bibitem [\protect \citeauthoryear {%
Rousseau%
\ \protect \BOthers {.}}{%
Rousseau%
\ \protect \BOthers {.}}{%
{\protect \APACyear {2018}}%
}]{%
Rousseau+2018}
\APACinsertmetastar {%
Rousseau+2018}%
\begin{APACrefauthors}%
Rousseau, B.%
, {\'E}rard, S.%
, Beck, P.%
, Quirico, {\'E}.%
, Schmitt, B.%
, Brissaud, O.%
\BDBL {}others%
\end{APACrefauthors}%
\unskip\
\newblock
\APACrefYearMonthDay{2018}{}{}.
\newblock
{\BBOQ}\APACrefatitle {Laboratory simulations of the {V}is-{NIR} spectra of
  comet 67{P} using sub-$\mu$m sized cosmochemical analogues} {Laboratory
  simulations of the {V}is-{NIR} spectra of comet 67{P} using sub-$\mu$m sized
  cosmochemical analogues}.{\BBCQ}
\newblock
\APACjournalVolNumPages{Icarus}{306}{}{306--318}.
\PrintBackRefs{\CurrentBib}

\bibitem [\protect \citeauthoryear {%
Rubin%
, Trigo-Rodr{\'\i}guez%
, Huber%
\BCBL {}\ \BBA {} Wasson%
}{%
Rubin%
\ \protect \BOthers {.}}{%
{\protect \APACyear {2007}}%
}]{%
Rubin+2007}
\APACinsertmetastar {%
Rubin+2007}%
\begin{APACrefauthors}%
Rubin, A\BPBI E.%
, Trigo-Rodr{\'\i}guez, J\BPBI M.%
, Huber, H.%
\BCBL {}\ \BBA {} Wasson, J\BPBI T.%
\end{APACrefauthors}%
\unskip\
\newblock
\APACrefYearMonthDay{2007}{}{}.
\newblock
{\BBOQ}\APACrefatitle {Progressive aqueous alteration of {CM} carbonaceous
  chondrites} {Progressive aqueous alteration of {CM} carbonaceous
  chondrites}.{\BBCQ}
\newblock
\APACjournalVolNumPages{Geochimica et Cosmochimica Acta}{71}{9}{2361--2382}.
\PrintBackRefs{\CurrentBib}

\bibitem [\protect \citeauthoryear {%
Schaepman-Strub%
, Schaepman%
, Painter%
, Dangel%
\BCBL {}\ \BBA {} Martonchik%
}{%
Schaepman-Strub%
\ \protect \BOthers {.}}{%
{\protect \APACyear {2006}}%
}]{%
Schaepman-Strub+2006}
\APACinsertmetastar {%
Schaepman-Strub+2006}%
\begin{APACrefauthors}%
Schaepman-Strub, G.%
, Schaepman, M.%
, Painter, T\BPBI H.%
, Dangel, S.%
\BCBL {}\ \BBA {} Martonchik, J\BPBI V.%
\end{APACrefauthors}%
\unskip\
\newblock
\APACrefYearMonthDay{2006}{}{}.
\newblock
{\BBOQ}\APACrefatitle {Reflectance quantities in optical remote
  sensing--{D}efinitions and case studies} {Reflectance quantities in optical
  remote sensing--{D}efinitions and case studies}.{\BBCQ}
\newblock
\APACjournalVolNumPages{Remote sensing of environment}{103}{1}{27--42}.
\PrintBackRefs{\CurrentBib}

\bibitem [\protect \citeauthoryear {%
Sch{\"a}fer%
\ \protect \BOthers {.}}{%
Sch{\"a}fer%
\ \protect \BOthers {.}}{%
{\protect \APACyear {2018}}%
}]{%
Schafer+2018}
\APACinsertmetastar {%
Schafer+2018}%
\begin{APACrefauthors}%
Sch{\"a}fer, M.%
, Sch{\"a}fer, T.%
, Izawa, M\BPBI R.%
, Cloutis, E\BPBI A.%
, Schr{\"o}der, S\BPBI E.%
, Roatsch, T.%
\BDBL {}others%
\end{APACrefauthors}%
\unskip\
\newblock
\APACrefYearMonthDay{2018}{}{}.
\newblock
{\BBOQ}\APACrefatitle {Ceres' spectral link to carbonaceous
  chondrites―Analysis of the dark background materials} {Ceres' spectral link
  to carbonaceous chondrites―analysis of the dark background
  materials}.{\BBCQ}
\newblock
\APACjournalVolNumPages{Meteoritics \& Planetary Science}{}{}{}.
\PrintBackRefs{\CurrentBib}

\bibitem [\protect \citeauthoryear {%
Shkuratov%
, Starukhina%
, Hoffmann%
\BCBL {}\ \BBA {} Arnold%
}{%
Shkuratov%
\ \protect \BOthers {.}}{%
{\protect \APACyear {1999}}%
}]{%
Shkuratov+1999}
\APACinsertmetastar {%
Shkuratov+1999}%
\begin{APACrefauthors}%
Shkuratov, Y.%
, Starukhina, L.%
, Hoffmann, H.%
\BCBL {}\ \BBA {} Arnold, G.%
\end{APACrefauthors}%
\unskip\
\newblock
\APACrefYearMonthDay{1999}{}{}.
\newblock
{\BBOQ}\APACrefatitle {A model of spectral albedo of particulate surfaces:
  Implications for optical properties of the {M}oon} {A model of spectral
  albedo of particulate surfaces: Implications for optical properties of the
  {M}oon}.{\BBCQ}
\newblock
\APACjournalVolNumPages{Icarus}{137}{2}{235--246}.
\PrintBackRefs{\CurrentBib}

\bibitem [\protect \citeauthoryear {%
Stein%
\ \protect \BOthers {.}}{%
Stein%
\ \protect \BOthers {.}}{%
{\protect \APACyear {2017}}%
}]{%
Stein+2017}
\APACinsertmetastar {%
Stein+2017}%
\begin{APACrefauthors}%
Stein, N.%
, Ehlmann, B.%
, Palomba, E.%
, De~Sanctis, M.%
, Nathues, A.%
, Hiesinger, H.%
\BDBL {}others%
\end{APACrefauthors}%
\unskip\
\newblock
\APACrefYearMonthDay{2017}{}{}.
\newblock
{\BBOQ}\APACrefatitle {The formation and evolution of bright spots on {C}eres}
  {The formation and evolution of bright spots on {C}eres}.{\BBCQ}
\newblock
\APACjournalVolNumPages{Icarus}{}{}{}.
\PrintBackRefs{\CurrentBib}

\bibitem [\protect \citeauthoryear {%
Sugita%
\ \protect \BOthers {.}}{%
Sugita%
\ \protect \BOthers {.}}{%
{\protect \APACyear {2019}}%
}]{%
Sugita+2019}
\APACinsertmetastar {%
Sugita+2019}%
\begin{APACrefauthors}%
Sugita, S.%
, Honda, R.%
, Morota, T.%
, Kameda, S.%
, Sawada, H.%
, Tatsumi, E.%
\BDBL {}others%
\end{APACrefauthors}%
\unskip\
\newblock
\APACrefYearMonthDay{2019}{}{}.
\newblock
{\BBOQ}\APACrefatitle {The geomorphology, color, and thermal properties of
  {R}yugu: Implications for parent-body processes} {The geomorphology, color,
  and thermal properties of {R}yugu: Implications for parent-body
  processes}.{\BBCQ}
\newblock
\APACjournalVolNumPages{Science}{364}{6437}{}.
\PrintBackRefs{\CurrentBib}

\bibitem [\protect \citeauthoryear {%
Takir%
\ \BBA {} Emery%
}{%
Takir%
\ \BBA {} Emery%
}{%
{\protect \APACyear {2012}}%
}]{%
Takir+Emery2012}
\APACinsertmetastar {%
Takir+Emery2012}%
\begin{APACrefauthors}%
Takir, D.%
\BCBT {}\ \BBA {} Emery, J\BPBI P.%
\end{APACrefauthors}%
\unskip\
\newblock
\APACrefYearMonthDay{2012}{}{}.
\newblock
{\BBOQ}\APACrefatitle {Outer main belt asteroids: Identification and
  distribution of four 3-$\mu$m spectral groups} {Outer main belt asteroids:
  Identification and distribution of four 3-$\mu$m spectral groups}.{\BBCQ}
\newblock
\APACjournalVolNumPages{Icarus}{219}{2}{641--654}.
\PrintBackRefs{\CurrentBib}

\bibitem [\protect \citeauthoryear {%
Takir%
\ \protect \BOthers {.}}{%
Takir%
\ \protect \BOthers {.}}{%
{\protect \APACyear {2013}}%
}]{%
Takir+2013}
\APACinsertmetastar {%
Takir+2013}%
\begin{APACrefauthors}%
Takir, D.%
, Emery, J\BPBI P.%
, Mcsween~Jr, H\BPBI Y.%
, Hibbitts, C\BPBI A.%
, Clark, R\BPBI N.%
, Pearson, N.%
\BCBL {}\ \BBA {} Wang, A.%
\end{APACrefauthors}%
\unskip\
\newblock
\APACrefYearMonthDay{2013}{}{}.
\newblock
{\BBOQ}\APACrefatitle {Nature and degree of aqueous alteration in {CM} and {CI}
  carbonaceous chondrites} {Nature and degree of aqueous alteration in {CM} and
  {CI} carbonaceous chondrites}.{\BBCQ}
\newblock
\APACjournalVolNumPages{Meteoritics \& Planetary Science}{48}{9}{1618--1637}.
\PrintBackRefs{\CurrentBib}

\bibitem [\protect \citeauthoryear {%
Vokrouhlick{\`y}%
, Bottke%
\BCBL {}\ \BBA {} Nesvorn{\`y}%
}{%
Vokrouhlick{\`y}%
\ \protect \BOthers {.}}{%
{\protect \APACyear {2016}}%
}]{%
Vokrouhlicky+2016}
\APACinsertmetastar {%
Vokrouhlicky+2016}%
\begin{APACrefauthors}%
Vokrouhlick{\`y}, D.%
, Bottke, W\BPBI F.%
\BCBL {}\ \BBA {} Nesvorn{\`y}, D.%
\end{APACrefauthors}%
\unskip\
\newblock
\APACrefYearMonthDay{2016}{}{}.
\newblock
{\BBOQ}\APACrefatitle {Capture of trans-{N}eptunian planetesimals in the main
  asteroid belt} {Capture of trans-{N}eptunian planetesimals in the main
  asteroid belt}.{\BBCQ}
\newblock
\APACjournalVolNumPages{The Astronomical Journal}{152}{2}{39}.
\PrintBackRefs{\CurrentBib}

\bibitem [\protect \citeauthoryear {%
K.~Walsh%
\ \protect \BOthers {.}}{%
K.~Walsh%
\ \protect \BOthers {.}}{%
{\protect \APACyear {2019}}%
}]{%
Walsh+2019}
\APACinsertmetastar {%
Walsh+2019}%
\begin{APACrefauthors}%
Walsh, K.%
, Jawin, E.%
, Ballouz, R\BHBI L.%
, Barnouin, O.%
, Bierhaus, E.%
, Connolly, H.%
\BDBL {}others%
\end{APACrefauthors}%
\unskip\
\newblock
\APACrefYearMonthDay{2019}{}{}.
\newblock
{\BBOQ}\APACrefatitle {Craters, boulders and regolith of (101955) {B}ennu
  indicative of an old and dynamic surface} {Craters, boulders and regolith of
  (101955) {B}ennu indicative of an old and dynamic surface}.{\BBCQ}
\newblock
\APACjournalVolNumPages{Nature Geoscience}{12}{4}{242--246}.
\PrintBackRefs{\CurrentBib}

\bibitem [\protect \citeauthoryear {%
K\BPBI J.~Walsh%
, Morbidelli%
, Raymond%
, O'brien%
\BCBL {}\ \BBA {} Mandell%
}{%
K\BPBI J.~Walsh%
\ \protect \BOthers {.}}{%
{\protect \APACyear {2011}}%
}]{%
Walsh+2011}
\APACinsertmetastar {%
Walsh+2011}%
\begin{APACrefauthors}%
Walsh, K\BPBI J.%
, Morbidelli, A.%
, Raymond, S\BPBI N.%
, O'brien, D\BPBI P.%
\BCBL {}\ \BBA {} Mandell, A\BPBI M.%
\end{APACrefauthors}%
\unskip\
\newblock
\APACrefYearMonthDay{2011}{}{}.
\newblock
{\BBOQ}\APACrefatitle {A low mass for {M}ars from {J}upiter's early gas-driven
  migration} {A low mass for {M}ars from {J}upiter's early gas-driven
  migration}.{\BBCQ}
\newblock
\APACjournalVolNumPages{Nature}{475}{7355}{206}.
\PrintBackRefs{\CurrentBib}

\bibitem [\protect \citeauthoryear {%
Watanabe%
\ \protect \BOthers {.}}{%
Watanabe%
\ \protect \BOthers {.}}{%
{\protect \APACyear {2019}}%
}]{%
Watanabe+2019}
\APACinsertmetastar {%
Watanabe+2019}%
\begin{APACrefauthors}%
Watanabe, S.%
, Hirabayashi, M.%
, Hirata, N.%
, Hirata, N.%
, Noguchi, R.%
, Shimaki, Y.%
\BDBL {}others%
\end{APACrefauthors}%
\unskip\
\newblock
\APACrefYearMonthDay{2019}{}{}.
\newblock
{\BBOQ}\APACrefatitle {Hayabusa2 arrives at the carbonaceous asteroid 162173
  {R}yugu―A spinning top--shaped rubble pile} {Hayabusa2 arrives at the
  carbonaceous asteroid 162173 {R}yugu―a spinning top--shaped rubble
  pile}.{\BBCQ}
\newblock
\APACjournalVolNumPages{Science}{364}{6437}{268--272}.
\PrintBackRefs{\CurrentBib}

\bibitem [\protect \citeauthoryear {%
Wolf%
\ \BBA {} Palme%
}{%
Wolf%
\ \BBA {} Palme%
}{%
{\protect \APACyear {2001}}%
}]{%
Wolf+Palme2001}
\APACinsertmetastar {%
Wolf+Palme2001}%
\begin{APACrefauthors}%
Wolf, D.%
\BCBT {}\ \BBA {} Palme, H.%
\end{APACrefauthors}%
\unskip\
\newblock
\APACrefYearMonthDay{2001}{}{}.
\newblock
{\BBOQ}\APACrefatitle {The solar system abundances of phosphorus and titanium
  and the nebular volatility of phosphorus} {The solar system abundances of
  phosphorus and titanium and the nebular volatility of phosphorus}.{\BBCQ}
\newblock
\APACjournalVolNumPages{Meteoritics \& Planetary Science}{36}{4}{559--571}.
\PrintBackRefs{\CurrentBib}

\bibitem [\protect \citeauthoryear {%
Zolensky%
\ \protect \BOthers {.}}{%
Zolensky%
\ \protect \BOthers {.}}{%
{\protect \APACyear {2002}}%
}]{%
Zolensky+2002}
\APACinsertmetastar {%
Zolensky+2002}%
\begin{APACrefauthors}%
Zolensky, M.%
, Nakamura, K.%
, Gounelle, M.%
, Mikouchi, T.%
, Kasama, T.%
, Tachikawa, O.%
\BCBL {}\ \BBA {} Tonui, E.%
\end{APACrefauthors}%
\unskip\
\newblock
\APACrefYearMonthDay{2002}{}{}.
\newblock
{\BBOQ}\APACrefatitle {Mineralogy of {T}agish {L}ake: An ungrouped type 2
  carbonaceous chondrite} {Mineralogy of {T}agish {L}ake: An ungrouped type 2
  carbonaceous chondrite}.{\BBCQ}
\newblock
\APACjournalVolNumPages{Meteoritics \& Planetary Science}{37}{5}{737--761}.
\PrintBackRefs{\CurrentBib}

\bibitem [\protect \citeauthoryear {%
Zubko%
, Mennella%
, Colangeli%
\BCBL {}\ \BBA {} Bussoletti%
}{%
Zubko%
\ \protect \BOthers {.}}{%
{\protect \APACyear {1996}}%
}]{%
Zubko+1996}
\APACinsertmetastar {%
Zubko+1996}%
\begin{APACrefauthors}%
Zubko, V.%
, Mennella, V.%
, Colangeli, L.%
\BCBL {}\ \BBA {} Bussoletti, E.%
\end{APACrefauthors}%
\unskip\
\newblock
\APACrefYearMonthDay{1996}{}{}.
\newblock
{\BBOQ}\APACrefatitle {Optical constants of cosmic carbon analogue grains --
  {I}. {S}imulation of clustering by a modified continuous distribution of
  ellipsoids} {Optical constants of cosmic carbon analogue grains -- {I}.
  {S}imulation of clustering by a modified continuous distribution of
  ellipsoids}.{\BBCQ}
\newblock
\APACjournalVolNumPages{Monthly Notices of the Royal Astronomical
  Society}{282}{4}{1321--1329}.
\PrintBackRefs{\CurrentBib}

\end{thebibliography}

\end{document}